\documentclass[prb,twocolumn,preprintnumbers,amsmath,amssymb,superscriptaddress]{revtex4}
\usepackage{graphicx}
\usepackage{latexsym}
\usepackage{amsmath}
\usepackage{graphics}
\usepackage{amssymb}
\usepackage{layout}
\usepackage{verbatim}
\usepackage{amsfonts,epsfig}
\newcommand{\ket}[1]{\left|#1\right>}
\newcommand{\bra}[1]{\left< #1 \right|}
\newcommand{\beq}{\begin{equation}}
\newcommand{\eeq}{\end{equation}}
\newcommand{\bea}{\begin{eqnarray}}
\newcommand{\eea}{\end{eqnarray}}
\newcommand{\nn}{\nonumber}
\newcommand{\tr}{\hbox{Tr}}

\newcommand{\mean}[1]{\langle{#1}\rangle{}}

\newcommand{\A}{\mathcal{A}}

\begin{document}

\title{Master equation approach to the central spin decoherence problem: the uniform coupling model and the role of projection operators}
\author{Edwin Barnes}
\email{barnes@umd.edu}
\affiliation{Condensed Matter Theory Center, Department of Physics, University of Maryland, College Park, MD 20742-4111, USA}
\author{{\L}ukasz Cywi{\'n}ski}
\email{lcyw@ifpan.edu.pl}
\affiliation{Institute of Physics, Polish Academy of Sciences, al.~Lotnik{\'o}w 32/46, PL 02-668 Warszawa, Poland}
\author{S. Das Sarma}
\affiliation{Condensed Matter Theory Center, Department of Physics, University of Maryland, College Park, MD 20742-4111, USA}
\date{\today }

\begin{abstract}
The generalized Master equation of the Nakajima-Zwanzig (NZ) type has been used extensively to investigate the coherence dynamics of the central spin model with the nuclear bath in a narrowed state characterized by a well defined value of the Overhauser field. We revisit the perturbative NZ approach and apply it to the exactly solvable case of a system with uniform hyperfine couplings. This is motivated by the fact that the effective Hamiltonian-based theory suggests that the dynamics of the realistic system at low magnetic fields and short times can be mapped onto the uniform coupling model. We show that the standard NZ approach fails to reproduce the exact solution of this model beyond very short times, while the effective Hamiltonian calculation agrees very well with the exact result on timescales during which most of the coherence is lost.
Our key finding is that in order to extend the timescale of applicability of the NZ approach in this case, instead of using a single projection operator one has to use a set of correlated projection operators which properly reflect the symmetries of the problem and greatly improve the convergence of the theory. This suggests that the correlated projection operators are crucial for a proper description of narrowed state free induction decay at short times and low magnetic fields.
Our results thus provide important insights toward the development of a more complete theory of central spin decoherence applicable in a broader regime of timescales and magnetic fields.
\end{abstract}

\maketitle

\section{Introduction}
Qubits based on spins of single electrons confined in a semiconductor environment\cite{Hanson_RMP07,Liu_AP10} are one of most promising platforms for quantum computation. When the host semiconductor is a III-V compound material (i.e.~GaAs or InAs), the presence of the nuclear spins, coupled to the electron (or a hole) by hyperfine (hf) interaction, is unavoidable.\cite{Coish_pssb09,Cywinski_APPA11} The qubit decoherence induced by hf interaction with such a nuclear bath has been a subject of many theoretical works. For this paper, the directly relevant ones are those focusing on purely hf-induced  dynamics,\cite{Khaetskii_PRL02,Schliemann_PRB02,Khaetskii_PRB03,Dobrovitski_PRE03,Coish_PRB04,Erlingsson_PRB04,Shenvi_scaling_PRB05,Yao_PRB06,Zhang_PRB06,Al_Hassanieh_PRL06,Deng_PRB06,Liu_NJP07,Saikin_PRB07,Chen_PRB07,Bortz_PRB07,Bortz_JSM07,Deng_PRB08,Fischer_PRA07,Coish_PRB08,Ferraro_PRB08,Cywinski_PRL09,Cywinski_PRB09,Cywinski_PRB10,Coish_PRB10,Fischer_PRL10,Bortz_PRB10,Neder_PRB11}
 i.e.~on the problem described by the central spin Hamiltonian
\beq
\hat{H} = \sum_{k}A_{k} \mathbf{S}\cdot \mathbf{I}_{k} + \Omega S^{z} + \sum_{k} \omega_{k} I^{z}_{k} \,\, ,  \label{eq:H}
\eeq
with $\mathbf{S}$ denoting the central spin,  $\mathbf{I}_{k}$ the nuclear spins, $A_{k}$ the hf couplings, and $\Omega$ and $\omega_{k}$ the central spin and nuclear spin Zeeman energies, respectively. This Hamiltonian can be solved exactly via Bethe ansatz\cite{Gaudin,Bortz_PRB07,Faribault_PRB11} for the case of uniform $\omega_k \! = \! \omega$, where $\omega$ can be removed by going to a rotating frame.\cite{Coish_PRB04} However, extracting potentially experimentally relevant information about the dynamics of the system, for example the time dependence of the reduced density matrix of the electron spin after initializing it in a pure state, is very hard when starting from the exact eigenstates,\cite{Yuzbashyan_JPA05,Bortz_PRB07,Bortz_PRB10} and the largest system sizes considered had up to only $N\! = \! 30$ nuclear spins.\cite{Bortz_PRB07}
Numerical calculations of the system's evolution can be also obtained using the Chebyshev polynomial method,\cite{Dobrovitski_PRE03,Zhang_JPC07} which has been used to calculate the free evolution\cite{Zhang_PRB06} and spin echo\cite{Cywinski_PRB10} signals for $N \! \leq \! 20$. Larger systems with $N\! \sim \! 10^{4}$ were investigated using an approximate spin-coherent-state P representation,\cite{Al_Hassanieh_PRL06,Zhang_JPC07} which, however, can only be used in the case of an unpolarized thermal nuclear bath, which is too restrictive for many applications including the considerations presented here.
In order to calculate the dynamics of the central spin on experimentally relevant timescales, and for experimentally relevant system sizes ($N \! \sim \! 10^{4} - 10^{6}$ in III-V compound quantum dots), one has to resort to approximate analytical methods, which include the use of the Generalized Master Equation (GME) of the Nakajima-Zwanzig (NZ) type,\cite{Coish_PRB04,Fischer_PRA07,Coish_PRB08,Coish_PRB10} the GME of the time-convolutionless (TCL) kind,\cite{Fischer_PRA07,Ferraro_PRB08,Ferraro_PS10} and the cluster-expansion type theories using the effective pure dephasing Hamiltonian,\cite{Yao_PRB06,Cywinski_PRL09,Cywinski_PRB09,Cywinski_PRB10,Cywinski_APPA11,Ferraro_PS10}  obtained perturbatively from the original Hamiltonian of Eq.~(\ref{eq:H}) by an appropriate canonical transformation.

In this paper we focus on the NZ theory applied to the case of \emph{narrowed state} free induction decay (NFID), i.e.~on the situation in which the nuclear system is prepared in an eigenstate of the Overhauser operator $h^{z} \equiv \sum_{k}A_{k}I^{z}_{k}$. This case is both important for quantum computation applications and experimentally relevant, since significant progress in nuclear state narrowing has been made recently\cite{Greilich_Science06,Latta_NP09,Barthel_PRL09,Vink_NP09,Foletti_NP09} (see also Ref.~\onlinecite{Coish_pssb09} and references therein). Furthermore, the high magnetic field regime of NFID (defined by the condition of $\Omega \! > \! \mathcal{A}$ where $\mathcal{A} \! \equiv \! \sum_{k}A_{k} \! \approx \! 90$ $\mu$eV in GaAs) was extensively investigated using the NZ approach.\cite{Coish_PRB04,Coish_PRB08,Coish_PRB10} Let us note that the  other GME approaches, such as the TCL formalism and, more importantly in the context of this paper, the master equations employing the correlated projection operators,\cite{Fischer_PRA07,Ferraro_PRB08,Ferraro_PS10} have been until now only employed in the case of thermal (non-narrowed) nuclear baths.

Let us briefly recount the main results of the NZ theory for the case of NFID. In the NZ approach, the equation of motion for the reduced density matrix of the central spin has an integro-differential form \cite{Fick,Breuer} with the memory kernel explicitly showing the influence of the history of the spin on its dynamics at a given point in time (i.e.~the non-Markovian effects). While this feature is physically appealing and offers a natural setting for investigation of the Markovian approximation and the corrections to it arising from the temporal nonlocality of the memory kernel, the NZ approach is technically very demanding. In the case of the central spin model, the memory kernel is expanded in powers of the flip-flop operator, $V_{\text{ff}} \sim S^{\pm}I^{\mp}$, and the expansion becomes very complicated with increasing order of perturbation theory. It also does not have any diagrammatic structure since it is not based on expanding a generalized (i.e.~time- or contour-ordered) exponential, which would allow one to write down more easily higher-order contributions and identify the most important classes of terms at each order. Furthermore, the integro-differential equation is solved via Laplace transform, and the final result is obtained by inverse transform of a rather complicated expression.

Despite these obstacles, large progress has been made with this method. The 2nd order solution\cite{Coish_PRB04} was shown to reproduce the exact solution for the fully polarized bath case,\cite{Khaetskii_PRB03} signifying the important role played by nuclear polarization (which limits the phase space for flip-flops) in improving the convergence of the NZ approach. For a general polarization, the 2nd order result was shown to lead to a very small coherence decay at high $B$ fields satisfying $\mathcal{A}/\Omega \! \ll \! 1$. Specifically, the fraction of coherence lost was only about $\mathcal{A}^{2}/N\Omega^2 \! \ll \! 1$. A complete decay of the transverse electron spin components was obtained after going to the 4th order of the flip-flop expansion.\cite{Coish_PRB08,Coish_PRB10} There, a controlled solution was obtained for $\mathcal{A}/\Omega \! < \! 1$ in the case of an unpolarized bath: after an initial quadratic decay shoulder the decay was of the exponential form $\exp(-t/T_{2})$, with $T_{2} \! \sim \! N\Omega^{2}/\mathcal{A}^3$, with the $1/t^2$ tail appearing at very long times. Recently it was shown\cite{Coish_PRB10} that with finite nuclear polarization, and for nuclear spin $I\! <\! 1$, it is possible to extend the 4th order solution to lower $B$ fields (down to $\mathcal{A}/\Omega \approx 1$ for unpolarized nuclei). As $\Omega$ is decreased, corrections to the above formula for $T_{2}$ appear, and at $\mathcal{A}/\Omega \! \approx \! 1$ a minimum of $T_{2}$ and a new kind of coherence envelope oscillation were predicted.\cite{Coish_PRB10}

It has to be stressed that most of the above features of the high-field NZ solution, especially the exponential decay, arise due to the inhomogeneous coupling of the nuclei to the central spin.
In the 4th order in electron-nuclear flip-flops, the processes in which pairs of remote nuclei flip-flop among each other contribute to the system's dynamics. In the effective Hamiltonian language,\cite{Yao_PRB06,Liu_NJP07,Coish_PRB08,Cywinski_PRL09,Cywinski_PRB09} one can say that the Overhauser field fluctuates due to electron-mediated flip-flops of these remote nuclei. At \emph{short times} defined by $t \! \ll \! N/\mathcal{A} \sim 1/A_{k}$ the differences of Knight shifts $A_{kl}\! \equiv \! A_{k}- A_{l}$ of these nuclei are unimportant due to the energy-time uncertainty, leading to the independence of the quadratic decay shoulder on the wavefunction shape.\cite{Coish_PRB10} At longer times, the Knight shifts must fulfill more stringent energy conservation conditions, giving rise to the long-time Markovian exponential decay, with the $T_{2}$ time strongly dependent on the shape of the wavefunction.\cite{Coish_PRB08} (Compare also the expressions for $T_{2}$ from Refs.~\onlinecite{Coish_PRB08} and \onlinecite{Cywinski_PRB09} which correspond, respectively, to the two-dimensional Gaussian $e^{-(x^2+y^2)/a^2}$ and the same function modified by a cosine form factor in the $z$ direction).

An insight into the importance of the short-time regime is provided by theories based on an effective-Hamiltonian, specifically the Ring Diagram Theory (RDT), in which a class of diagrams of leading order in $1/N$ is resummed in the cluster expansion of the electron's density matrix.\cite{Cywinski_PRL09,Cywinski_PRB09,Cywinski_APPA11} According to these theories, for low fields such that $\mathcal{A}/\sqrt{N} \! \ll \! \Omega \! < \! \mathcal{A}$,  most of the coherence decay occurs at short times, at which the distribution of the hf couplings does not matter, and only the quantities $\mathcal{A}$ and $N$ are important. This suggests that the low-field dynamics of NFID in a realistic system is closely related to dynamics in a system with \emph{uniform} hf couplings.

In RDT, one starts with an approximate effective Hamiltonian derived in the 2nd order of perturbation theory with respect to the flip-flop operator.\cite{Shenvi_scaling_PRB05,Yao_PRB06,Coish_PRB08,Cywinski_PRB09} While this step makes it harder to precisely ascertain the limits of applicability of this approach, it allows for a very convenient formal simplification of the problem. The effective Hamiltonian is of the pure dephasing form, which allows for the use of standard tools of non-equilibrium quantum dynamics, such as the closed time-loop contour and the cluster expansion of the bath average of a generalized exponent.\cite{Saikin_PRB07,Cywinski_PRL09,Cywinski_PRB09} Such methods were previously used in order to calculate decoherence for many other models of quantum baths,\cite{Makhlin_PRL04,Grishin_PRB05,Lutchyn_PRB08} however in the spin bath case there are additional complications due to the lack of a simple Wick's theorem for spin operators. (See Refs.~\onlinecite{Saikin_PRB07} and \onlinecite{Yang_PRB08} for the derivation of all the diagrams in the 4th order of perturbation theory with respect to inter-nuclear interactions.) However, for the electron-mediated (or hf-mediated) interactions, due to the fact that they couple all the $N$ nuclear spins with comparable strength, one can resum all the leading diagrams in the $1/N$ expansion, which amounts to calculating only one diagram at each order of the cluster expansion.

For NFID at large $B$ fields ($\mathcal{A}/\Omega \! \ll \! 1$), this resummation leads to an exponential decay with the same $T_{2}$ as the one obtained from the NZ approach\cite{Coish_PRB08,Coish_PRB10} (albeit without the the $1/t^2$ long-time tail), down to $\mathcal{A}/\Omega \! \approx \! 1$ at which the minimum of $T_{2}$ is not reproduced. Within RDT one can calculate the short-time decoherence also at low magnetic fields obeying only the condition $\mathcal{A}/\sqrt{N}\Omega \! \ll \! 1$.
The spin echo signal predicted by RDT\cite{Cywinski_PRL09,Cywinski_PRB09} was later seen in experiments on double GaAs quantum dots,\cite{Bluhm_NP11} and it was recently rederived using an explicitly semi-classical approach.\cite{Neder_PRB11} For NFID, a decay envelope of the form $(1+(t/\tau)^{2})^{-1/2}$, with $\tau \! \sim \! N\Omega/\mathcal{A}^{2}$, and a $\pi/2$ phase shift of the electron precession for $t \! \gg \! \tau$ were predicted.\cite{Cywinski_PRB09} As we show explicitly in this paper, this result arises also as the large $N$ limit of the exact solution of the fully quantum calculation for the system with uniform hf couplings.

In this paper we apply the standard 4th order NZ theory\cite{Coish_PRB04,Coish_PRB08,Coish_PRB10} to the otherwise exactly solvable model with uniform hf couplings. We find that the NZ calculation of NFID disagrees with the exact result beyond a very short timescale (at which the coherence barely decays), while the RDT calculation reproduces the envelope of the NFID signal very well. We show that in order to improve the performance of the NZ theory, one needs to replace the commonly used single projection operator by a family of projection operators on nuclear subspaces which are singled out by the electron-nuclear coupling.
Such an approach was used previously in both the NZ and TCL generalized master equation theories in the simpler case of a thermal nuclear bath,\cite{Fischer_PRA07,Ferraro_PRB08,Ferraro_PS10} but it has not been used until now for the \emph{narrowed} FID case. This result strongly suggests that the main physics of the short-time and long-time central spin decoherence is significantly different, with simple dephasing by distinct nuclear states dominating at short times, and nontrivial dynamics of inhomogeneously coupled nuclei dominating at long times. The use of the correlated projection operators allows the NZ theory to fully capture the short-time dynamics in a wide range of magnetic fields.

Let us briefly note that there exists a family of theories investigating the problem of dipolarly-induced electron spin dephasing,\cite{deSousa_PRB03,Witzel_PRB05,Witzel_PRB06,Yao_PRB06,Saikin_PRB07,Witzel_PRB08,Yang_PRB08,Witzel_PRL10}  in which the dipolarly-induced flip-flops between the nuclear spins lead to fluctuations of the Overhauser field. These theories, which are all based on some version of a linked cluster (cumulant expansion) theorem, are in agreement with the spin echo measurements in phosphorus doped silicon,\cite{Abe_PRB10} bismuth doped silicon,\cite{George_PRL10} and in GaAs singlet-triplet qubits at magnetic fields higher than $B \approx \! 0.3$ T.\cite{Bluhm_NP11} The dipolar processes are indeed expected to dominate the decoherence at high $B$ since the increasing qubit energy splitting is suppressing the dynamical processes due to hf interaction, while the  dipolar flip-flops of the nuclei are practically unaffected by $B$.
These experiments suggest that the window of parameters in which pure hf interactions dominate the electron spin decoherence is rather small in GaAs and Si. In fact, the spin echo results in large GaAs dots\cite{Bluhm_NP11} suggest that purely hf effects could be seen in lateral GaAs dots only at $B\! < \! 0.3$ T.
However, it was suggested that in small and strained InGaAs quantum dots the dipolar interactions between the nuclei might be suppressed by strongly inhomogeneous Knight shifts and by the quadrupolar interactions.\cite{Coish_PRB10} Experimental results suggesting suppression of dipolarly-induced spin diffusion in such dots have appeared recently.\cite{Latta_arxiv11}

Nothwithstanding the future experimental developments, which hopefully will allow for measurements of purely hf-induced decoherence in a wide range of $B$ fields, spanning the short-time semiclassical  and long-time Markovian regimes (with possible nontrivial crossover between them suggested by results of Ref.~\onlinecite{Coish_PRB10}), the central spin decoherence problem remains theoretically interesting in itself. Its further investigation will hopefully lead to establishing deeper connections between various approaches known from the theory of open quantum systems, and the properties of the exact Bethe ansatz solution.

The paper is organized in the following way. In Section \ref{sec:H_NZ}, we introduce the hyperfine Hamiltonian and give the explicit form of the initial nuclear density matrix characterizing NFID. In Section \ref{sec:exact_uniform}, we present the exact solution of NFID for the uniform coupling model. Section \ref{sec:NZ} offers an extensive review of the Nakajima-Zwanzig Master equation treatment of the central spin model, with the electron spin coherence computed up to 4th order in the hyperfine flip-flop interaction. This section, together with detailed appendices \ref{app:4thorderSigma} and \ref{app:correlators}, can be mostly skipped by a reader deeply familiar with previous derivations of this theory.\cite{Coish_PRB04,Coish_PRB10} However, we hope that our derivations, apart from making the paper self-contained, are useful additions to the existing works, clarifying some technical issues and expanding the discussion of certain aspects of a highly technical theory. In the subsequent Section \ref{sec:NZ_box}, we present a derivation of NFID in the uniform coupling model using the 4th order NZ approach and show that the results disagree with the exact solution except at very early times. We demonstrate that the strong discrepancy between the exact and NZ solutions can be lifted by replacing the standard single projection operator in the NZ theory with a family of correlated projection (CP) operators, and that the modified NZ theory (which we refer to as the NZ-CP theory) then gives results which are indistinguishable from the exact solution. We further show that the NZ-CP result is closely related to and improves upon that which is obtained by applying RDT to the uniform coupling model. The reasons for this behavior of the NZ theory, together with the implications of these findings for the coherence decay in realistic systems with non-uniform couplings, are discussed in Section \ref{sec:discussion}.

%%%%%%%%%%%%%%%%%%%%%%%%%%%%%%%%%%%%
%%% HAMILTONIAN AND NUCLEAR BATH STATE
%%%%%%%%%%%%%%%%%%%%%%%%%%%%%%%%%%%%
\section{HAMILTONIAN AND INITIAL NUCLEAR BATH STATE} \label{sec:H_NZ}
The Hamiltonian for a single electron spin coupled to a nuclear spin bath via the hyperfine interaction has the form
\bea
H&=&H_0+V_{\text{ff}},\label{ham}\\
H_0&=&\Omega S^z+\sum_k\omega_k I_k^z+S^zh^z,\label{Hzero}\\
V_{\text{ff}}&=&{1\over2}\left(h^+S^-+h^-S^+\right),\label{Vff}
\eea
with
\beq
h^i\equiv \sum_k A_k I_k^i.\label{hi}
\eeq
The $S^i$ are the components of the electron spin operator, while the $I^i_k$ are the components of the spin operator for the $k$th nucleus. The electron spin raising and lowering operators $S^{\pm}$ and their nuclear counterparts are defined in the usual way:
\beq
S^{\pm}=S^x\pm i S^y,\qquad I_k^{\pm}=I_k^x\pm i I_k^y.
\eeq
We have included a magnetic field which points in the $z$-direction, and $\Omega$ and $\omega_k$ are the Zeeman energies of the electron and nuclei. In order to properly describe the nuclear bath in III-V semiconductors, we should allow for several different isotopic species of nuclei. Rather than introduce an extra index to label distinct nuclear species, we have absorbed this index into the site index $k$ on the hyperfine couplings $A_k$ and nuclear Zeeman energies $\omega_k$. These quantities will in general depend on the nuclear species since different species will have different gyromagnetic factors. For example, the hyperfine couplings are given by $A_k=\nu_0{\cal A}_\alpha |\Psi(r_k)|^2$, where $\nu_0$ is the volume of the primitive unit cell, $\Psi(r_k)$ is the electron envelope wavefunction evaluated at the position $r_k$ of the $k$th nucleus, and ${\cal A}_\alpha$ is the total hyperfine interaction energy between the electron and a nucleus of species $\alpha$ and depends on the electron and nuclear spin gyromagnetic factors.

The sums in Eqs. (\ref{ham})-(\ref{hi}) range over some number $N$ of nuclei which comprise the bath. Quantum dots residing in III-V materials like GaAs and InAs typically contain $10^4$-$10^6$ nuclei which interact appreciably with the electron, so physical values of $N$ should lie in this range, although in analytical calculations often one simply performs a sum over an infinite number of spins (out of which only about $N$ are appreciably coupled to the electron spin).
It is often useful to define the ``effective number of nuclei appreciably interacting with the electron'' $N$:
\beq
N\equiv {{\cal A}^2\over \sum_k A_k^2},\label{defN}
\eeq
where
\beq
{\cal A}\equiv \sum_k A_k\label{defcalA}
\eeq
is the total hyperfine interaction energy between the electron and the nuclear bath. For the most part, we will leave $N$ arbitrary in subsequent sections, but when a more precise definition is necessary, we will use Eq. (\ref{defN}).

Let us quote some numbers in order to make a connection to experimentally relevant timescales and magnetic field regimes. In GaAs, $\mathcal{A} \! \approx \! 90$ $\mu$eV (see e.g.~Ref.~\onlinecite{Coish_pssb09} and references therein). Using the effective $g$-factor $g_{\text{eff}} \! \approx \! 0.5$ of an electron in this material, $\Omega \approx \mathcal{A}$ for a magnetic field of $B \! \approx \! 3$ T.  Another characteristic magnetic field in the central spin problem is that corresponding to the typical Overhauser field for a thermal nuclear bath, $\mathcal{A}/\sqrt{N}$, which is on the order of a few mT. Note that the maximal hf coupling of a nuclear spin is $A_{k} \! \sim \! \mathcal{A}/N$, which is much smaller than $\Omega$ for the magnetic fields considered in this paper (i.e.~fields such that $\Omega>\mathcal{A}/\sqrt{N}$). An important quantity is also the timescale of $N/\mathcal{A}$, defining here the boundary of the short-time regime. For $N \! = \! 10^4$ - $10^6$ this corresponds to timescales of $100$ ns - $10$ $\mu$s.

It is difficult to obtain explicit results from the NZ GME unless strong assumptions are made about the form of the initial nuclear density matrix, which we denote $\rho_I(0)$. Fortunately, the NZ GME seems most tractable in the case of NFID, which is the primary focus of the present work. As we develop the structure of the NZ GME adapted to the central spin problem in Section \ref{sec:NZ}, we will attempt to keep the form of $\rho_I(0)$ as general as possible at each stage of the calculation and make clear the points at which it is necessary to further specify $\rho_I(0)$. For the sake of clarity, we will state here the precise final form that we will impose on $\rho_I(0)$ in order to obtain explicit results for the electron spin coherence in the NFID case.

In discussing the initial density matrix of the nuclei, we have already alluded to the first assumption we will make about the initial conditions, namely that the initial density matrix for the total system is separable:
\beq
\rho(0)=\rho_e(0)\otimes\rho_I(0).
\eeq
Here, $\rho_e$ is the reduced density matrix describing the electron spin degrees of freedom. This assumption is quite reasonable since, in quantum computing applications, it is generally desirable to initialize the electron spin to some well defined state which is independent of the nuclei.

In contrast, a much stronger assumption we will make is that $\rho_I(0)$ is diagonal in the basis of $h^z$ eigenstates. As in Refs. \onlinecite{Coish_PRB04,Coish_PRB10}, we will denote these states by $\ket{n_i}$ with $i=1...2^N$; they are just tensor products of eigenstates of the nuclear spin operators $I_k^z$:
\beq
\ket{n_i}=\bigotimes_k \ket{I_k,m_k^i},\qquad I_k^z\ket{I_k,m_k^i}=m_k^i\ket{I_k,m_k^i},
\eeq
where $I_k$ is the spin of the $k$th nucleus. We may write
\beq
h^z\ket{n_i}=h^z_{n_i}\ket{n_i},\qquad h^z_{n_i}\equiv \sum_kA_km_k^i.
\eeq
We will further assume that $\rho_I(0)$ only contains components which share the same $h^z$ eigenvalue, denoted $h^z_n$:
\beq
\rho_I(0)=\sum_{i=1}^g \rho_{ii}\ket{n_i}\bra{n_i},\label{narrowedstates}
\eeq
where
\beq
h^z\ket{n_i}=h^z_n\ket{n_i},\quad \forall i\in1...g,\label{hzeigen}
\eeq
The form of the density matrix given in (\ref{narrowedstates}-\ref{hzeigen}) describes a particular ``narrowed" set of allowed states for the nuclear bath.\cite{Coish_PRB04} In the so-called ``box model" limit where the hyperfine couplings are all equal, $A_k=A={\cal A}/N$, this amounts to restricting the possible nuclear states to a set which contains only states with a fixed net polarization along the magnetic field direction. We will see that the NZ GME approach enjoys many simplifications when the nuclear bath states are constrained in this way.

For some of the more explicit results we will obtain, we will also follow Refs.~\onlinecite{Coish_PRB04,Coish_PRB10} in assuming that the nuclear spin bath is uniformly polarized. By this, we mean that all traces of nuclear spin operators are independent of the nuclear site indices. For example, the correlator
\beq
\tr\left\{I_p^+I_q^-\rho_I(0)I_k^-I_\ell^+\right\}
\eeq
is independent of $k,\ell,p,q$ under this supposition. The uniform polarization assumption is less crucial than the narrowed-state condition of Eq. (\ref{narrowedstates}) and can be lifted in most cases, albeit at the expense of having to deal with more complicated algebraic expressions. For the purposes of elucidating more general features of the NZ GME approach, it is often a useful assumption to make. Throughout this paper we make explicit the points at which we invoke the uniform polarization condition.

%%%%%%%%%%%%%%%%%%%%%%%%%%%%%
%%% UNIFORM COUPLING EXACT SOLUTION
%%%%%%%%%%%%%%%%%%%%%%%%%%%%%
\section{Uniform coupling model}  \label{sec:exact_uniform}
In this section, we will consider the uniform-coupling (also referred to as `box model') limit in which all the hyperfine couplings are equal,
\beq
A_{k} = A \equiv \frac{\mathcal{A}}{N} \,\, .
\eeq
This is a semiclassical limit of the central spin problem since we have a large collective nuclear spin $\mathbf{I} \! = \! \sum_{k} \mathbf{I}_{k}$ coupled to the electron spin---the dynamics is mostly classical for large $N$. This is also very close to the static bath limit: for $\Omega \gg \mathcal{A}/\sqrt{N}$ and for a typical collective spin magnitude $|I|\approx \sqrt{N}$, $\mathbf{I}$ is almost static as it cannot follow the quickly precessing electron spin vector (if we neglect the nuclear Larmor precession, which is crucial for the case of spin echo decay,\cite{Cywinski_PRL09,Cywinski_PRB09,Bluhm_NP11,Neder_PRB11} but which is of much smaller importance for NFID).

The limit of uniform hyperfine couplings greatly simplifies the electron-nuclear dynamics, making this a natural limit in which to compare various approaches to the problem. In particular, this limit is exactly solvable,\cite{Khaetskii_PRB03,Zhang_PRB06,Bortz_JSM07,Coish_JAP07,Fischer_PRA07,Cywinski_PRB10} and we will begin by reviewing this solution for the NFID case. In Section \ref{sec:NZ_box}, we will use this solution to test how well the NZ approach (both a standard one, and the one using correlated projection operators) can describe the dynamics of the box model.

Restricting attention to the homonuclear case, $\omega_k=\omega$, with all the nuclei having spin $I=1/2$, we can solve the uniform coupling model exactly by working in the basis of eigenstates of the total nuclear spin operator $I^z=\sum_k I^z_k$; we denote these states by $|jm\rangle$ with $I^z|jm\rangle=m|jm\rangle$. In this basis, we may write the expectation value $\mean{S^+(t)}$ in terms of the appropriate matrix element of the reduced density matrix evolved with respect to the full Hamiltonian:
\beq
\mean{S^+(t)}=\sum_{jm}n_j \bra{\downarrow,j,m}e^{-iHt}\rho(0)e^{iHt}\ket{\uparrow,j,m}.
\eeq
The $\uparrow,\downarrow$ inside the kets denote the eigenstates of the electron spin operator $S^z$. $\rho(0)$ is the full density matrix at time $t=0$, and $n_j$ is the degeneracy of nuclear states with a fixed $j$ and $m$,\cite{Melikidze_PRB04}
\beq
n_j={N!\over({N\over2}-j)!({N\over2}+j)!}{2j+1\over{N\over2}+j+1}.
\eeq
In order to facilitate the computation, we will assume that the initial nuclear density matrix is diagonal in the $|jm\rangle$ basis. In fact, in order to compare with the NZ GME result computed in Sec.~\ref{sec:NZ_box}, we will assume that the initial nuclear density matrix is proportional to the identity in a subspace of fixed $m$ and is zero outside this subspace. This is equivalent to the form of $\rho_I(0)$ that we assumed in Eqs. (\ref{narrowedstates})-(\ref{hzeigen}), if in the latter we make the further assumption that the nuclei are uniformly polarized, $\rho_{ii}=1/Z$ with $Z=\sum_jn_j$. When the density matrix has this form, our formula becomes
\beq
\mean{S^+(t)}={1\over Z}\sum_{j=|m|}^{N/2}n_j \bra{\downarrow,j,m}e^{-iHt}\rho_e(0)e^{iHt}\ket{\uparrow,j,m},
\eeq
where $\rho_e(0)$ is the initial electron density matrix.

In the case where all hyperfine couplings are equal, $A_k=A$, it is straightforward to work out the action of the evolution operator $e^{-iHt}$ on the $\ket{\uparrow/\downarrow, j,m}$ states:\cite{Fischer_PRA07,Cywinski_PRB10}
\bea
e^{-iHt}\ket{\uparrow,j,m}&=&a_{jm}\ket{\uparrow,j,m}+b_{jm}\ket{\downarrow,j,m+1},\nn\\
e^{-iHt}\ket{\downarrow,j,m}&=&c_{jm}\ket{\downarrow,j,m}+d_{jm}\ket{\uparrow,j,m-1},\label{evolofjmstates}
\eea
with
\bea
a_{jm}&=&e^{-iE_m^+t}\left[\cos\left({N_{jm}^+\over2}t\right)-i{Z_m^+\over N_{jm}^+}\sin\left({N_{jm}^+\over2}t\right)\right],\nn\\
b_{jm}&=&-ie^{-iE_m^+t}{X_{jm}^+\over N_{jm}^+}\sin\left({N_{jm}^+\over2}t\right),\nn\\
c_{jm}&=&e^{-iE_m^-t}\left[\cos\left({N_{jm}^-\over2}t\right)-i{Z_m^-\over N_{jm}^-}\sin\left({N_{jm}^-\over2}t\right)\right],\nn\\
d_{jm}&=&-ie^{-iE_m^-t}{X_{jm}^-\over N_{jm}^-}\sin\left({N_{jm}^-\over2}t\right),\label{abcddefs}
\eea
and
\bea
E_m^\pm&=&(m\pm 1/2)\omega-\A/(4N),\nn\\
X_{jm}^\pm&=&\A\sqrt{j(j+1)-m(m\pm1)}/N,\nn\\
Z_m^\pm&=&\pm\left[\Omega-\omega+\A(m\pm1/2)/N\right],\nn\\
N_{jm}^\pm&=&\sqrt{(X_{jm}^\pm)^2+(Z_m^\pm)^2}.\label{XZNdefs}
\eea
Using (\ref{evolofjmstates}) and expanding the initial electron density matrix as
\beq
\rho_e(0)={1\over2}\mathbf{1}+2\mean{S^z(0)} S^z+\mean{S^+(0)} S^-+\mean{S^-(0)} S^+,
\eeq
one finds that only the $S^-$ component of $\rho_e(0)$ contributes to $\mean{S^+(t)}$:
\bea
\mean{S^+}&=&{\mean{S^+(0)}\over Z}\sum_{j=|m|}^{N/2}n_j \bra{\downarrow,j,m}e^{-iHt}S^-e^{iHt}\ket{\uparrow,j,m}\nn\\
&=&{\mean{S^+(0)}\over Z}\sum_{j=|m|}^{N/2}n_j a_{jm}^*c_{jm}.
\eea
For later comparison, we transform this result to a certain rotating frame defined with respect to the frequency $\Omega_n+\Delta\Omega$, where
\beq
\Omega_n\equiv \Omega+h^z_n,
\eeq
and the ``Lamb shift" $\Delta\Omega$ will be defined later on. In the rotating frame, the exact solution to the uniform coupling model is then given by
\beq
x_{exact}(t)={x_0\over Z}e^{-i(\Omega+\A m/N+\Delta\Omega)t}\sum_{j=|m|}^{N/2}n_j a_{jm}^*c_{jm}.\label{exactbox}
\eeq
To obtain this expression, we have used that $h^z_n=\A m/N$ in the case of uniform coupling.

%%%%%%%%%%%%%%%%%%%%%%%%%
%%% NAKAJIMA ZWANZIG MASTER EQUATION
%%%%%%%%%%%%%%%%%%%%%%%%%
\section{Nakajima-Zwanzig Master Equation and its perturbative expansion}\label{sec:NZ}
The first step in constructing the NZ GME is to separate the density matrix for the total system into two parts, often referred to as the relevant and the irrelevant parts:\cite{Fick,Breuer}
\beq
\rho=\rho_{rel}+\rho_{irr}.\label{dmpartition}
\eeq
The relevant part, $\rho_{rel}$, is a density matrix for the degrees of freedom whose evolution one wishes to compute, while the irrelevant part, $\rho_{irr}$, contains the remaining degrees of freedom in the total system. While these remaining degrees of freedom will influence the evolution of $\rho_{rel}$, their own evolution is not of direct concern and need not be computed explicitly. In applications where one considers the dynamics of a system coupled to a bath, $\rho_{rel}$ is typically a density matrix describing the degrees of freedom of the system. In this case, note that strictly speaking, $\rho_{irr}$ is not a density matrix since it does not satisfy $\tr\rho_{irr}=1$.

The partition described in Eq. (\ref{dmpartition}) is implemented by introducing projection superoperators $P$ and $Q$ such that
\beq
P\rho=\rho_{rel},\quad Q\rho=\rho_{irr},\quad P+Q=1,\quad PQ=0.
\eeq
The last two relations above ensure that $P$ and $Q$ are true projectors in the sense that $P^2=P$ and $Q^2=Q$. With these operators, the Liouville equation for $\rho$ can be transformed into an exact equation for the evolution of $\rho_{rel}$, which in the case of a time-independent Hamiltonian $H$ has the form:\cite{Fick,Breuer}
\beq
P\dot\rho(t)=-iPLP\rho(t)-i\int_0^tdt'\widehat{\Sigma}(t-t')P\rho(t'),\label{genNZ}
\eeq
\beq
\widehat{\Sigma}(t)\equiv -iPLQe^{-iLQt}QLP.\label{genMemKernel}
\eeq
The Liouvillian superoperator $L$ implements the evolution of the total system and is defined to act on an arbitrary operator ${\cal O}$ according to
\beq
L{\cal O}=[H,{\cal O}].
\eeq
The superoperator $\widehat\Sigma$ is referred to as the memory kernel; this quantity contains the full dynamics of the bath and controls how these dynamics affect the evolution of the system. It is important to note that the particular form of the NZ equation given in Eq. (\ref{genNZ}) assumes that the initial density matrix satisfies the condition $Q\rho(0)=0$.

In the case of our central spin model, Eqs. (\ref{ham})-(\ref{hi}), we choose $\rho_{rel}$ to be essentially the reduced density matrix for the electron spin, $\rho_e$. This means that the projection operator $P$ involves a trace over the nuclear bath:\cite{Coish_PRB04,Coish_PRB08,Coish_PRB10}
\beq
P\rho=\rho_I(0)\tr_I\rho=\rho_e\otimes\rho_I(0).\label{NZprojector}
\eeq
We have introduced a factor of the initial nuclear density matrix $\rho_I(0)$ in the definition of $P$ to satisfy the contraint $P^2=P$. Note that the condition $Q\rho(0)=0$ is tantamount to assuming that $\rho(0)$ is separable:
\beq
\rho(0)=\rho_e(0)\otimes\rho_I(0).
\eeq
This is an assumption we will make throughout this work.

It is important to stress that Eq. (\ref{NZprojector}) is not the only possible choice for the projector $P$. For example, it is possible to instead define $P$ as a sum over many projection operators which project onto various subspaces of the nuclear bath state space, and the choice of $P$ can strongly influence the convergence properties of the resulting theory.\cite{Fischer_PRA07,Ferraro_PRB08} In fact, we will see later on in the context of the uniform coupling model that the choice made in Eq. (\ref{NZprojector}) is far from ideal, and that a more sophisticated choice leads to a vast improvement in the convergence of the theory. Given the connection between the box model and the short-time, low B-field regime of the real system with non-uniform couplings, we will argue that this observation carries important consequences for the development of a more complete theory of the central spin problem. For now, we will keep the definition of $P$ in Eq. (\ref{NZprojector}) and continue our construction of the ``standard" NZ theory.

The form of the NZ equation given in Eq. (\ref{genNZ}) is quite difficult to work with. In order to reduce it to a more tractable form, we will place restrictions on the structure of the initial nuclear density matrix, $\rho_I(0)$. In particular, we will assume that $\rho_I(0)$ is diagonal in the $\ket{n_i}$ basis: $\rho_I(0)=\sum_i\rho_{ii}\ket{n_i}\bra{n_i}$. If we then multiply both sides of Eq. (\ref{genNZ}) by the operator $S^+$ and take the trace, we obtain an equation for $\mean{S^+}$ only:
\beq
{d\over dt}\mean{S^+(t)}=i\Omega_n\mean{S^+(t)}-i\int_0^tdt'\Sigma(t-t')\mean{S^+(t')},\label{NZ}
\eeq
where $\Omega_n\equiv\Omega+h^z_n$ with
\beq
h^z_n\equiv \tr_I[h^z\rho_I(0)]=\sum_i \rho_{ii}h^z_{n_i}.\label{defhzn}
\eeq
Note that this definition of $h^z_n$ is more general than the meaning we gave this symbol in Eq.~(\ref{hzeigen}) in the context of the fully restricted nuclear density matrix. The two meanings coincide when $\rho_I(0)$ has the form of Eq. (\ref{narrowedstates}). The memory kernel is now a function instead of an operator:
\beq
\Sigma(t)\equiv-i\tr\left[S^+PLQe^{-iLQt}QLPS^-\rho_I(0)\right].\label{memKernel}
\eeq
Without the assumption that $\rho_I(0)$ is diagonal in the $\ket{n_i}$ basis, additional terms involving $\mean{S^z(t)}$ and $\mean{S^-(t)}$ appear in Eq. (\ref{NZ}), and the problem becomes considerably more complicated. Note that we did not have to assume the fully restricted form for $\rho_I(0)$ quoted in Eq. (\ref{narrowedstates}) where the sum is only over $\ket{n_i}$ states corresponding to the same Overhauser field. Eq. (\ref{NZ}) is valid when the sum contains other states as well, i.e.~when the nuclear state is not narrowed.

Eq. (\ref{NZ}) is an integro-differential equation which can easily be solved by performing a Laplace transform, after which the equation becomes algebraic with the solution
\beq
\mean{S^+(s)}=\int_0^\infty dt e^{-st}\mean{S^+(t)}={\mean{S^+(t=0)}\over s-i\Omega_n+i\Sigma(s)}.
\eeq
The solution in the time domain is then obtained by computing the Bromwich inversion integral,
\beq
\mean{S^+(t)}={1\over2\pi i}\int_{\gamma-i\infty}^{\gamma+i\infty} ds e^{st}\mean{S^+(s)},\label{bromwich}
\eeq
where the contour defined by the real number $\gamma$ must be chosen such that it lies to the right of all the poles of $\mean{S^+(s)}$. Therefore, solving for $\mean{S^+(t)}$ requires solving for the Laplace transform of the memory kernel:
\bea
\Sigma(s)&=&\int_0^\infty dt e^{-st}\Sigma(t)\nn\\&=&-i\tr\left[S^+PLQ{1\over s+iLQ}QLPS^-\rho_I(0)\right] \,\ , \label{memKernel2}
\eea
Computing the memory kernel exactly is a difficult problem except in very simple cases, so we will proceed to calculate it perturbatively in the following section.

\subsection{The expansion of the memory kernel}   \label{sec:expansion}
In order to make further progress, it is necessary to construct a perturbative expansion of the memory kernel given in Eq.~(\ref{memKernel2}). We will expand this quantity in a power series in the number of flip-flops, i.e. in powers of $V_{\text{ff}}$:
\beq
\Sigma(s)=\Sigma^{(2)}(s)+\Sigma^{(4)}(s)+O(V_{\text{ff}}^6).\label{memkernelexpansion}
\eeq
We can only have even terms in this expansion since the flip-flops are virtual in the physical limit $\Omega\gg\omega_k$. For nuclear density matrices $\rho_I(0)$ which are diagonal in the $\ket{n_i}$ basis, the odd terms are strictly zero due to the structure of the memory kernel, Eq. (\ref{memKernel2}). We will see shortly why a zeroth order term has not been included in Eq. (\ref{memkernelexpansion}).

To facilitate the expansion, it is convenient to define the following superoperators:
\beq
L_0{\cal O}=[H_0,{\cal O}],
\eeq
\beq
L_V{\cal O}=[V_{\text{ff}},{\cal O}].
\eeq
Replacing $L\to L_0+L_V$ in (\ref{memKernel2}), we can then recast our perturbative flip-flop expansion as an expansion in $L_V$. The expansion of the memory kernel is tantamount to an expansion of the operator ${1\over s+i(L_0+L_V)Q}$ in powers of $L_V$, and this series is straightforward to construct:
\bea
&&{1\over s+iLQ}=\big\{1-i G_Q(s)L_VQ-G_Q(s)L_VQG_Q(s)L_VQ\nn\\&&+iG_Q(s)L_VQG_Q(s)L_VQG_Q(s)L_VQ\big\}G_Q(s)+O(L_V^4),\nn\\&&\label{expandedprop}
\eea
where
\beq
G_Q(s)\equiv{1\over s+iL_0Q}.
\eeq
Using this expansion, we will compute the memory kernel up to fourth order.

Before we proceed to compute the terms of the series, we first pause to write down some identities which will be useful at various stages of the expansion. When $\rho_I(0)$ is diagonal in the $\ket{n_i}$ basis, the following identity holds for any integers $k$ and $\ell$:
\beq
PL_0^kL_VL_0^\ell P=0.\label{PLVPidentity}
\eeq
This identity can be further generalized to include any combination of $L_0$'s and an odd number of $L_V$'s sandwiched between two $P$'s. When $\rho_I(0)$ is further restricted to the form given in Eq. (\ref{narrowedstates}), we have an additional identity (expressed in two equivalent ways):
\beq
QL_0P=0,\qquad QL_0Q=QL_0.\label{narrowedPQidentity}
\eeq
The first form of this identity allows us to replace one of the $L$'s in Eq.~(\ref{memKernel2}) with $L_V$:
\beq
\Sigma(t)=-i\tr[S^+PLQ {1\over s+iLQ}QL_VPS^-\rho_I(0)],\label{memKernel3}
\eeq
which indicates that the zeroth order memory kernel vanishes, as was already presumed in (\ref{memkernelexpansion}). Note that in using Eq.~(\ref{narrowedPQidentity}) to arrive at Eq.~(\ref{memKernel3}), we are assuming the fully restricted form of $\rho_I(0)$ stated in Eq.~(\ref{narrowedstates})-(\ref{hzeigen}), whereas up until now, we have only needed to assume that $\rho_I(0)$ is diagonal in the $\ket{n_i}$ basis. In fact, at this point it is not really necessary to impose Eq. (\ref{narrowedstates}), and we could instead continue to suppose only that $\rho_I(0)$ is diagonal, in which case we would find a non-vanishing zeroth-order term in the memory kernel expansion. However, slightly further into the calculation we will impose Eqs.~(\ref{narrowedstates})-(\ref{hzeigen}), and so at this stage we may as well make the simplifications that this form of $\rho_I(0)$ brings.

\subsection{Second-order memory kernel}
Inserting Eq. (\ref{expandedprop}) into Eq. (\ref{memKernel3}), we find at second order
\bea
\Sigma^{(2)}(s)&=&-i\tr[S^+PL_VQG_Q(s)QL_VPS^-\rho_I(0)]\nn\\&-&\tr[S^+PL_0QG_Q(s)L_VQG_Q(s)QL_VPS^-\rho_I(0)].\nn\\&&
\eea
The second term on the right-hand side can be shown to vanish identically for the central spin Hamiltonian. Focusing then on the first term, this expression is somewhat complicated by the dependence of $G_Q(s)$ on $Q$. Instead of working with $G_Q(s)$, we choose to work with its time-domain counterpart, $e^{-iL_0Qt}$, in terms of which we have
\beq
\Sigma^{(2)}(t)=-i\tr[S^+PL_VQe^{-iL_0Qt}QL_VPS^-\rho_I(0)].\label{2ndorderMK}
\eeq
It is easy to show that the identity (\ref{narrowedPQidentity}) enables the extraction of the projection operator $Q$ from the exponent:
\beq
Qe^{-iL_0Qt}=Qe^{-iL_0t}\equiv Q G(t).\label{GQidentity}
\eeq
The superoperator $G(t)$ is simply the evolution operator corresponding to the unperturbed part of the Hamiltonian, $H_0$:
\beq
G(t)\rho(0)=e^{-i H_0t}\rho(0)e^{iH_0t}.
\eeq
Also notice that (\ref{GQidentity}) immediately implies
\beq
QG_Q(s)=QG(s),
\eeq
with the ``propagator" defined as
\beq
G(s)\equiv {1\over s+iL_0}.
\eeq
At this point, we will invoke (by using Eq.~(\ref{narrowedPQidentity})) the most constrained form of the initial nuclear density matrix, Eq.~(\ref{narrowedstates})-(\ref{hzeigen}), as it is not clear how to proceed without factoring the projector $Q$ out of the propagator $G(s)$. If one could proceed without performing this factorization, then it would suffice to assume only that $\rho_I(0)$ is diagonal in the $\ket{n_i}$ basis, albeit several more terms would contribute at each order of the memory kernel expansion. In any case, we are ultimately interested in applying this formalism to the case of NFID, which is defined precisely by Eqs.~(\ref{narrowedstates})-(\ref{hzeigen}).

Plugging Eq.~(\ref{GQidentity}) into Eq.~(\ref{2ndorderMK}), eliminating factors of $Q$ with applications of the identity given in Eq.~(\ref{PLVPidentity}), and dropping factors of $P$ using the observations that $PS^-\rho_I(0)=S^-\rho_I(0)$ and $\tr[S^+P{\cal O}]=\tr[S^+{\cal O}]$ for any operator ${\cal O}$, we arrive at
\beq
\Sigma^{(2)}(t)=-i\tr[S^+L_VG(t)L_VS^-\rho_I(0)].\label{2ndorderMK2}
\eeq
It remains to plug in explicit expressions for the various operators and perform the trace. Defining the following set of operators which act in the nuclear subspace,
\beq
U_{\pm}(t)\equiv e^{\mp {i\over2}(\Omega+h^z)t-i\sum_k\omega_k I_k^zt},\label{Upm}
\eeq
we can express the action of $G(t)$ on a matrix $\rho$ as
\beq
G\rho=\left(\begin{matrix}U_+\rho_{\uparrow\uparrow}U_+^\dag& U_+\rho_{\uparrow\downarrow}U_-^\dag\cr U_-\rho_{\downarrow\uparrow}U_+^\dag & U_-\rho_{\downarrow\downarrow}U_-^\dag\end{matrix}\right),\label{Gaction}
\eeq
while the action of $L_V$ has the explicit form
\beq
L_V\rho={1\over2}\left(
\begin{matrix}
h^-\rho_{\downarrow\uparrow}-\rho_{\uparrow\downarrow}h^+& h^-\rho_{\downarrow\downarrow}-\rho_{\uparrow\uparrow}h^-\cr h^+\rho_{\uparrow\uparrow}-\rho_{\downarrow\downarrow}h^+& h^+\rho_{\uparrow\downarrow}-\rho_{\downarrow\uparrow}h^-
\end{matrix}\right),\label{LVaction}
\eeq
where for instance $\rho_{\uparrow\downarrow}=\bra{\uparrow}\rho\ket{\downarrow}$, with $\ket{\uparrow}$ and $\ket{\downarrow}$ denoting the eigenstates of the electron spin operator $S^z$. After a bit of algebra, we find
\beq
\Sigma^{(2)}(t)={1\over4i}\sum_kA_k^2e^{i\omega_kt}\left[c_k^-e^{i{A_k\over2}t}+c_k^+e^{-i{A_k\over2}t}\right],
\eeq
where
\beq
c_k^\pm\equiv \tr\left\{I_k^\mp I_k^\pm \rho_I(0)\right\}.
\eeq
The Laplace transform of this is
\beq
\Sigma^{(2)}(s)={1\over4i}\sum_k A_k^2\left[{c_k^+\over s-i(\omega_k-{A_k\over2})}+{c_k^-\over s-i(\omega_k+{A_k\over2})}\right].\label{2ndorderMK3}
\eeq

\subsection{Lamb shift and rotating frame}
The effect of $\Sigma^{(2)}$ on $\mean{S^+(t)}$ was studied extensively in Ref.~\onlinecite{Coish_PRB04}, where it was shown that it leads to both a shift in the precession frequency of the electron spin (Lamb shift) as well as to a small decay of $\mean{S^+(t)}$ at short times referred to as ``visibility loss". In the limit of large $N$, the visibility loss effect is suppressed,\cite{Coish_PRB04,Coish_PRB10} and one can make the approximation that the only role of the second-order memory kernel is to generate the Lamb shift. We will refer to this as the ``Lamb shift approximation."

To make the definition of the Lamb shift more precise, first recall the Bromwich integral formula from Eq. (\ref{bromwich}):
\beq
\mean{S^+(t)}={\mean{S^+(t=0)}\over2\pi i}\int_{\gamma-i\infty}^{\gamma+i\infty} ds e^{st}{1\over s-i\Omega_n+i\Sigma(s)}.\label{bromwich2}
\eeq
Since $\Sigma(s)$ vanishes at zero hyperfine coupling, it is clear from this formula that the zeroth-order behavior of $\mean{S^+(t)}$ is just a precession with frequency $\Omega_n$. At non-zero coupling, this precession is shifted by an amount determined by the real part of the memory kernel. We define the Lamb shift, $\Delta\Omega$, as the shift in precession frequency that would occur if the imaginary part of the memory kernel (the part which produces decay) were zero. We can read off a self-consistent equation for $\Delta\Omega$ from Eq. (\ref{bromwich2}) by supposing $\hbox{Im}[\Sigma(s)]=0$ and requiring that the integrand have a pole at $s=i(\Omega_n+\Delta\Omega)$:
\beq
\Delta\Omega= -\hbox{Re}[\Sigma(i\Omega_n+i\Delta\Omega)].
\eeq
In our flip-flop expansion, this can be approximated by
\beq
\Delta\Omega\approx -\hbox{Re}[\Sigma^{(2)}(i\Omega_n+i\Delta\Omega)].\label{lambshift}
\eeq
Using this formula in conjuction with Eq. (\ref{2ndorderMK3}), we then find that the Lamb shift is given by
\beq
\Delta\Omega\approx {1\over4\Omega_n}\sum_k A_k^2(c_k^++c_k^-),\label{lambshift2}
\eeq
in the limit $\Omega_n\gg\omega_k,A_k$.

Now that we have defined the Lamb shift more precisely, we should also clarify the manner in which the Lamb shift approximation is implemented. This approximation consists of replacing $\Sigma^{(2)}(s+i\Omega_n+i\Delta\Omega)\to-\Delta\Omega$ in the expansion of the memory kernel. In order to make this replacement in (\ref{bromwich2}), we need to shift the integration variable $s\to s+i\Omega_n+i\Delta\Omega$. In the Lamb shift approximation, we then have
\beq
\mean{S^+}={\mean{S^+(t=0)}\over2\pi i}e^{i(\Omega_n+\Delta\Omega)t}\int_{\gamma-i\infty}^{\gamma+i\infty} ds e^{st}{1\over s+i\widetilde{\Sigma}^{(4)}(s)},\label{bromwichLSA}
\eeq
where we have kept up to fourth order in the memory kernel and defined
\beq
\widetilde{\Sigma}^{(4)}(s)\equiv \Sigma^{(4)}(s+i\Omega_n+i\Delta\Omega).
\eeq

It is often useful\cite{Coish_PRB10} to remove the high-frequency oscillations arising from the prefactor $e^{i(\Omega_n+\Delta\Omega)t}$ in Eq.~(\ref{bromwichLSA}) by introducing a ``co-rotating" coherence measure $x$:
\beq
x(t)\equiv 2e^{-i(\Omega_n+\Delta\Omega)t}\mean{S^+(t)}.
\eeq
Note that $\widetilde{\Sigma}^{(4)}(s)$ is the fourth-order term of the Laplace transform of the memory kernel in the rotating frame:
\beq
\widetilde{\Sigma}(t)=e^{-i(\Omega_n+\Delta\Omega)t}\Sigma(t).
\eeq
This quantity serves as the kernel in the integro-differential equation governing the evolution of $x(t)$:
\beq
\dot x(t)=-i\Delta\Omega x(t)-i\int_0^t dt'\widetilde{\Sigma}(t-t')x(t').\label{rotatingNZGME}
\eeq
The rotating frame renders some of the more subtle features arising from the hyperfine flip-flops more transparent, and we will make extensive use of it when we solve the uniform coupling model using the NZ GME later on.

\subsection{Fourth-order memory kernel}
The fourth-order terms which emerge from inserting Eq.~(\ref{expandedprop}) into Eq.~(\ref{memKernel3}) are
\bea
\Sigma^{(4)}(s)&=&i\tr\{S^+P[1-iL_0QG_Q(s)]L_VQG_Q(s)L_VQ\nn\\&\times&G_Q(s)L_VQG_Q(s)QL_VPS^-\rho_I(0)\}.
\eea
We can again use Eq. (\ref{narrowedPQidentity}) to replace $QG_Q(s)$ with $QG(s)$ and to discard most of the projection operators, yielding
\bea
\Sigma^{(4)}(s)&=&i\tr\{S^+[1-iL_0QG(s)]L_VG(s)L_VQ\nn\\&\times&G(s)L_VG(s)L_VS^-\rho_I(0)\}.\label{4thorderMK}
\eea
If we then replace each instance of the propagator $G(s)$ with $\int_0^\infty dt e^{-st}G(t)$, then the remaining steps can be performed in a manner quite similar to the treatment we have given for the second-order memory kernel. This time, however, the expressions are much more complicated, and we relegate them to Appendix \ref{app:4thorderSigma}.

\subsection{High-frequency limit}\label{highfreqlimit}
From the explicit expression for the fourth-order memory kernel given in Appendix \ref{app:4thorderSigma}, it is clear that the dominant contribution will come from the region $s\approx-i\Omega_n$. This is because the fourth-order memory kernel is a sum of many terms where each term is a product of three simple poles. Some of these poles are located at values of $s$ which only depend on the $k$th nuclear Zeeman energy and hf coupling. However, most of the terms contain a pole located in the vicinity of $s\approx-i\Omega_n$. Due to this structure, $x(s)$ has poles at both  $s\approx-i\Omega_n$ and at low frequencies, but the residues at the latter poles are strongly suppressed compared to the former ones. This means that if we consider the function $\bar{\Sigma}^{(4)}(s)\equiv \Sigma^{(4)}(s+i\Omega_n)$, we can neglect $s$, $\omega_k$, and $A_k$ relative to $\Omega_n$. The resulting approximate expression for $\bar{\Sigma}^{(4)}(s)$ is given in Appendix \ref{app:4thorderSigma}. The approximate expression for $\bar{\Sigma}^{(4)}(s)$ can be simplified by using the results of Appendix \ref{app:correlators} to evaluate bath correlators, with the result
\bea
\bar{\Sigma}^{(4)}&\approx& {-i\over16\Omega_n^2}\sum_{k\ne\ell}A_k^2A_\ell^2\sum_i\rho_{ii}c_k^{(i)-}c_\ell^{(i)+}{2s+i(A_k-A_\ell)\over s+i(A_k-A_\ell)}\nn\\&\times&\Bigg\{{1\over s-i(\omega_k-\omega_\ell-{1\over2}(A_k-A_\ell))}\nn\\&+&{1\over s+i(\omega_k-\omega_\ell+{1\over2}(A_k-A_\ell))}\Bigg\}\nn\\&-& {i\over16\Omega_n^2s}\sum_i\rho_{ii}\left(\sum_{k}A_k^2\left[c_k^{(i)-}+c_k^{(i)+}\right]\right)^2 \nn\\&+& {i\over16\Omega_n^2s}\left(\sum_{k}A_k^2\left[c_k^-+c_k^+\right]\right)^2.\label{largefreqlimit}
\eea
The symbols $c_k^{(i)\pm}$ are defined in Appendix \ref{app:correlators}. The fourth-order memory kernel given in Eq.~(\ref{largefreqlimit}) simplifies considerably if we assume a uniformly polarized nuclear bath. In this case, the last two terms in Eq.~(\ref{largefreqlimit}) cancel each other. This follows from the fact that $\sum_i\rho_{ii}c_k^{(i)\pm}c_\ell^{(i)\pm}=c^\pm c^\pm$ for such a bath; this identity is proven in Appendix \ref{app:correlators}. (Here, $c_k^\pm=c^\pm$ is independent of $k$ by definition.) In using this form of the identity, we are also assuming a homonuclear bath since we have thrown away the species information in discarding the site indices $k$ and $\ell$. This assumption could easily be lifted by introducing additional indices, but we will not do this for the sake of simplicity. Therefore, for a uniformly polarized homonuclear spin bath, we have
\beq
\bar{\Sigma}^{(4)}\approx {-ic^+c^-\over4\Omega_n^2}\sum_{k\ne\ell}{A_k^2A_\ell^2\over s+i(A_k-A_\ell)}.\label{largefreqlimit2}
\eeq

Implementing the same procedure for obtaining the high-frequency limit on the second-order memory kernel, Eq. (\ref{2ndorderMK3}), we find that the result is a constant:
\beq
\bar{\Sigma}^{(2)}\approx-{c^++c^-\over4\Omega_n}\sum_kA_k^2\approx-\Delta\Omega.
\eeq
In the last equality, we have pointed out that the high-frequency limit of $-\bar{\Sigma}^{(2)}$ is just the Lamb shift we have already computed. This was given in Eq.~(\ref{lambshift2}) in the context of a more general bath (not necessarily uniformly polarized). We see that the Lamb shift approximation is automatically incorporated into the high-frequency approximation.

%%%%%%%%%%%%%%%%%%
%%% NAKAJIMA-ZWANZIG BOX
%%%%%%%%%%%%%%%%%%
\section{The NZ solution for the uniform coupling model} \label{sec:NZ_box}

We proceed to solve the uniform coupling model within the NZ framework in the case where all nuclei have spin 1/2. We will find that the solution is structurally incompatible with the exact solution reviewed in Section \ref{sec:exact_uniform}, except at very early times. Let us define the characteristic time
\beq
\tau \equiv 4 \frac{\Omega_{n}}{\A} \frac{N}{\A} \,\, .
\eeq
Note that $\tau \! \approx \! \Delta\Omega^{-1}$. For times $t\lesssim \tau$, the NZ solution agrees quite well with the exact solution; however, we will show that beyond this time scale, the NZ result rapidly breaks down. We will argue that this failure is an unavoidable consequence of the basic structure of the perturbative NZ approach with projector as defined in Eq. (\ref{NZprojector}), suggesting that this approach is inappropriate in the case at hand. A remedy for this problem will be presented in Section \ref{sec:correlated_box}.

Starting from the expressions for the second and fourth-order memory kernels, Eqs. (\ref{2ndorderMK3}) and (\ref{4thorderMK3}), assuming a homonuclear bath, $\omega_k=\omega$, and setting $A_k=A = \mathcal{A}/N$, we find in the rotating frame
\beq
\widetilde\Sigma^{(2)}=-{i\mu c^-\over s+i\varpi_1}-{i\mu c^+\over s+i\varpi_2},\label{2ndorderbox}
\eeq
\beq
\widetilde\Sigma^{(4)}=\left[{1\over s+i\varpi_1}+{1\over s+i\varpi_2}\right]^2\left[{2i\mu^2c^+c^-\over s+i\varpi_3}+{i\mu^2c^+c^-\over s+i\varpi_4}\right],\label{4thorderbox}
\eeq
with
\bea
\varpi_1&=& \Omega_n+\Delta\Omega-\omega-\A/(2N),\nn\\ \varpi_2&=&\Omega_n+\Delta\Omega-\omega+\A/(2N),\nn\\ \varpi_3&=&2\Omega_n-2\omega+\Delta\Omega,\nn\\ \varpi_4&=&\Delta\Omega,\label{varpidef}
\eea
and
\beq
\mu\equiv {\A^2 \over4N}={\Omega_n\over\tau}.\label{mudef}
\eeq
In the limit of uniform couplings with spin 1/2 nuclei, the Lamb shift reduces to
\beq
\Delta\Omega\approx {\A^2\over4N\Omega_n}={\mu\over\Omega_n}.\label{boxLS}
\eeq
In the above results, we have used the fact that we are restricting to the case of a uniformly polarized nuclear bath so that the bath correlators $c_k^\pm=c^\pm$ are independent of the nuclear site index $k$. Further details about how the dependence on $c^\pm$ arises in (\ref{2ndorderbox}) and (\ref{4thorderbox}) are given in Appendix \ref{app:correlators}. For the remainder of this section, we will absorb $\omega$ into the definition of $\Omega_n$. In the rotating frame, $x(t)$ must depend on $\omega$ only in the combination $\Omega-\omega$, so it is simple to restore the explicit $\omega$-dependence if desired. In any case, it is generally safe to neglect $\omega$ since for any finite magnetic field $\omega\ll\Omega$.

The solution to Eq.~(\ref{rotatingNZGME}) for the electron spin coherence $x(t)$ in terms of a Bromwich inversion integral is
\beq
x(t)={1\over2\pi i}\int_{\gamma-i\infty}^{\gamma+i\infty}ds e^{st} {x_0\over s+i\Delta\Omega+i\widetilde\Sigma^{(2)}(s)+i\widetilde\Sigma^{(4)}(s)},\label{xoft}
\eeq
where $x_0\! \equiv \! x(t \!= \!0)$ is the initial condition in the time domain. Note that we are not making the Lamb shift approximation here since we are keeping the full $\widetilde{\Sigma}^{(2)}(s)$. The Laplace transform of $x(t)$ is a rational function:
\beq
x(s)=x_0{R(s)\over \Phi(s)},
\eeq
where the polynomials $R(s)$ and $\Phi(s)$ are given by
\beq
R(s)\equiv (s+i\varpi_1)^2(s+i\varpi_2)^2(s+i\varpi_3)(s+i\varpi_4),
\eeq
\bea
\Phi(s)&\equiv& (s+i\Delta\Omega)R(s)+\mu c^-{R(s)\over s+i\varpi_1}+\mu c^+{R(s)\over s+i\varpi_2}\nn\\&-&\mu^2c^+c^-\left[2s+i(\varpi_1+\varpi_2)\right]^2\left[3s+i\varpi_3+2i\varpi_4\right].\nn\\&&
\eea
The fact that $x(s)$ is a rational function guarantees that the sum of the residues equals $x_0$ so that $x_0$ is indeed the initial value of $x(t)$. Denoting the seven zeros of $\Phi(s)$ by $s_i$, we may write
\beq
\Phi(s)=\prod_{i=0}^6(s-s_i).
\eeq
We can solve for the $s_i$ perturbatively in the small parameter ${\A^2\over N\Omega_n^2}\ll1$ (equivalently, $\Omega_n\tau\gg1$). This essentially amounts to an expansion in small $\mu$ about the zeros of $sR(s)$, which are 0, $-i\varpi_1$, $-i\varpi_2$, $-i\varpi_3$, $-i\varpi_4$. Since $-i\varpi_1$ and $-i\varpi_2$ are each zeros of $sR(s)$ with multiplicity 2, they will each give rise to two separate zeros of $\Phi(s)$. We must also keep in mind that the Lamb shift is linear in $\mu$ (see Eq.~(\ref{boxLS})). We have
\bea
\Phi(s)&=&\left(s+i\eta{\mu\over\Omega_n}\right)R(s)\nn\\&+&\eta\mu c^-{R(s)\over s+i\varpi_1}+\eta\mu c^+{R(s)\over s+i\varpi_2}\\&-&\mu^2c^+c^-\left[2s+i(\varpi_1+\varpi_2)\right]^2\left[3s+i\varpi_3+2i\varpi_4\right].\nn
\eea
We have also introduced the parameter $\eta$ so that we may consider the effect of making the Lamb shift approximation, which would amount to neglecting both the $\Delta\Omega$ and $\widetilde{\Sigma}^{(2)}(s)$ (under the assumption that they cancel one another) in the denominator of Eq. (\ref{xoft}). This approximation is implemented by setting $\eta=0$.

Having introduced all the bookkeeping devices we will need, we can proceed to compute the zeros of $\Phi(s)$ approximately by finding the $u_i$ which solve the equation
\beq
\Phi(s_{R,i}+\mu u_i)=0
\eeq
to leading order in $\mu$, where $s_{R,i}$ satisfies
\beq
s_{R,i}R(s_{R,i})=0,
\eeq
Following this recipe, the $s_i$ are found to be
\beq
s_0=-i{1\over2\tau}\left[1+\sqrt{1+16c^+c^-}\right],\nn
\eeq
\beq
s_1=-i{1\over2\tau}\left[1-\sqrt{1+16c^+c^-}\right],\nn
\eeq
\beq
s_2=-i\varpi_1+{c^-\over2\tau}\left[-i\eta-\sqrt{4c^+/c^--\eta}\right],\nn
\eeq
\beq
s_3=-i\varpi_1+{c^-\over2\tau}\left[-i\eta+\sqrt{4c^+/c^--\eta}\right],\nn
\eeq
\beq
s_4=-i\varpi_2+{c^+\over2\tau}\left[-i\eta-\sqrt{4c^-/c^+-\eta}\right],\nn
\eeq
\beq
s_5=-i\varpi_2+{c^+\over2\tau}\left[-i\eta+\sqrt{4c^-/c^+-\eta}\right],\nn
\eeq
\beq
s_6=-i\varpi_3.\label{thesi}
\eeq
We have only kept up to first order in $\A^2/N\Omega_n^2=(\Omega_n\tau)^{-1}$ in the above expressions for the $s_i$. Notice that $s_6=-i\varpi_3$, so that this root of $\Phi(s)$ is also a root of $R(s)$ and thus is not a pole of $x(s)$. The residues of $x(s)$ at the poles ($s_i$ for $i=0...5$) are $2\pi ir_i$ where
\beq
r_0={x_0\over2}\left(-1+{\eta\over\Omega_n\tau}\right){1-\sqrt{1+16c^+c^-}\over\sqrt{1+16c^+c^-}},\nn
\eeq
\beq
r_1={x_0\over2}\left(1-{\eta\over\Omega_n\tau}\right){1+\sqrt{1+16c^+c^-}\over\sqrt{1+16c^+c^-}},\nn
\eeq
\beq
r_2=-ix_0{c^-\over4\Omega_n\tau}{\left(i\eta+\sqrt{4c^+/c^--\eta}\right)^2\over\sqrt{4c^+/c^--\eta}},\nn
\eeq
\beq
r_3=ix_0{c^-\over4\Omega_n\tau}{\left(-i\eta+\sqrt{4c^+/c^--\eta}\right)^2\over\sqrt{4c^+/c^--\eta}},\nn
\eeq
\beq
r_4=-ix_0{c^+\over4\Omega_n\tau}{\left(i\eta+\sqrt{4c^-/c^+-\eta}\right)^2\over\sqrt{4c^-/c^+-\eta}},\nn
\eeq
\beq
r_5=ix_0{c^+\over4\Omega_n\tau}{\left(-i\eta+\sqrt{4c^-/c^+-\eta}\right)^2\over\sqrt{4c^-/c^+-\eta}}.\label{theri}
\eeq
The electron spin coherence in the rotating frame is then
\beq
x_{NZ}(t)=\sum_{i=0}^5 r_ie^{s_it}.\label{GMEbox}
\eeq

\subsection{Time scale for validity of the NZ GME uniform coupling model solution}

Notice that two of the exponentials contributing to $x(t)$ always diverge in the large-time limit. The terms associated with the poles $s_3$ and $s_5$ exhibit this divergence for nearly all nuclear polarizations. The only exception is the case of maximal polarization, for which $c^+c^-=0$; we will return to this special case shortly. The fact that divergences arise suggests that the NZ  result for the uniform coupling model breaks down on a timescale given roughly by $\tau \sim N\Omega/\A^2$. The positive real parts of the $s_i$ are direct contributions of the fourth-order memory kernel. Even though these contributions appear to introduce a time scale cutoff for the solution, their presence actually extends the validity of the NZ result to larger times as will be shown in the next section when we compare the NZ  solution with the exact solution.

Although the large-time divergences generated by the positive real parts of the $s_i$ indicate that there must be a time scale cutoff, they are not really responsible for this cutoff. In particular, even if no divergences were present, the result we have obtained for $x(t)$ in the uniform coupling model would still be invalid beyond the time scale of $\tau$. This is because we are solving perturbatively for the phases $\hbox{Im}(s_i)$ in the phase factors $\exp[i\hbox{Im}(s_i)t]$ appearing in Eq.~(\ref{GMEbox}). If we solved for $\hbox{Im}(s_i)$ to order $\A^2/N\Omega_n\sim1/\tau$, then the result would be valid up to time scales which are inversely related to the next order in the flip-flop expansion ($t\sim N^2\Omega_n^3/\A^4$) since corrections to the periodic phase factor $\exp[i\hbox{Im}(s_i)t]$ resulting from a small correction $\delta s_i$ will become significant when $\hbox{Im}(\delta s_i)t\sim1$ regardless of the fact that $\A/\sqrt{N}\Omega_n\ll1$. However, it is important to note that we have not obtained the full $O(\A^2/N\Omega_n)$ contributions to the $s_i$ as is suggested by the particular form of the expressions given in Eq.~(\ref{thesi}). Specifically, if we focus only on the fourth-order memory kernel contributions by setting $\eta=0$, it is evident that the $s_i$ still receive corrections at order $\A^2/N\Omega_n\sim1/\tau$. Therefore, we see that higher-order terms in the flip-flop expansion modify the lowest-order corrections to the $s_i$, and we would need to include all terms in the expansion of the memory kernel in order to obtain the full $O(\A^2/N\Omega_n)$ corrections to the $s_i$. Of course, keeping all these terms would yield the exact $s_i$ and $r_i$ and not just the $O(\A^2/N\Omega_n)$ corrections. We conclude that the NZ approach with the simple projector defined in Eq. (\ref{NZprojector}) and in which the memory kernel is computed up to some finite order in flip-flops will yield a solution for the uniform coupling model which is valid only up to the time scale of $t \! < \! \tau$. We have checked that even if one computes the poles of $\Phi(s)$ exactly (numerically), the NZ  result still breaks down at the time scale $t\sim\tau$, in further support of the conclusion that the breakdown on this time scale can only be avoided by keeping higher-order terms in the memory kernel expansion.

A notable exception to the above argument arises in the case of maximal nuclear spin polarization $m=\pm N/2$, where either $c^+=0$ or $c^-=0$. In this case, the fourth-order memory kernel vanishes identically and so does not modify the $s_i$, suggesting that  the NZ  result is valid on arbitrarily large time scales. We will see below that the NZ approach incorporating only the second-order memory kernel yields the exact solution\cite{Khaetskii_PRB03,Coish_PRB04} in this special case, confirming this conjecture.

Note that the above conclusions are compatible with the considerations of relevant timescales in a realistic, inhomogeneouly coupled system. There we expect the box model to apply at short times, $t \! \ll \! N/\A$. The previous works applying the NZ theory to such a problem were focusing on the high field regime, $\A/\Omega \! < \! 1$, in which the long-time solution could be controlled perturbatively. Under this condition, $\tau \! > \! N/\A$, and the NZ theory successfully describes the short-time coherence dynamics.\cite{Coish_PRB10} At low magnetic fields, however, we have $\tau \! < \! N/\A$, and the NZ approach is not expected to work properly at the short timescale.

\subsection{Comparison with the exact solution}
When the nuclear spin polarization is maximal, $m=\pm N/2$, the fourth-order memory kernel vanishes identically. For concreteness, we consider the case $m=N/2$ ($c^+=0$, $c^-=1$), in which the Laplace transform of the spin coherence becomes quite simple:
\beq
x(s)=x_0 {s+i\varpi_1\over s(s+i\varpi_1)+\mu}.
\eeq
Since there are only two poles, we can obtain these exactly,
\beq
s_\pm=-i{\varpi_1\over2}\pm {i\over2}\sqrt{\varpi_1^2+4\mu},
\eeq
and it is likewise a simple matter to compute the residues:
\beq
2\pi ir_\pm={x_0\over2}\left[1\pm {\varpi_1\over\sqrt{\varpi_1^2+4\mu}}\right].
\eeq
The electron spin coherence is then given by
\bea
x(t)=\sum_\pm r_\pm e^{s_\pm t}&=&x_0e^{-i\varpi_1t/2}\Big[\cos\left({t\over2}\sqrt{\varpi_1^2+4\mu}\right)\nn\\&+&i{\varpi_1\over\sqrt{\varpi_1^2+4\mu}}\sin\left({t\over2}\sqrt{\varpi_1^2+4\mu}\right)\Big].\nn\\&&
\eea
Plugging in the expressions for $\varpi_1$ and $\mu$, Eqs. (\ref{varpidef}) and (\ref{mudef}), we obtain the exact solution from Eq. (\ref{exactbox}) with $m=N/2$. The fact that the second-order memory kernel suffices for obtaining the exact solution\cite{Khaetskii_PRB03} for any distribution of $A_{k}$ in the case of maximal polarization was previously pointed out in Ref.~\onlinecite{Coish_PRB04}.

Returning to the case of arbitrary nuclear polarization $m$, it is straightforward to check that the NZ result agrees with the exact solution for short times. In particular, if we consider the case $t\ll\tau$, then we may approximate the poles as follows
\beq
s_0\approx s_1\approx 0,\quad s_2\approx s_3\approx -i\varpi_1,\quad s_4\approx s_5\approx -i\varpi_2,
\eeq
leading to the short-time behavior
\beq
x(t)\approx x_0+{x_0\over\Omega_n\tau}\left(c^-e^{-i\varpi_1t}+c^+e^{-i\varpi_2t}-1\right).\label{GMEboxshorttime}
\eeq
Now turning to the exact solution, Eq. (\ref{exactbox}), we can implement the same approximation by replacing $N_{jm}^+\approx \varpi_2$, $N_{jm}^-\approx \varpi_1$ in the arguments of the sines and cosines arising from $a^*_{jm}$ and $c_{jm}$. Furthermore, we expand the coefficients of the sines to second order in the hyperfine coupling:
\bea
{Z^+_{jm}\over N^+_{jm}}&\approx& 1-{\A^2\over2\varpi_2^2N^2}\left[j(j+1)-m(m+1)\right],\nn\\
{Z^-_{jm}\over N^-_{jm}}&\approx& 1-{\A^2\over2\varpi_1^2N^2}\left[j(j+1)-m(m-1)\right].\label{exacttoNZapproximations}
\eea
At least for low to moderate nuclear polarizations and considering that for typical values of $j$ we may write $j(j+1)\sim N$, this approximation is valid roughly in the limit $\Omega \! \gg \! \mathcal{A}/\sqrt{N}$. In the above expressions, we can replace $\varpi_1$ and $\varpi_2$ with $\Omega_n$ since the differences lead to higher order corrections in $\A/(\Omega N)$. After some algebra, we arrive at
\bea
x(t)&\approx& x_0\left\{1-{2\over\Omega_n\tau NZ}\sum_{j=|m|}^{N/2}n_j\left[j(j+1)-m^2\right]\right\}\nn\\
&+&{x_0\over\Omega_n\tau NZ}e^{-i\varpi_2t}\sum_{j=|m|}^{N/2}n_j\left[j(j+1)-m(m+1)\right]\nn\\
&+&{x_0\over\Omega_n\tau NZ}e^{-i\varpi_1t}\sum_{j=|m|}^{N/2}n_j\left[j(j+1)-m(m-1)\right].\nn\\&&
\eea
The identity
\beq
{2\over NZ}\sum_{j=|m|}^{N/2}n_j\left[j(j+1)-m^2\right]=1,
\eeq
immediately implies that (using also Eq. (\ref{uniformspinhalfcpm}))
\beq
{1\over NZ}\sum_{j=|m|}^{N/2}n_j\left[j(j+1)-m^2\pm m\right]=c^\mp,
\eeq
and we get back the short-time NZ  result, Eq. (\ref{GMEboxshorttime}).

\begin{figure}
\begin{center}
\includegraphics[width=3.0in]{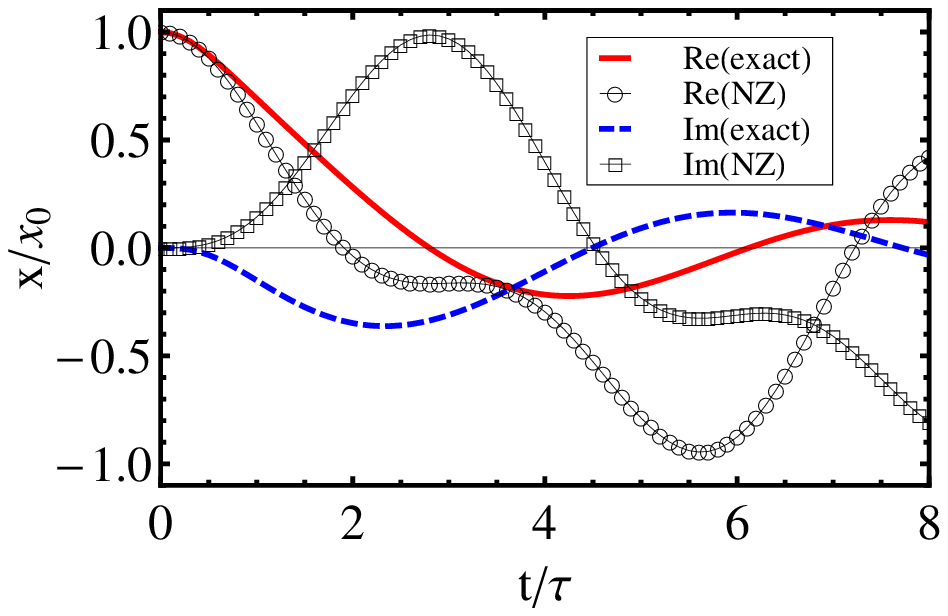} \\
\caption{\label{fig:exactvsGMEM100m0}Exact solution of the uniform coupling model (Eq. (\ref{exactbox})) vs. NZ GME result (Eq. (\ref{GMEbox})) for ${\cal A}=\Omega$, $\omega/\Omega=10^{-3}$, and $m=0$.}
\end{center}
\end{figure}
\begin{figure}
\begin{center}
\includegraphics[width=3.0in]{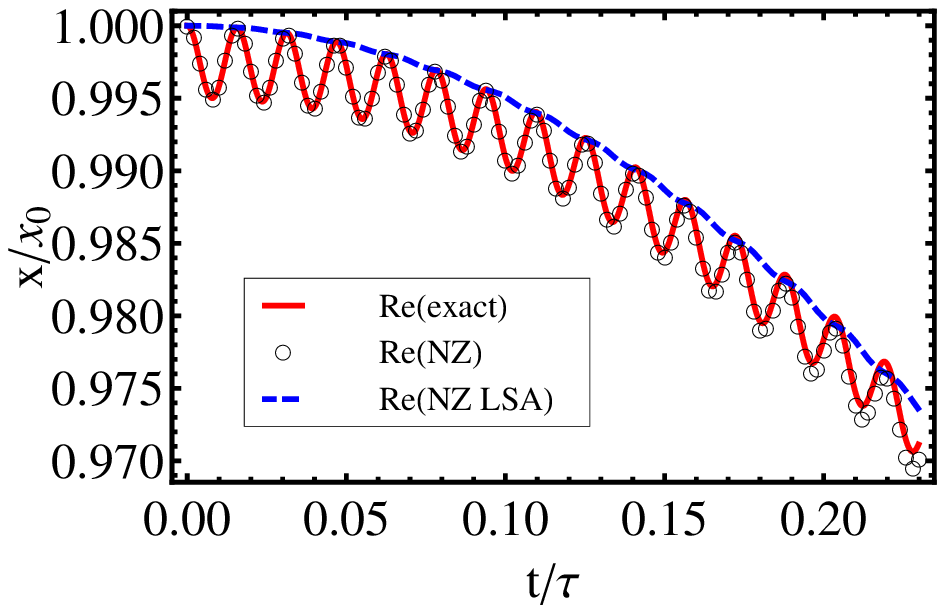} \\
\caption{\label{fig:exactvsGMEM100m0zoom}Zoomed in version of Fig.~\ref{fig:exactvsGMEM100m0} along with the Lamb shift approximation (LSA) of the NZ GME result for $N=100$.}
\end{center}
\end{figure}
\begin{figure}
\begin{center}
\includegraphics[width=3.0in]{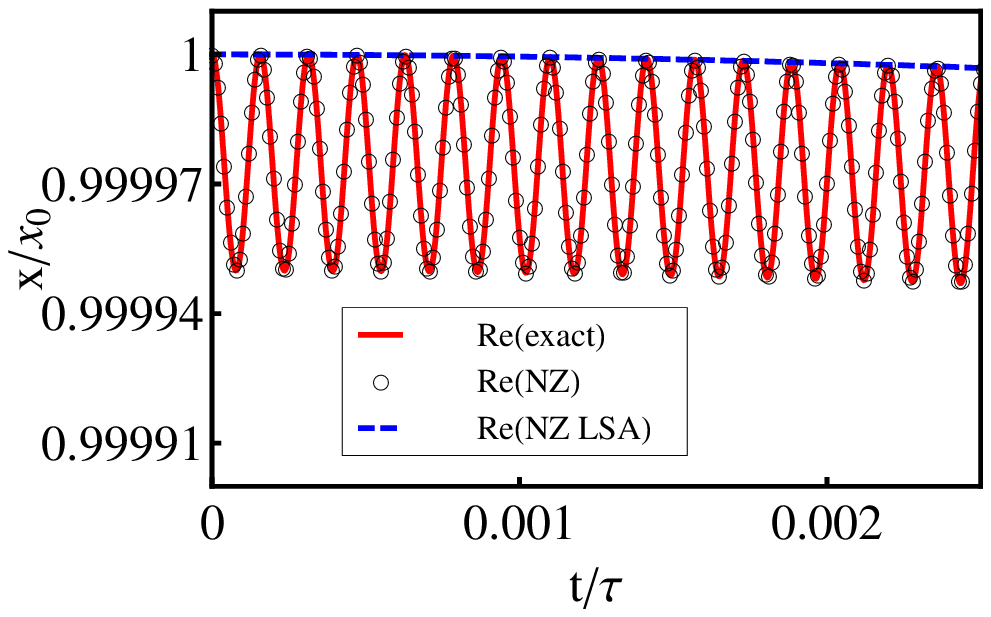} \\
\caption{\label{exactvsGMEM10000m0zoom4}Zoomed in version of Fig.~\ref{fig:exactvsGMEM100m0} along with the Lamb shift approximation (LSA) of the NZ GME result for $N=10^4$.}
\end{center}
\end{figure}
\begin{figure}
\begin{center}
\includegraphics[width=3.0in]{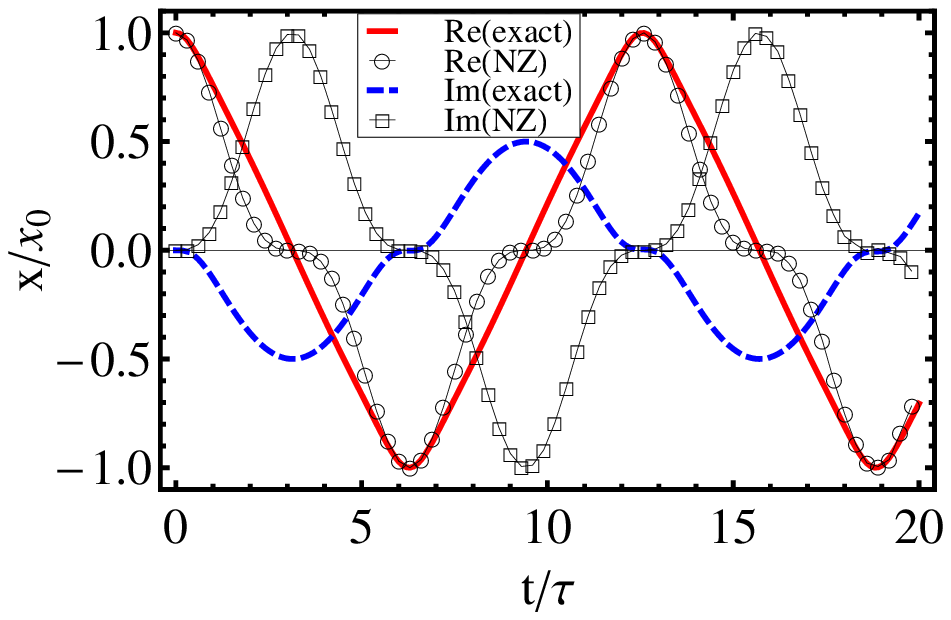} \\
\caption{\label{exactvsGMEM10000mMover4}Exact solution of the uniform coupling model (Eq. (\ref{exactbox})) vs. NZ GME result (Eq. (\ref{GMEbox})) for $N=10^4$, ${\cal A}=\Omega$, $\omega/\Omega=10^{-3}$ and $m=N/4$.}
\end{center}
\end{figure}
\begin{figure}
\begin{center}
\includegraphics[width=3.0in]{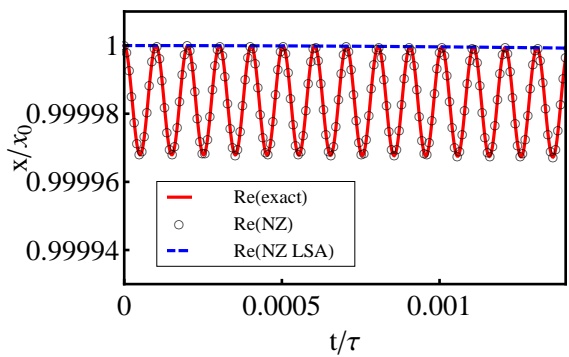} \\
\caption{\label{exactvsGMEM10000mMover4zoom} Zoomed in version of Fig.~\ref{exactvsGMEM10000mMover4} along with the Lamb shift approximation (LSA) of the NZ GME result for $N=10^4$.}
\end{center}
\end{figure}

In Fig.~\ref{fig:exactvsGMEM100m0}, we compare the NZ  result with the exact solution in the case of zero net nuclear polarization ($m=0$), and it is clear that the two agree only for short times $t\lesssim \tau/2$. We have not specified the number of nuclei in the caption of Fig.~\ref{fig:exactvsGMEM100m0} because the curves are valid for a wide range of $N$. This is because the envelope of $x(t/\tau)$ is essentially independent of $N$ as can be seen by noticing that the envelope arises from the first two contributions, $r_0e^{s_0t}+r_1e^{s_1t}$, and these contributions only depend on $N$ through their $\tau$-dependence.\footnote{In the case of very long times, the last four contributions to $x(t)$ become comparable to the first two, and all six terms generate the envelope, but this is well beyond the regime of validity $t\lesssim \tau$ of the NZ GME result. Also note that in addition to the dependence of the first two contributions on $N$ coming from $\tau$, there is a weak $N$-dependence coming from the $c^\pm$ factors in the case of non-zero and non-maximal nuclear polarization as well.} The remaining four contributions to $x(t)$ give rise to a small modulation which is visible in Figs.~\ref{fig:exactvsGMEM100m0zoom} and \ref{exactvsGMEM10000m0zoom4}. The frequency and amplitude of this modulation are given roughly by $\Omega_n\tau$ and $1/(\Omega_n\tau)$ respectively. Since $\Omega_n\tau=4N\Omega_n^2/{\cal A}^2$, we see that the frequency scales linearly with $N$, while the amplitude scales like $1/N$. This behavior is evident in a comparison of Figs.~\ref{fig:exactvsGMEM100m0zoom} and \ref{exactvsGMEM10000m0zoom4}.

It should be stressed that while the value of $\mathcal{A}/\Omega$ plays a crucial role for the long-time results in the non-uniform coupling case,\cite{Coish_PRB04,Coish_PRB10} this parameter does not have any particular significance in the uniform coupling case considered here. The amplitude of the fast oscillation in the exact and NZ results depends of course on the ratio of $\mathcal{A}/\Omega$, but for large $N$ the amplitude of this oscillation does not become significant in the range of $\Omega$ which we consider, i.e.~$\Omega \! \gg \! \mathcal{A}/\sqrt{N}$. On the other hand, the envelope of the NFID signal depends on $\Omega$ and $\mathcal{A}$ only through $\tau \approx (N/\mathcal{A})\cdot (\Omega/\mathcal{A})$. When varying the parameter $\mathcal{A}/\Omega$ between $0.1$ and $10$, the envelopes in Fig.~\ref{fig:exactvsGMEM100m0} do not change, only the fast oscillation components change; the (dis)agreement between the NZ solution and the exact result remains unchanged.

Both the envelope and modulation change under variations of the nuclear polarization $m$. The case $m=N/4$ is shown in Figs.~\ref{exactvsGMEM10000mMover4} and \ref{exactvsGMEM10000mMover4zoom}. Fig.~\ref{exactvsGMEM10000mMover4} reveals that $m=N/4$ is in some sense special because the NZ  approach is able to reproduce the oscillation period of the $x(t)$ envelope along with the amplitude of its real part. We have tried various nuclear polarizations and found that the NZ approach is only able to reproduce these features in the $m=N/4$ case. Generically, the single-projector NZ result visibly disagrees with the exact solution beyond $t\sim\tau$.

The fact that Figs.~\ref{fig:exactvsGMEM100m0}-\ref{exactvsGMEM10000mMover4zoom} show that the NZ solution, Eq.~(\ref{GMEbox}), disagrees with the exact answer in Eq.~(\ref{exactbox}) beyond time $t\sim\tau$ is not surprising in light of the fact that the exact solution is a sum over a large number of oscillatory functions, while the NZ result consists of a sum of only six exponentials.
In the case of a realistic wavefunction, the NZ approach can lead to more complicated behavior due to an appearance of branch cuts in the continuum limit \cite{Coish_PRB04,Coish_PRB10} It is should also be strongly stressed that for a realistic wavefunction, a new pole related to flip-flops between nuclei with similar, but distinct, values of $A_{k}$, appears. In fact, this pole dominates the high-field and long-time decay of coherence, and its influence is well-controlled for times much longer than $\tau$ and $T_{2} \approx \tau (\frac{\Omega}{\mathcal{A}})$, as discussed in Ref.~\onlinecite{Coish_PRB10}. This is in stark contrast to the behavior of a few poles determining the coherence dynamics in the uniform coupling case.

%%%%%%%%%%%%%%
%%% CORRELATED PROJECTORS
%%%%%%%%%%%%%%
\subsection{The NZ solution of the uniform coupling model using correlated projectors}  \label{sec:correlated_box}
We have seen that a straightforward application of the standard NZ GME to the uniform coupling model yields a result which is in good agreement with the exact solution only for very short times. From the technical point of view, this failure of the NZ approach could be traced back to not having enough poles in the Laplace transformed solution, i.e.~not having enough frequencies to sum over in the real-time expression. Clearly, in order to reproduce the exact answer one needs to find a way of bringing all these missing frequencies back into the theory.

The above NZ results were obtained using a standard choice for the projection operator $P$, which was defined in Eq.~(\ref{NZprojector}). However, it was shown in Ref.~\onlinecite{Fischer_PRA07} that the standard projection operator is far from being the best possible choice in contexts where the Hamiltonian exhibits a significant degree of symmetry. When symmetries are present, one can instead replace $P$ with a series of so-called {\sl correlated} projection (CP) operators which project onto invariant subspaces of state space, enabling one to expand the reduced density matrix for the system as a sum of matrices, each capturing the components of the state lying in a particular subspace. This greatly enhances the number of dynamical degrees of freedom, resulting in a remarkable improvement in the agreement between the NZ calculation and the exact result. In this section we will show that, whereas the standard fourth-order NZ  result agrees rather poorly with the exact solution, even at second order the NZ GME agrees extremely well with the exact answer when correlated projectors are employed.

\begin{figure}
\begin{center}
\includegraphics[width=3.0in]{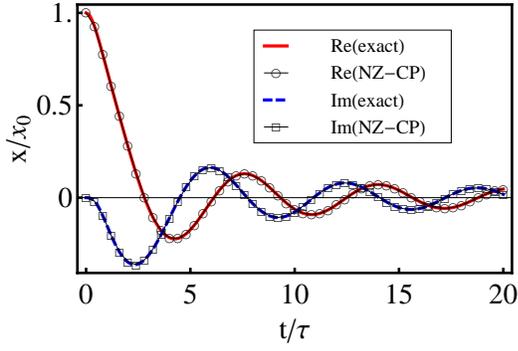}
\caption{Exact solution of the uniform coupling model (Eq. (\ref{exactbox})) vs. NZ-CP result (Eq. (\ref{xGMEcp})) for $N=10^4$, ${\cal A}=\Omega$, $\omega/\Omega=10^{-3}$ and $m=0$.} \label{fig:NZcp}
\end{center}
\end{figure}

\begin{figure}
\begin{center}
\includegraphics[width=3.0in]{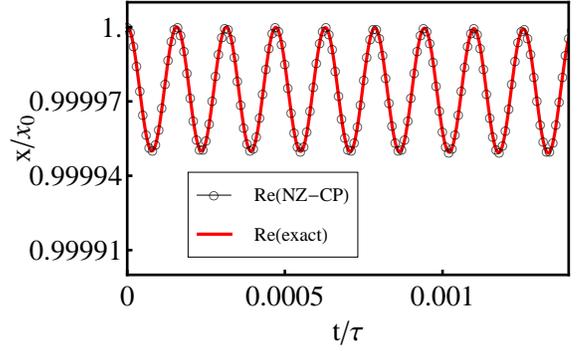} \\
\caption{ Zoomed in version of Fig.~\ref{fig:NZcp}.} \label{fig:NZcpzoom}
\end{center}
\end{figure}

For this calculation, we choose to work in the interaction picture defined with respect to the unperturbed (Zeeman and Overhauser) part of the Hamiltonian, $H_0$. In terms of the Liouville operator $L_I$ and total density matrix $\tilde\rho$ in the interaction picture, the second-order NZ GME equation for the electron spin coherence $\tilde\rho_{e,-+}$ becomes
\beq
\dot{\tilde\rho}_{e,-+}=-\int_0^t dt' \hbox{Tr}\left\{S^+PL_I(t)L_I(t')P\tilde\rho(t')\right\}.\label{NZinteqn}
\eeq
For the uniform coupling model, $L_I$ acts on an arbitrary density matrix as
\bea
&&L_I\left(\begin{matrix} \rho_{++}& \rho_{+-} \cr \rho_{-+}&\rho_{--}\end{matrix}\right)\nn\\&&={1\over2}\left(\begin{matrix} h_{+-}\rho_{-+}-\rho_{+-}h_{-+} \,\,\, & h_{+-}\rho_{--}-\rho_{++}h_{+-}\cr h_{-+}\rho_{++}-\rho_{--}h_{-+} \,\,\, & h_{-+}\rho_{+-}-\rho_{-+}h_{+-}\end{matrix}\right),\nn\\&&
\eea
with
\bea
h_{+-}&\equiv&(\A/N)e^{i(\Omega-\A/(2N))t} I^- e^{i(\A/N)t I^z},\nn\\
h_{-+}&\equiv&(\A/N)e^{-i(\Omega+\A/(2N))t} I^+ e^{-i(\A/N)t I^z}=h_{+-}^\dagger,\nn\\&&
\eea
where $I^\pm=\sum_kI^\pm_k$ are the total nuclear creation and annihilation operators. At this point, we are ready to define the correlated projection operators appropriate for the uniform coupling model. In Section \ref{sec:exact_uniform}, we already made use of the fact that in the case of uniform couplings, the total nuclear angular momentum is conserved, and we may work in the $\ket{jm}$ basis of nuclear states. Defining the nuclear operator $\Pi_{jm}$ to be the projector onto the subspace of fixed $j$ and $m$ quantum numbers, we choose the superprojector $P$ to be such that it acts on the total density matrix as\cite{Fischer_PRA07}
\beq
P\tilde\rho=\sum_{jm}\hbox{Tr}_I(\Pi_{jm}\tilde\rho)\otimes {1\over n_j}\Pi_{jm}\equiv\sum_{jm}\tilde\rho^{jm}_e\otimes {1\over n_j}\Pi_{jm}.\label{corproj}
\eeq
The $\tilde\rho^{jm}_e$ are a set of matrices which sum to give the reduced density matrix for the electron spin:
\beq
\tilde\rho_e=\sum_{jm}\tilde\rho^{jm}_e=\sum_{m=-N/2}^{N/2}\sum_{j=|m|}^{N/2}n_j\tilde\rho^{jm}_e.
\eeq
Inserting Eq. (\ref{corproj}) into Eq. (\ref{NZinteqn}), we obtain after some algebra
\bea
\dot{\tilde\rho}_{e,-+}^{jm}(t)&=&-{1\over4}\int_0^tdt'\Big[e^{-iZ^+_m(t-t')}(X^+_{jm})^2 \nn\\&+& e^{iZ^-_m(t-t')}(X^-_{jm})^2\Big]{\tilde\rho}_{e,-+}^{jm}(t').\nn\\&&\label{NZinteqnii}
\eea
where $X^\pm_{jm}$ and $Z^\pm_m$ were defined earlier in Eq. (\ref{XZNdefs}). Taking the Laplace transform of both sides of Eq. (\ref{NZinteqnii}), we obtain an algebraic equation for the Laplace transform of ${\tilde\rho}_{e,-+}^{jm}(t)$ (which we call $\tilde\rho(s)^{jm}$ for brevity) which is readily solved:
\bea
\tilde{\rho}^{jm}(s)&=&\rho^{jm}_0(s+iZ^+_m)(s-iZ^-_m)\bigg\{s^3+i(Z^+_m-Z^-_m)s^2\nn\\&+& [Z^+_mZ^-_m+{1\over4}(X^+_{jm})^2+{1\over4}(X^-_{jm})^2]s\nn\\&+&{i\over4}[Z^+_m(X^-_{jm})^2-Z^-_m(X^+_{jm})^2]\bigg\}^{-1}.
\eea
$\rho^{jm}_0$ is the intial value of ${\tilde\rho}_{e,-+}^{jm}(t)$. To obtain ${\tilde\rho}_{e,-+}^{jm}(t)$, we must use the Bromwich inversion formula:
\beq
{\tilde\rho}_{e,-+}^{jm}(t)={\rho_0\over2\pi i}\lim_{\gamma\to0}\int_{\gamma-i\infty}^{\gamma+i\infty}ds e^{st}\tilde{\rho}^{jm}(s).
\eeq
The value of this integral is determined by the three poles of $\tilde{\rho}^{jm}(s)$, which we will call $s^{jm}_1$, $s^{jm}_2$, $s^{jm}_3$, and by their associated residues $r^{jm}_1$, $r^{jm}_2$, $r^{jm}_3$. In terms of these poles, the solution is
\beq
{\tilde\rho}_{e,-+}^{jm}(t)=\rho^{jm}_0\sum_{i=1}^3 r^{jm}_i e^{s^{jm}_it},
\eeq
where the residues are given by
\bea
r^{jm}_1&=& {(s^{jm}_1+iZ^+_m)(s^{jm}_1-iZ^-_m)\over(s^{jm}_1-s^{jm}_2)(s^{jm}_1-s^{jm}_3)},\nn\\
r^{jm}_2&=& {(s^{jm}_2+iZ^+_m)(s^{jm}_2-iZ^-_m)\over(s^{jm}_2-s^{jm}_1)(s^{jm}_2-s^{jm}_3)},\nn\\
r^{jm}_3&=& {(s^{jm}_3+iZ^+_m)(s^{jm}_3-iZ^-_m)\over(s^{jm}_3-s^{jm}_1)(s^{jm}_3-s^{jm}_2)}.
\eea
To complete the solution, we must give the explicit forms of the $s^{jm}_i$:
\bea
s^{jm}_1&=&-{a\over3}-{2^{1/3}\over 3d}(3b-a^2)+{d\over 2^{1/3}3},\nn\\
s^{jm}_2&=&-{a\over3}+{(1+i\sqrt{3})(3b-a^2)\over2^{2/3}3d}-{(1-i\sqrt{3})d\over2^{1/3}6},\nn\\
s^{jm}_3&=&-{a\over3}+{(1-i\sqrt{3})(3b-a^2)\over2^{2/3}3d}-{(1+i\sqrt{3})d\over2^{1/3}6},\nn\\&&
\eea
with
\bea
d&\equiv& \Big(-2a^3+9ab-27c\nn\\&+&3\sqrt{3}\sqrt{-a^2b^2+4b^3+4a^3c-18abc+27c^2}\Big)^{1/3},\nn\\&&
\eea
and
\bea
a&=&i(Z^+_m-Z^-_m),\nn\\
b&=&Z^+_mZ^-_m+{1\over4}(X^+_{jm})^2+{1\over4}(X^-_{jm})^2,\nn\\
c&=&{i\over4}[Z^+_m(X^-_{jm})^2-Z^-_m(X^+_{jm})^2].
\eea
Since the interaction picture can be thought of as the frame rotating with frequency $\Omega_n$, to switch to the rotating frame with frequency $\Omega_n+\Delta\Omega$ we simply need to multiply the result by $e^{-i\Delta\Omega t}$. The electron spin coherence in the rotating frame and in the presence of fixed nuclear polarization $m$ is then
\beq
x_{NZ-CP}(t)={x_0\over Z}e^{-i\Delta\Omega t}\sum_{j=|m|}^{N/2}n_j\sum_{i=1}^3 r^{jm}_i e^{s^{jm}_it}.\label{xGMEcp}
\eeq
This solution is plotted along with the exact solution in Figs.~\ref{fig:NZcp} and \ref{fig:NZcpzoom}. It is clear from these figures that the NZ solution with correlated projectors (NZ-CP) agrees remarkably well with the exact solution.

It is not difficult to extract analytically the envelope and modulation from the NZ-CP solution. To do this, it helps to first perform a large $N$ expansion, wherein the leading order terms in the $s^{jm}_i$ are
\bea
s^{jm}_1 &\approx& i Z^-_m-i {(X^-_{jm})^2\over 4\Omega_n},\nn\\
s^{jm}_2 &\approx& i {(X^+_{jm})^2+(X^-_{jm})^2\over4\Omega_n},\nn\\
s^{jm}_3 &\approx& -i Z^+_m-i{(X^+_{jm})^2\over 4\Omega_n},
\eea
while those of the $r^{jm}_i$ are
\bea
r^{jm}_1 &\approx& {(X^-_{jm})^2\over 4\Omega_n^2},\nn\\
r^{jm}_2 &\approx& 1-{(X^+_{jm})^2+(X^-_{jm})^2\over4\Omega_n^2},\nn\\
r^{jm}_3 &\approx& {(X^+_{jm})^2\over 4\Omega_n^2}.
\eea
In the limit of large $N$, the coherence therefore takes the form
\bea
x_{NZ-CP}(t)&\approx&{x_0\over Z}e^{-i\Delta\Omega t}\sum_{j=|m|}^{N/2}n_j e^{i{(X^+_{jm})^2+(X^-_{jm})^2\over4\Omega_n}t}\nn\\&\times&\Bigg\{1-{(X^+_{jm})^2+(X^-_{jm})^2\over4\Omega_n^2} \nn\\&+& {(X^+_{jm})^2\over4\Omega_n^2}e^{-i \left[Z^+_m+{(X^+_{jm})^2\over4\Omega_n}\right]t}\nn\\&+& {(X^-_{jm})^2\over4\Omega_n^2}e^{-i \left[-Z^-_m+{(X^-_{jm})^2\over4\Omega_n}\right]t}\Bigg\}.\label{approxGMEcp}
\eea
The factor in curly brackets in Eq. (\ref{approxGMEcp}) is precisely that which gives rise to the small modulation depicted in Fig. \ref{fig:NZcpzoom}, while the rest of the expression produces the envelope in Fig.~\ref{fig:NZcp}.

The above expression can be obtained from the exact solution, Eq. (\ref{exactbox}), by expanding in the limit $X^\pm_{jm}\ll Z^\pm_m$, in which case we find
\bea
a_{jm}^*&\approx&e^{i{N_{jm}^+\over2}t}\left[1-{(X^+_{jm})^2\over4(Z^+_m)^2}\left(1-e^{-iN_{jm}^+t}\right)\right],\nn\\
c_{jm}&\approx&e^{i{N_{jm}^-\over2}t}\left[1-{(X^-_{jm})^2\over4(Z^-_m)^2}\left(1-e^{-iN_{jm}^-t}\right)\right],
\eea
where $a_{jm}$, $c_{jm}$ and $N_{jm}^\pm$ were defined in Eqs. (\ref{abcddefs}) and (\ref{XZNdefs}). These expressions result from expanding only the $Z^\pm_m/N^\pm_{jm}$ factors appearing in $a_{jm}$ and $c_{jm}$. If we furthermore expand the $N_{jm}^\pm$ factors in the temporal exponents as
\beq
N_{jm}^\pm\approx |Z_m^\pm|+{(X^\pm_{jm})^2\over4|Z^\pm_m|},\label{approxNjm}
\eeq
and also make the following approximation in the denominators,
\beq
Z^\pm_m\approx \Omega_n,
\eeq
then we get back Eq. (\ref{approxGMEcp}).

If we assume typical values of $j$ and $m$ with $j\gg m$, then we have roughly $X^\pm_{jm}\sim {\cal A}/\sqrt{N}$, and the above approximations are valid in the limit $\Omega\gg{\cal A}/\sqrt{N}$. Recall that this is the same condition we had for the validity of the NZ solution without correlated projectors. In fact, the approximations we have made in the exact solution to arrive at the NZ-CP result are essentially the same ones we made to relate the exact and NZ solutions (see Eq. (\ref{exacttoNZapproximations})), the only difference being that here we have kept some of the $j$-dependence in the exponents when evaluating the sum over $j$.  Moreover, note that the expansion of the $N^\pm_{jm}$ factor appearing in the temporal exponent of the exact solution introduces a time scale on which the solution is valid. It is somewhat difficult to say precisely what the time scale of validity is, however, because of the sum over $j$. To obtain a very rough estimate, we can again consider a typical value of $j$ and write $X^\pm_{jm}\sim {\cal A}/\sqrt{N}$. The timescale should roughly correspond to the inverse of the error introduced by expanding $N^\pm_{jm}$ in the temporal exponent. This error is on the order of
\beq
N_{jm}^\pm-|Z_m^\pm|-{(X^\pm_{jm})^2\over4|Z^\pm_m|}\sim{(X^\pm_{jm})^4\over |Z^\pm_m|^3}\sim {{\cal A}^4\over N^2\Omega^3},
\eeq
(assuming that $j\gg m$), indicating a time scale
\beq
t\sim {N^2\Omega^3\over {\cal A}^4}.
\eeq

\subsection{Comparison with the RDT solution of the uniform coupling model}
We now review the effective Hamiltonian RDT solution to the box model and show how it is related to the exact solution and the NZ-CP result of the previous section. At second order in the flip-flop expansion, we have processes in which the electron spin flip-flops with the $k$th nucleus and then subsequently flip-flops with the $\ell$th nucleus. This leads to an effective flip-flop interaction between the $k$th and $\ell$th nuclear spins which is mediated by the hyperfine interaction. We can derive the effective Hamiltonian describing this internuclear flip-flop process by performing a canonical transformation on the hyperfine Hamiltonian in Eqs. (\ref{ham})-(\ref{hi}): $H_{eff}=e^{-{\cal S}}He^{{\cal S}}$, where the unitary operator $e^{-{\cal S}}$ is chosen in such a way as to eliminate the original $V_{\text{ff}}$ interaction.\cite{Shenvi_scaling_PRB05,Yao_PRB06,Coish_PRB08,Cywinski_PRB09} It is important to note that the transformation on states, $\ket{\psi}\to e^{-{\cal S}}\ket{\psi}$, is neglected in the derivation of $H_{eff}$. The dominant terms in the resulting effective Hamiltonian are
\bea
H_{eff}&=&-\sum_k {A_k^2\over4\Omega}I_k^z+S^z\sum_k{A_k^2\over2\Omega}\left(I_k^2-(I_k^z)^2\right)\nn\\&+& S^z\sum_{k\ne \ell}{A_kA_\ell\over2\Omega}I_k^+I_\ell^-.\label{hameff}
\eea

\begin{figure}
\begin{center}
\includegraphics[width=3.0in]{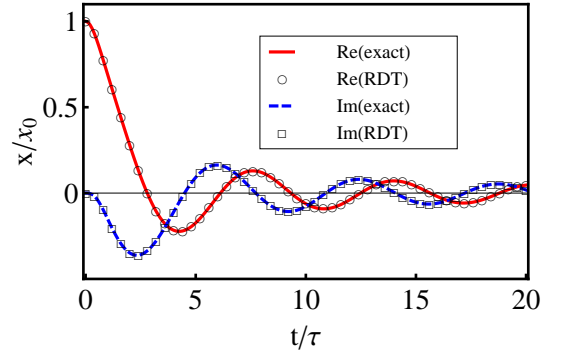} \\
\caption{\label{fig:exactvsRDTM10000m0}Exact solution of the uniform coupling model (Eq. (\ref{exactbox})) vs. RDT result (Eq. (\ref{RDTbox})) for $N=10^4$, ${\cal A}=\Omega$, $\omega/\Omega=10^{-3}$ and $m=0$.}
\end{center}
\end{figure}

\begin{figure}
\begin{center}
\includegraphics[width=3.0in]{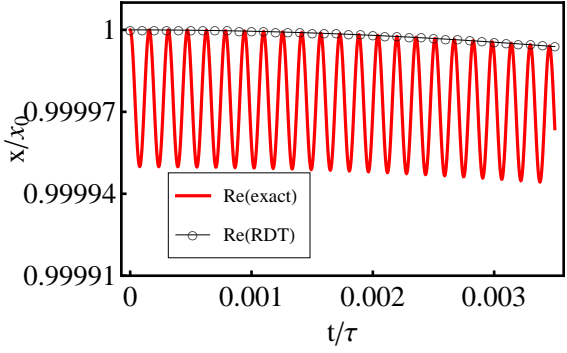} \\
\end{center}
\caption{\label{fig:exactvsRDTM10000m0zoom}Zoomed in version of Fig.~\ref{fig:exactvsRDTM10000m0}.}
\end{figure}

The effective Hamiltonian solution for NFID was obtained in Ref.~\onlinecite{Cywinski_PRB09} in the case of zero nuclear polarization, $m=0$. In the rotating frame, it is given by
\beq
x(t)=x_0e^{-i\Delta\Omega t}{\exp[i\arctan(t/\tau)]\over\sqrt{1+(t/\tau)^2}}={x_0e^{-i\Delta\Omega t}\over1-it/\tau}\,\, . \label{RDTbox}
\eeq
This result is compared with the exact solution, Eq.~(\ref{exactbox}), in Figs.~\ref{fig:exactvsRDTM10000m0} and \ref{fig:exactvsRDTM10000m0zoom}. To generate these figures, we have used the high-frequency form of the Lamb shift, Eq.~(\ref{lambshift2}), where for zero polarization, $c^\pm_k=1/2$ (see Appendix \ref{app:correlators}). It is clear from Fig.~\ref{fig:exactvsRDTM10000m0} that the envelope of the RDT solution agrees very well with that of the exact solution up to time scales well beyond $\tau$. The fact that the two envelopes agree even at very large time scales is particularly interesting given that, in the case of RDT, $x\sim 1/t$ for $t\gg\tau$; this $1/t$ tail is evidently reproduced by the sum of many exponential terms comprising the exact solution. On the other hand, Fig.~\ref{fig:exactvsRDTM10000m0zoom} reveals that RDT does not succeed in capturing the fast small-amplitude modulation of the coherence.

It is not difficult to extract the RDT solution as a limit of the exact solution. First recall that the RDT result using the effective Hamiltonian is expected to be valid\cite{Cywinski_PRB09} for $\Omega \! \gg \! {\cal A}/\sqrt{N}$ and in the limit of large $N$. Starting from the exact solution, Eq.~(\ref{exactbox}), we can apply the first of these approximations along with the fact that for the $j$ states which contribute the most to the sums we have $j(j+1)\sim N$, to write
\beq
Z_m^\pm\approx\pm\Omega,\qquad N_{jm}^\pm\approx\Omega\left[1+{\A^2\over2\Omega^2N^2}j(j+1)\right],\label{approxZN}
\eeq
so that we obtain
\beq
x(t)\approx x_0e^{-i\Delta\Omega t}{1\over Z}\sum_{j=0}^{N/2}n_je^{i{2\over N}j(j+1)t/\tau}.\label{approxexact}
\eeq
Expanding the exponential and using that in the large $N$ limit,
\beq
{1\over Z}\sum_{j=0}^{N/2}n_jj^k(j+1)^k={k!N^k\over2^k}+O(N^{k-1}),
\eeq
we arrive at the RDT solution:
\beq
x(t)\approx x_0e^{-i\Delta\Omega t}\sum_{k=0}^\infty \left({it\over\tau}\right)^k={x_0e^{-i\Delta\Omega t}\over1-it/\tau}.\label{RDTfromexact}
\eeq
Strictly speaking, this derivation is only valid for $t<\tau$ since otherwise the infinite series in Eq.~(\ref{RDTfromexact}) is not defined, and the last line in Eq.~(\ref{RDTfromexact}) should be treated as an analytical continuation of the previous expression along the real axis (a similar kind of analytical continuation was in fact encountered in the derivation of the RDT result in Ref.~\onlinecite{Cywinski_PRB09}).
The correctness of the analytical continuation can be easily checked by comparing a numerical evaluation of Eq.~(\ref{approxexact}) with Eq.~(\ref{RDTbox})  in the limit $N\to\infty$ for all times $t$.

Notice that the approximations we have made in the exact solution to arrive at the RDT result are essentially the same ones we made to relate the exact and NZ-CP solutions (see Eq. (\ref{approxNjm})), the only difference being that in the context of NZ-CP, we kept one more order in the expansion, yielding the additional factor in curly brackets in Eq. (\ref{approxGMEcp}). This additional factor is precisely that which gives rise to the small modulation depicted in Fig. \ref{fig:NZcpzoom}. Thus, the RDT and NZ-CP solutions of the box model belong to the same expansion of the exact solution. Moreover, note that it is the expansion of the $N^\pm_{jm}$ factor appearing in the temporal exponent of the exact solution that introduces a time scale on which the NZ-CP solution is valid. The fact that we are making the very same expansion here implies that the RDT solution will be valid on a time scale similar to that of the NZ-CP result.

%%%%%%%%%%%%%%
%%% DISCUSSION
%%%%%%%%%%%%%%
\section{DISCUSSION}  \label{sec:discussion}
In this Section, we aim to understand the reasons for the failure of the standard (single-projector) NZ approach discussed in Sec.~\ref{sec:NZ_box}, and for the success of the correlated projector approach from the previous Section.

\subsection{Non-narrowed Gaussian bath}
It is instructive to take a small detour, and discuss first the NZ solution for the case of Ising coupling to the bath spins,\cite{Krovi_PRA07} for which $V_{\text{ff}} \! = \! 0$. Furthermore, let us abandon for a moment the assumption that the state of the bath is narrowed, and take $\rho_{I} \sim 1$.

This case is of course exactly solvable,\cite{Krovi_PRA07} and in the large $N$ limit the results are equivalent to the classical calculation,\cite{Merkulov_PRB02}  in which the bath is assumed to be static, and the central spin dynamics is calculated by averaging the spin precession over a Gaussian ensemble of nuclear polarization vectors $\mathbf{B}_{I}$ of the form $P(\mathbf{B}^{2}_{I}) \sim  \exp \left( -\mathbf{B}^{2}_{I}/2\sigma^2 \right )$, where $\sigma^{2} \! =\! \frac{1}{3} \sum_{k} I_{k}(I_{k}+1) A^{2}_{k}$. A simple Gaussian integral gives the well-known $\exp[ -(t/T^{*}_{2})^2 ]$ decay of the envelope of the FID signal, with $T_{2}^{*} \! = \! \sqrt{2}/\sigma$. For large $N$, this is practically equivalent to the evaluation of the quantum expression
\beq
\langle S^{+}(t) \rangle =  \frac{1}{2^{N}} e^{i\Omega t} \sum_{n_{i}} \exp ( i h^{z}_{n_{i}}t ) \langle S^{+}(0) \rangle \,\, . \label{eq:Ising}
\eeq
Note that in the above expression, every $\ket{n_{i}}$ state present in $\rho_{I}(0)$ contributes a frequency $h^{z}_{n_{i}}$, and the averaging of these frequencies leads to dephasing (inhomogeneous broadening).

It was noticed before that the standard NZ approach, when carried out to a finite order in the expansion (in powers of $h_{z}$ in this case), cannot reproduce the above simple solution.\cite{Krovi_PRA07} On the other hand, the TCL approach gives the exact result already in the 2nd order of expansion. This feature can be traced back to the fact that in the NZ expansion, one obtains so-called \emph{partial cumulants}\cite{Breuer,Krovi_PRA07} of the bath operators (i.e.~$\langle h^{4}_{z} \rangle_{pc}  \! =\! \langle h^{4}_{z} \rangle - \langle h^{2}_{z}\rangle^2$ appears in the 4th order), while in the TCL calculation one encounters the \emph{ordered cumulants} (i.e.~$\langle h^{4}_{z} \rangle_{oc}  \! =\! \langle h^{4}_{z} \rangle - 3\langle h^{2}_{z}\rangle^2$), which are clearly closely related to the usual cumulants. For the Gaussian bath, the ordered cumulants beyond 2nd order are identically zero, and the TCL approach succeeds because it captures the crucial statistical properties of the bath already in the 2nd order of the expansion.\cite{Doll_CP08} The NZ approach, on the other hand, is incompatible with the structure of the bath correlators, and it has to be carried out to infinite order to recover the Gaussian decay of the transverse spin.

\subsection{Uniform coupling model as a classical non-Gaussian bath}
The effective Hamiltonian-based RDT solution for dephasing in the uniform coupling model, Eq.~(\ref{RDTbox}), which is the same as the large $N$ limit of the exact solution (see the discussion leading to Eq.~(\ref{RDTfromexact})), can also be obtained from a classical calculation.\cite{Cywinski_APPA11} We simply fix the $B^{z}_{I}$ component of the nuclear field, perform the Gaussian average of the classical equations of motion for $S^{x,y}$, and recover the RDT results. The fast oscillation of the exact solution is clearly an effect which is being missed by the classical large-$N$ limit. The same result is obtained when one employs the classical limit of the effective Hamiltonian, $S^{z} [ (B^{x}_{I})^2 +(B^{y}_{I})^2 ]/2\Omega$ (as was done for the spin echo case in Ref.~\onlinecite{Neder_PRB11}), and performs the Gaussian average over all the precession frequencies due to the possible magnitudes of the transverse nuclear field. This shows that the RDT (at short-times in the general case) and the uniform coupling calculations are essentially equivalent to classical averaging over a \emph{square} of a Gaussian-distributed classical variable. The latter approach was succesfully applied to the calculation of the Rabi oscillations decay of spin qubits.\cite{Dobrovitski_01,Taylor_QIP06,Koppens_PRL07,Hanson_Science08}

In the case of a bath operator coupling which is the square of a Gaussian-distributed variable, the bath correlators are such that cumulants of all orders are nonzero. However, these cumulants are simply the ring diagrams,\cite{Makhlin_PRL04,Cywinski_PRB09} and the summation of all of them can be performed. In order to achieve this, it is crucial to use a theoretical approach in which the standard, i.e.~ordered, cumulants appear in a natural fashion, as it occurs in the effective Hamiltonian-based RDT calculation, or in the TCL approach.\cite{Breuer}

As we discussed in the previous section, the standard (single-projector) NZ theory does not have a natural relation with the statistical properties of the bath, specifically with the structure of the ordered cumulants of the bath variables.  Furthermore, in the standard NZ approach we project on a single tensor product state of the system and a fixed state of the environment (chosen here, as it is usually done, to be $\rho_{I}(0)$). The Ising coupling case clearly shows that this can a sub-optimal approximation: in this case, the central spin coherence decays \emph{only} due to the fact that the exact density matrix at finite times possesses a nontrivial structure deriving from the evolution of each of the nuclear states present in $\rho_{I}(0)$, i.e.~in Eq.~(\ref{eq:Ising}) we have a sum over all the $\ket{n_{i}}$ states, and each state contributes a different frequency. What is explicitly shown in the exact solution of the uniform coupling model is the fact that, even for the narrowed state of the bath, this is the case. The exact solution in Eq.~(\ref{exactbox}) involves a sum over the $j$ quantum numbers: in order to obtain the correct result, one has to preserve the structure of the nuclear density matrix which is singled out by its coupling to the central spin (i.e.~the nontrivial dynamics in the $j,m,m\pm 1$ subspaces). In the classical limit this averaging over the nuclear states present in the narrowed density matrix is expressed by integration over the $B^{x}_{I}$ and $B^{y}_{I}$ components of the nuclear field. The RDT calculation captures correctly this averaging by a proper resummation of all the diagrams relevant in the large $N$ limit. The NZ-CP calculation simply preserves the key structure of the total density matrix during the system's evolution, and in this way recovers the classical averaging limit.

\subsection{Consequences for the non-uniform couplings theory}
As noted in Ref.~\onlinecite{Fischer_PRA07}, the single-projector approach is expected to be valid when the system-environment coupling is weak in some sense. As shown in Ref.~\onlinecite{Coish_PRB04}, \onlinecite{Coish_PRB08}, and \onlinecite{Coish_PRB10}, in the case of the hf-coupled spin bath this ``weak coupling'' condition is $\mathcal{A}/\Omega \! < \! 1$, and it apparently guarantees that the single projector approach is valid at all times. Under this condition the short time ($t \! \ll \! N/\mathcal{A}$) decay is very small, i.e.~a quadratic initial decay, $x(t) \! \approx \! 1 - 2(t/\tau)^2$ with $t \! \ll \! \tau$. This result \emph{does not} depend on the shape of the electron wavefuntion.
The decay at longer times is due to the fluctuations of the Overhauser field induced by flip-flops involving nuclei with different couplings to the central spin; the inhomogeneity of couplings is crucial at this timescale. Most of the decay for $\mathcal{A}/\Omega \! \ll \! 1$ is of the exponential form recovered by the RDT calculation. With decreasing $\Omega$, the non-exponential features of the single-projector NZ solution become more prominent, but the form of the NFID at low fields $\Omega < \mathcal{A}$ cannot be determined in a controlled fashion,\cite{Coish_PRB10} at least for realistic values of nuclear spin $I>1/2$ and for zero or small nuclear polarization.

At short times, the RDT solution remains controlled (as long as one believes in the robustness of the transformation leading to the effective Hamiltonian \emph{on this timescale}) when $\Omega  \! \gg \! \mathcal{A}/\sqrt{N}$. In this regime, the RDT solution is explicitly independent of the shape of the electron wavefunction,\cite{Cywinski_PRL09,Cywinski_PRB09} as was the case for the short-time NZ result.\cite{Coish_PRB10} This agrees with the intuition that at this timescale all the inter-nuclear flip-flops should be virtual, and all of them should be equally important. It suggests that on short timescales, it should be possible to map the dynamics of the real system onto the dynamics of a model system with uniform couplings.\cite{Cywinski_PRB09,Ferraro_PS10} As we have shown here, the RDT reproduces the envelope of the exact result in this case, while the standard NZ approach fails beyond very short times, $t < \! \tau \!  = \! 4N\Omega/\A^2$  (at which the envelope is well described by the quadratic decay), suggesting that the latter approach is indeed unable to describe the low-field decay, most of which occurs at short times ($t \! < \! N/\A$).
This is consistent with the results of Refs.~\onlinecite{Coish_PRB04,Coish_PRB08,Coish_PRB10}, in which only high fields were considered (or a large nuclear polarization was assumed), and the single-projector NZ theory was shown to be very hard to control at low fields. It should be however noted that the broadening of the energy bandwidth available for multiple inter-nuclear flip flops in higher orders in $V_{\text{ff}}$, discussed briefly in Ref.~\onlinecite{Coish_PRB10}, shows that the boundary of the short-time regime can move to even shorter times as one goes to a higher order in the expansion of the NZ memory kernel. It is nevertheless unclear what happens to this trend as the expansion is continued to higher orders. The previous discussion suggests that in order for a single-projector NZ approach to recover the uniform coupling bath limit one has to go to an infinite order in the memory kernel expansion. This makes it hard to perturbatively delineate, within the single-projector NZ approach, the timescale on which the uniform coupling model gives a good approximation to the real problem. A full-Hamiltonian approach starting from an assumption that the box model holds on some timescale, and correctly describing the decoherence in a broad range of magnetic fields, should shed more light on these issues.

The use of correlated projection operators\cite{Breuer,Fischer_PRA07,Ferraro_PRB08} makes the 2nd order NZ result agree very well with both the envelope and the small oscillations of the exact result for NFID decay in the uniform coupling case. This strongly suggests that the correlated projection operator technique should be an important element of a theory of central spin decoherence encompassing all the regimes of magnetic fields and timescales.

\section{CONCLUSIONS}
In this paper, we have applied the Nakajima-Zwanzig  generalized Master equation, originally developed for the hyperfine-coupled central spin problem in Refs.~\onlinecite{Coish_PRB04,Coish_PRB08,Coish_PRB10}, to the exactly solvable case of uniform hyperfine couplings. We have shown that the NZ calculation of narrowed state free induction decay (NFID), i.e.~the evolution of the electron coupled to a nuclear bath acting on it with  a well defined Overhauser field, fails very quickly in this case, and can only account for the initial decoherence dynamics. We have traced the origin of this failure to the fact that, in this NZ approach, a single operator was used to project the total density matrix onto the tensor product of the central spin and bath density matrices. While this approach works well at high magnetic fields, at which the coherence decays at long times, and the inhomogeneity of the electron's wavefunction (i.e.~the inhomogeneity of the hf couplings) is crucial for the coherence decay,\cite{Coish_PRB08,Coish_PRB10} it fails at short times and low fields, at which the shape of the wavefunction should be irrelevant, and the uniform coupling model is expected to be applicable. In  the latter situation, one has to modify the NZ approach by introducing a family of \emph{correlated projection operators},\cite{Fischer_PRA07,Ferraro_PRB08} with which one can capture the essential features of the exact dynamics: the electron loses its coherence by a rather simple dephasing process, in which the electron spin states acquire a different phase when interacting with nuclear states from different subspaces, and the coherence is lost due to averaging over the initial nuclear states. With this modification, the NZ GME reproduces very well the exact solution for the uniform coupling model.

The exact result for NFID in the uniform coupling model is also reproduced by the effective Hamiltonian-based solution from Refs.~\onlinecite{Cywinski_PRL09, Cywinski_PRB09} on a timescale long enough to capture the full coherence decay. This is another example of the effective Hamiltonian-based theory\cite{Cywinski_PRL09, Cywinski_PRB09}  reproducing the semiclassical limit (i.e.~the quantum calculation being equivalent to an average over classical nuclear fields)  of the central spin dephasing problem.\cite{Neder_PRB11}

According to the effective Hamiltonian-based theory, NFID should be described by Eq.~(\ref{RDTbox}) up to a timescale of $\sim \! 10 $ $\mu$s in GaAs dots (with $N \! \approx \! 10^{6}$), and this decay will be significant for magnetic fields smaller than $1$ T. The NFID measurements in this parameter range are within reach of the fast measurement techniques developed for gated GaAs dots.\cite{Barthel_PRL09} Upon increasing the magnetic field, the decoherence times calculated from the purely hf central spin Hamiltonian quickly become comparable to the predicted decay times due to dipolarly-induced spectral diffusion,\cite{Yao_PRB06,Witzel_PRB08} which were obtained with the cluster methods succesfully describing the high-field decay of the spin echo signal in GaAs.\cite{Bluhm_NP11} This makes the long-time regime (the exponential decay) probably very hard to observe in large GaAs dots. On the other hand, in the InGaAs dots, especially the smallest ones ($N\! \approx \! 10^{4}$), the short-time regime extends to at most $\sim \! 100$ ns, and the low-field requirement might be incompatible with the spin splitting needed for the optical manipulation of the qubit. These dots seem to be more suited for experimentally investigating the high-field and long-time regime of exponential decay since the predicted\cite{Witzel_PRB08} decoherence times due to the nuclear dipolar interactions are on the order of 5-10 $\mu$s (which is a lower bound since quadrupolar interactions were not included in that theory), giving a window of timescales in which the decay due to the hf interaction only could be observed. The exponential decay was indeed seen in such dots,\cite{Greilich_Science06} but more research is needed to ascertain its magnetic field dependence. If the nuclear state narrowing could be done in InGaAs dots at lower magnetic fields (e.g.~around $1$ T and possibly below), the crossover between the short and long timescales should be seen, making it possible to check the timescale at which the uniform coupling model predictions hold and to verify the predictions of Ref.~\onlinecite{Coish_PRB10} in the $\Omega \! \approx \! \mathcal{A}$ regime, in which the single-projector NZ results differ significantly from the effective Hamiltonian results.

The results of this paper make a strong statement regarding the structure of a theory which could describe the electron spin decoherence without the use of an effective Hamiltonian in a broad range of magnetic fields. In order to properly describe the short time ($t\! \ll \! N/\mathcal{A}$) regime, in which the coherence is expected to decay at low magnetic fields, one should employ the correlated projection operator approach in the derivation of a generalized Master equation. This was done employing both the NZ and the time-convolutionless (TCL) methods in the uniform \cite{Fischer_PRA07} and non-uniform\cite{Ferraro_PRB08} coupling cases (in the latter case only using the projectors of spaces of fixed $m$, not the $jm$ spaces used here), albeit only for a thermal nuclear bath. In this work we have shown that this feature of the theory is crucial also in the case of NFID and uniform couplings. The case of low-field NFID in a realistic, inhomogenously-coupled system remains to be further investigated. While we have focused here on the NZ approach, which was the subject of intense research\cite{Coish_PRB04,Coish_PRB08,Coish_PRB10} in the context of NFID, it might turn out that the TCL approach (untested yet for NFID) will be both easier to implement and more natural. The latter statement is supported by the fact that the structure of TCL is closer to the cumulant expansion structure which underpins the effective Hamiltonian solution from Refs.~\onlinecite{Cywinski_PRL09,Cywinski_PRB09}. These theories not only practically agree with the NZ approach at high fields in the non-uniform coupling case, but they also capture correctly the exact solution in the uniform coupling case, strongly suggesting that they correctly describe the low-field and short-time decay of NFID, just as they correctly described the spin echo signal.\cite{Bluhm_NP11,Neder_PRB11} However, the nature of the crossover between the high-field/long-time, and low-field/short-time decay behaviors remains to be elucidated using a theory capable of treating both regimes on equal footing.

\section{Acknowledgements}
This work is supported by LPS-NSA.
{\L}C acknowledges support from the Homing programme of the Foundation for Polish Science supported by the EEA Financial Mechanism. {\L}C also acknowledges enlightening discussions with V.V.~Dobrovitski regarding the semiclassical limit of the central spin problem, and with W.A.~Coish on the relationship between the uniform and non-uniform coupling models.

\appendix

%%%%%%%%%%%%%%%%
%%%    4TH ORDER SIGMA
%%%%%%%%%%%%%%%%
\section{The full expression for the fourth-order memory kernel} \label{app:4thorderSigma}
In this section, we will work out the explicit form for the Laplace transform of the fourth-order memory kernel. We will not make any assumptions about the distribution of hyperfine couplings or about the magnitude of $\Omega$ relative to other scales in the problem. The only assumption is that the initial nuclear density matrix has the form shown in Eq.~(\ref{narrowedstates}). Starting from Eq.~(\ref{4thorderMK}), we replace each occurrence of $G(s)$ with $\int_0^\infty dt e^{-st}G(t)$ to obtain
\bea
\Sigma^{(4)}(s)&=&i\int_0^\infty \prod_{i=1}^4dt_ie^{-s\sum_it_i}\tr\{S^+[s-iL_0QG(t_1)]\nn\\&\times&L_VG(t_2)L_VQG(t_3)L_VG(t_4)L_VS^-\rho_I(0)\}.\nn\\&&\label{4thorderMK2}
\eea
This step allows us to work with $G(t)$ as opposed to its Laplace transform, making it easier to find explicit expressions. We will evaluate the string of operators appearing inside the trace in two stages. First consider
\beq
T\equiv G(t_3)L_VG(t_4)L_VS^-\rho_I(0).
\eeq
We can evaluate this using Eqs. (\ref{Gaction}) and (\ref{LVaction}), finding
\beq
T=\left(\begin{matrix}0&T_{\uparrow\downarrow}\cr T_{\downarrow\uparrow}&0\end{matrix}\right),\label{Tstructure}
\eeq
with
\bea
T_{\uparrow\downarrow}&=&\sum_{k\ell}{\tilde a}_{k\ell}I_\ell^-\rho_I(0)I_k^-,\nn\\
T_{\downarrow\uparrow}&=&\sum_{k\ell}\left[{\tilde b}_{k\ell}I_\ell^+I_k^-\rho_I(0)+{\tilde c}_{k\ell}\rho_I(0)I_k^-I_\ell^+\right],
\eea
where
\bea
{\tilde a}_{k\ell}&=&-{1\over4}A_kA_\ell e^{-i(\Omega_n-\omega_k-\omega_\ell+{1\over2}(A_k-A_\ell))t_3}\nn\\&\times&\left[e^{i(\omega_k-{A_k\over2})t_4}+e^{i(\omega_\ell+{A_\ell\over2})t_4}\right],\nn\\
{\tilde b}_{k\ell}&=&{1\over4}A_kA_\ell e^{i(\Omega_n+\omega_k-\omega_\ell-{1\over2}(A_k-A_\ell))t_3}e^{i(\omega_k+{A_k\over2})t_4},\nn\\
{\tilde c}_{k\ell}&=&{1\over4}A_kA_\ell e^{i(\Omega_n+\omega_k-\omega_\ell+{1\over2}(A_k-A_\ell))t_3}e^{i(\omega_k-{A_k\over2})t_4}.\nn\\&&
\eea
For the second stage of the evaluation, we define
\beq
R\equiv L_VG(t_2)L_VQT,
\eeq
Given the structure of $T$, Eq. (\ref{Tstructure}), it is not difficult to show that $R$ has a similar form:
\beq
R=\left(\begin{matrix}0&R_{\uparrow\downarrow}\cr R_{\downarrow\uparrow}&0\end{matrix}\right),\label{Rstructure}
\eeq
In terms of $R$, we may write the trace in Eq. (\ref{4thorderMK2}) as
\beq
\tr\left\{S^+[s-iL_0QG(t_1)]R\right\}=\tr\left\{\Delta_1 R_{\downarrow\uparrow}\right\},\label{thetrace}
\eeq
with
\beq
\Delta_1\equiv s+i(h^z-h^z_n)e^{i(\Omega+h^z)t_1}.
\eeq
Therefore, it is only necessary to compute one of the components of $R$, and this can be expressed in terms of $T$ according to
\bea
R_{\downarrow\uparrow}&=&{1\over4}h^+U_+(t_2)\Big\{h^-T_{\downarrow\uparrow}-h^-\rho_I(0)\tr[T_{\downarrow\uparrow}]\nn\\&-&T_{\uparrow\downarrow}h^+\Big\} U_+^\dag(t_2)-{1\over4}U_-(t_2)\Big\{h^+T_{\uparrow\downarrow}\nn\\&-&T_{\downarrow\uparrow}h^-+\rho_I(0)\tr[T_{\downarrow\uparrow}]h^-\Big\}U_-^\dag(t_2)h^+,
\eea
where $U_\pm(t)$ were defined in Eq. (\ref{Upm}). The explicit form of the trace, Eq. (\ref{thetrace}), is rather messy, so we first break it up into four parts in an effort to improve readability:
\beq
\tr\left\{\Delta_1 R_{\downarrow\uparrow}\right\}=\sum_{i=1}^4 X_i,
\eeq
\bea
X_1&=&-{1\over4}\tr\left\{\Delta_1 h^+U_+(t_2)T_{\uparrow\downarrow}h^+U_+^\dag(t_2)\right\}\nn\\&-&{1\over4}\tr\left\{\Delta_1 U_-(t_2)h^+T_{\uparrow\downarrow}U_-^\dag(t_2)h^+\right\},\nn\\
X_2&=&-{1\over4}\tr[T_{\downarrow\uparrow}]\tr\left\{\Delta_1 h^+U_+(t_2)h^-\rho_I(0)U_+^\dag(t_2)\right\}\nn\\&-&{1\over4}\tr[T_{\downarrow\uparrow}]\tr\left\{
\Delta_1 U_-(t_2)\rho_I(0)h^-U_-^\dag(t_2)h^+\right\},\nn\\
X_3&=&{1\over4}\tr\left\{\Delta_1 h^+U_+(t_2)h^-T_{\downarrow\uparrow}U_+^\dag(t_2)\right\},\nn\\
X_4&=&{1\over4}\tr\left\{\Delta_1 U_-(t_2)T_{\downarrow\uparrow}h^-U_-^\dag(t_2)h^+\right\}.
\eea
These evaluate to
\bea
X_1&=&-{1\over4}\sum_{k\ell pq}{\tilde a}_{k\ell}A_pA_q\left[e^{i(\omega_p+{A_p\over2})t_2}+e^{i(\omega_q-{A_q\over2})t_2}\right] \nn\\&\times& \left[s+i(A_p-A_\ell)e^{i(\Omega_n+A_p-A_\ell)t_1}\right]\nn\\&\times&\tr\left\{I_p^+I_\ell^-\rho_I(0)I_k^-I_q^+\right\},
\eea
\bea
X_2&=&-{s\over16}e^{i\Omega_nt_3}\sum_{k\ell}A_k^2A_\ell^2e^{i\omega_\ell t_2}e^{i\omega_k t_4}\nn\\&\times&\left[c_\ell^-e^{i{A_\ell\over2}t_2}+c_\ell^+e^{-i{A_\ell\over2}t_2}\right]\nn\\&\times&\left[c_k^-e^{i{A_k\over2}t_4}+c_k^+e^{-i{A_k\over2}t_4}\right],
\eea
\bea
X_3&=&{1\over4}\sum_{k\ell pq}A_pA_qe^{i(\omega_p+{A_p\over2})t_2}\bigg\{s{\tilde b}_{k\ell}\tr\left\{I_\ell^+I_k^-\rho_I(0)I_p^+I_q^-\right\}\nn\\&+&{\tilde c}_{k\ell}\left[s+i(A_p-A_q) e^{i(\Omega_n+A_p-A_q)t_1}\right]\nn\\&\times&\tr\left\{I_p^+I_q^-\rho_I(0)I_k^-I_\ell^+\right\}\bigg\},
\eea
\bea
X_4&=&{1\over4}\sum_{k\ell pq}A_pA_qe^{i(\omega_p-{A_p\over2})t_2}\bigg\{s{\tilde c}_{k\ell}\tr\left\{I_q^-I_p^+\rho_I(0)I_k^-I_\ell^+\right\}\nn\\&+& {\tilde b}_{k\ell}\left[s+i(A_\ell-A_k)e^{i(\Omega_n+A_\ell-A_k)t_1}\right]\nn\\&\times&\tr\left\{I_\ell^+I_k^-\rho_I(0)I_q^-I_p^+\right\}\bigg\}.
\eea
Performing the four-fold Laplace transform in Eq. (\ref{4thorderMK2}) on each of the $X_i$ and denoting the results by $Y_i$, we find
\bea
Y_1&=&{1\over16}\sum_{k\ell pq}A_kA_\ell A_pA_q\nn\\&\times&{\tr\left\{I_p^+I_\ell^-\rho_I(0)I_k^-I_q^+\right\}\over s+i(\Omega_n-\omega_k-\omega_\ell+{1\over2}(A_k-A_\ell))}\nn\\&\times&\left[1+i{A_p-A_\ell\over s-i(\Omega_n+A_p-A_\ell)}\right]\nn\\&\times&\left[{1\over s-i(\omega_p+{A_p\over2})} +{1\over s-i(\omega_q-{A_q\over2})}\right]\nn\\&\times&\left[{1\over s-i(\omega_\ell+{A_\ell\over2})} +{1\over s-i(\omega_k-{A_k\over2})}\right],
\eea
\bea
Y_2&=&-{1\over16}\sum_{k\ell}A_k^2A_\ell^2{1\over s-i\Omega_n}\nn\\&\times&\left[{c_k^-\over s-i(\omega_k+{A_k\over2})}+{c_k^+\over s-i(\omega_k-{A_k\over2})}\right]\nn\\&\times& \left[{c_\ell^-\over s-i(\omega_\ell+{A_\ell\over2})}+{c_\ell^+\over s-i(\omega_\ell-{A_\ell\over2})}\right],
\eea
\bea
Y_3&=&{1\over16}\sum_{k\ell pq}A_kA_\ell A_pA_q{1\over s-i(\omega_p+{A_p\over2})}\Bigg\{\nn\\&&{1\over s-i(\omega_k+{A_k\over2})}{\tr\left\{I_\ell^+I_k^-\rho_I(0)I_p^+I_q^-\right\}\over s-i(\Omega_n+\omega_k-\omega_\ell-{1\over2}(A_k-A_\ell))}\nn\\&+&{1\over s-i(\omega_k-{A_k\over2})}{\tr\left\{I_p^+I_q^-\rho_I(0)I_k^-I_\ell^+\right\}\over s-i(\Omega_n+\omega_k-\omega_\ell+{1\over2}(A_k-A_\ell))}\nn\\&\times&\left[1+i{A_p-A_q\over s-i(\Omega_n+A_p-A_q)}\right]\Bigg\},
\eea
\bea
Y_4&=&{1\over16}\sum_{k\ell pq}A_kA_\ell A_pA_q{1\over s-i(\omega_p-{A_p\over2})}\Bigg\{\nn\\&&{1\over s-i(\omega_k-{A_k\over2})}{\tr\left\{I_q^-I_p^+\rho_I(0)I_k^-I_\ell^+\right\}\over s-i(\Omega_n+\omega_k-\omega_\ell+{1\over2}(A_k-A_\ell))}\nn\\&+&{1\over s-i(\omega_k+{A_k\over2})}{\tr\left\{I_\ell^+I_k^-\rho_I(0)I_q^-I_p^+\right\}\over s-i(\Omega_n+\omega_k-\omega_\ell-{1\over2}(A_k-A_\ell))}\nn\\&\times&\left[1+i{A_\ell-A_k\over s-i(\Omega_n+A_\ell-A_k)}\right]\Bigg\}.
\eea
The Laplace transform of the fourth-order memory kernel is given by
\beq
\Sigma^{(4)}(s)=i\sum_{i=1}^4Y_i(s).\label{4thorderMK3}
\eeq

In the high-frequency limit, we have in the rotating frame defined by $\Omega_n$ ($\bar{Y}_i=Y_i(s+i\Omega_n)$)
\beq
\bar{Y}_1\approx 0,
\eeq
\beq
\bar{Y}_2\approx {1\over16\Omega_n^2}\sum_{k\ell}A_k^2A_\ell^2\left[c_k^-+c_k^+\right]\left[c_\ell^-+c_\ell^+\right]{1\over s},
\eeq
\bea
\bar{Y}_3&\approx&-{1\over16\Omega_n^2}\sum_{k\ell pq}A_kA_\ell A_pA_q \nn\\&\times&\Bigg[{\tr\left\{I_\ell^+I_k^-\rho_I(0)I_p^+I_q^-\right\}\over s-i(\omega_k-\omega_\ell-{1\over2}(A_k-A_\ell))}\nn\\&+& {s\over s-i(A_p-A_q)}{\tr\left\{I_p^+I_q^-\rho_I(0)I_k^-I_\ell^+\right\}\over s-i(\omega_k-\omega_\ell+{1\over2}(A_k-A_\ell))}\Bigg],\nn\\&&\label{Y3approx}
\eea
\bea
\bar{Y}_4&\approx&-{1\over16\Omega_n^2}\sum_{k\ell pq}A_kA_\ell A_pA_q\nn\\&\times&\Bigg[{\tr\left\{I_q^-I_p^+\rho_I(0)I_k^-I_\ell^+\right\}\over s-i(\omega_k-\omega_\ell+{1\over2}(A_k-A_\ell))}\nn\\&+& {s\over s-i(A_\ell-A_k)}{\tr\left\{I_\ell^+I_k^-\rho_I(0)I_q^-I_p^+\right\}\over s-i(\omega_k-\omega_\ell-{1\over2}(A_k-A_\ell))}\Bigg],\nn\\&&\label{Y4approx}
\eea
and
\beq
\Sigma^{(4)}(s+i\Omega_n)=\bar{\Sigma}(s)=i\sum_{i=1}^4 \bar{Y}_i(s).
\eeq

%%%%%%%%%%%%%%%%%%%%%%%
%%%% NUCLEAR BATH CORRELATORS
%%%%%%%%%%%%%%%%%%%%%%%
\section{Nuclear bath correlators}\label{app:correlators}
In this section, we evaluate the nuclear bath correlators which arise in the expression for the fourth-order memory kernel. For example, in the expression given for $Y_3$ in Appendix (\ref{app:4thorderSigma}), there appears the correlator
\beq
\tr\left\{I_p^+I_q^-\rho_I(0)I_k^-I_\ell^+\right\}.
\eeq
Since $\rho_I(0)$ is assumed to be diagonal in the $\ket{n_i}$ basis, this correlator is only non-zero when each pair of raising and lowering operators act on the same nucleus, and we obtain
\bea
&&\!\!\!\!\tr\left\{I_p^+I_q^-\rho_I(0)I_k^-I_\ell^+\right\}\qquad\qquad\nn\\&&=[\delta_{k\ell}\delta_{pq}+\delta_{kp}\delta_{\ell q}(1-\delta_{kq})]\sum_i\rho_{ii}c_k^{(i)+}c_q^{(i)-},
\eea
where we have defined
\bea
c_k^{(i)\pm}&\equiv& \bra{n_i}I_k^\mp I_k^\pm \ket{n_i}\nn\\&=&I_k(I_k+1)-(m_k^i)^2\mp m_k^i.\label{defofckipm}
\eea
Recall that $I_k$ is the total spin of the $k$th nucleus and that $m_k^i$ is the eigenvalue of $I_k^z$ associated with the state $\ket{n_i}$. (The total spin of the $k$th nucleus depends on $k$ since we have incorporated information about different species into the nuclear site index $k$.) These quantities are related to the correlators $c_k^\pm$ according to
\beq
c_k^\pm\equiv \tr\left\{I_k^\mp I_k^\pm \rho_I(0)\right\}=\sum_i\rho_{ii}c^{(i)\pm}_k.
\eeq
We may express the other four correlators appearing in the $Y_i$ in terms of the $c_k^{(i)\pm}$:
\bea
&&\!\!\!\!\tr\left\{I_p^+I_\ell^-\rho_I(0)I_k^-I_q^+\right\}\qquad\qquad\nn\\&&=[\delta_{kq}\delta_{\ell p}+\delta_{kp}\delta_{\ell q}(1-\delta_{k\ell})]\sum_i\rho_{ii}c_k^{(i)+}c_\ell^{(i)-},\nn\\
&&\!\!\!\!\tr\left\{I_\ell^+I_k^-\rho_I(0)I_q^-I_p^+\right\}\qquad\qquad\nn\\&&=[\delta_{k\ell}\delta_{pq}+\delta_{kp}\delta_{\ell q}(1-\delta_{kq})]\sum_i\rho_{ii}c_k^{(i)-}c_q^{(i)+},\nn\\
&&\!\!\!\!\tr\left\{I_\ell^+I_k^-\rho_I(0)I_p^+I_q^-\right\}\qquad\qquad\nn\\&&=\sum_i\rho_{ii}[\delta_{k\ell}\delta_{pq}c_k^{(i)-}c_q^{(i)-}+\delta_{kp}\delta_{\ell q}(1-\delta_{kq})c_k^{(i)-}c_q^{(i)+}],\nn\\
&&\!\!\!\!\tr\left\{I_q^-I_p^+\rho_I(0)I_k^-I_\ell^+\right\}\qquad\qquad\nn\\&&=\sum_i\rho_{ii}[\delta_{k\ell}\delta_{pq}c_k^{(i)+}c_q^{(i)+}+\delta_{kp}\delta_{\ell q}(1-\delta_{kq})c_k^{(i)+}c_q^{(i)-}].\nn\\&&\label{4correlators}
\eea

As an aside, we note that in the case of a uniformly polarized homonuclear bath, the two-operator correlators become independent of $k$:
\beq
c^\pm=c^\pm_k=\sum_i\rho_{ii}c_k^{(i)\pm}.
\eeq
For the case of spin 1/2 nuclei, these are particularly simple,
\beq
c^\pm={1\over2}\mp\sum_i\rho_{ii}m_k^i={1\over2}\mp {m\over N},\label{uniformspinhalfcpm}
\eeq
where $m$ is the net polarization of all the nuclei defined by
\beq
m\equiv \sum_i\rho_{ii}\sum_k m_k^i.
\eeq
For a uniformly polarized nuclear bath, $m$ is closely related to the quantity $h^z_n$ defined in Eq. (\ref{defhzn}):
\beq
m={Nh^z_n\over{\cal A}}.\label{mhznrel}
\eeq

Next, we will prove a relation which is useful for simplifying the correlators in Eq. (\ref{4correlators}). In particular, we want to show that in the case of a uniformly polarized nuclear spin bath, we have
\bea
\sum_i\rho_{ii}c_k^{(i)+}c_\ell^{(i)-}&=&\left(\sum_i\rho_{ii}c_k^{(i)+}\right)\left(\sum_i\rho_{ii}c_\ell^{(i)-}\right)
\nn\\&=&c^+c^-.\label{uniformpolid}
\eea
In the final expression, we have used that the condition of uniform polarization implies that the result is independent of the nuclear site indices $k$ and $\ell$. We have also implicitly assumed a homonuclear bath since information about different nuclear species is incorporated into the site index. This last assumption can easily be relaxed by retaining this information in the form of an additional index; however, in the situations in this paper where Eq. (\ref{uniformpolid}) is employed, the homonuclear assumption is made anyway for the sake of simplicity. Also note that we are not assuming spin 1/2 nuclei in Eq. (\ref{uniformpolid}). Replacing the $c_k^{(i)\pm}$ with explicit expressions from Eq. (\ref{defofckipm}) and canceling several terms, we obtain
\bea
&&\!\!\sum_i\rho_{ii}c_k^{(i)+}c_\ell^{(i)-}-\left(\sum_i\rho_{ii}c_k^{(i)+}\right)\left(\sum_i\rho_{ii}c_\ell^{(i)-}\right)\nn\\&=&
\!\!\sum_i\rho_{ii}(m_k^i)^2(m_\ell^i)^2-\left(\sum_i\rho_{ii}(m_k^i)^2\right)\left(\sum_i\rho_{ii}(m_\ell^i)^2\right)\nn\\&-&
\!\!\sum_i\rho_{ii}(m_k^i)^2m_\ell^i+\left(\sum_i\rho_{ii}(m_k^i)^2\right)\left(\sum_i\rho_{ii}m_\ell^i\right)\nn\\&+&
\!\!\sum_i\rho_{ii}m_k^i(m_\ell^i)^2-\left(\sum_i\rho_{ii}m_k^i\right)\left(\sum_i\rho_{ii}(m_\ell^i)^2\right)\nn\\&-&
\!\!\sum_i\rho_{ii}m_k^im_\ell^i+\left(\sum_i\rho_{ii}m_k^i\right)\left(\sum_i\rho_{ii}m_\ell^i\right).\label{uniformpolid2}
\eea
Each term in Eq. (\ref{uniformpolid2}) is independent of $k$ and $\ell$ in the uniformly polarized case, so that the second and third lines on the right-hand side of the equation vanish identically. Consider the term $\sum_i\rho_{ii}m_k^im_\ell^i$ in the last line. We can multiply this by $A_kA_\ell$ and sum over $k$ and $\ell$. Since the $\ket{n_i}$ states all have the same $h^z$ eigenvalue, we know that
\beq
\sum_{k\ell}A_kA_\ell m_k^im_\ell^i=(h^z_n)^2
\eeq
is independent of $i$. On the other hand, $\sum_i\rho_{ii}m_k^im_\ell^i$ is independent of $k$ and $\ell$, so that multiplying by $A_kA_\ell$ and summing over $k$ and $\ell$ simply amounts to multiplying this expression by $\left(\sum_kA_k\right)^2={\cal A}^2$. These two observations together imply that
\beq
\sum_i\rho_{ii}m_k^im_\ell^i=\left(h^z_n\over {\cal A}\right)^2.
\eeq
From Eq. (\ref{mhznrel}) we have
\beq
\sum_i\rho_{ii}m_k^i={m\over N}={h^z_n\over{\cal A}},
\eeq
showing that the last two terms on the right-hand side of Eq. (\ref{uniformpolid2}) cancel each other.

To finish the proof of Eq. (\ref{uniformpolid}), it remains to show that the first line on the right-hand side of Eq. (\ref{uniformpolid2}) vanishes. This follows straight-forwardly from a simple generalization of the previous argument. We begin with the following expression
\beq
\sum_i\rho_{ii}m_k^im_\ell^im_p^im_q^i-\left(\sum_i\rho_{ii}m_k^im_p^i\right)\left(\sum_i\rho_{ii}m_\ell^im_q^i\right).\label{4ms}
\eeq
We can evaluate explicitly each of the two terms by multiplying by $A_kA_\ell A_pA_q/{\cal A}^4$ and summing over $k,\ell,p,q$. Since each term is independent of $k,\ell,p,q$ under the assumption of uniform polarization, this procedure should leave the terms unchanged. We find that the two terms evaluate to the same expression and thus cancel:
\bea
\sum_i\rho_{ii}m_k^im_\ell^im_p^im_q^i&-&\left(\sum_i\rho_{ii}m_k^im_p^i\right)\left(\sum_i\rho_{ii}m_\ell^im_q^i\right)
\nn\\&=&\left({h^z_n\over{\cal A}}\right)^4-\left({h^z_n\over{\cal A}}\right)^4=0.
\eea
The first line on the right-hand side of Eq. (\ref{uniformpolid2}) is just a special case of Eq. (\ref{4ms}) where $p=k$ and $q=\ell$, so we have also shown that this line vanishes, completing the proof of Eq. (\ref{uniformpolid}). Repeated application of these arguments can be used to show more generally that
\beq
\sum_i\rho_{ii}c_k^{(i)\pm}c_\ell^{(i)\pm}=c^\pm c^\pm,\label{uniformpolid3}
\eeq
where here the two sets of $\pm$ signs are independent of each other.

The identity in Eq. (\ref{uniformpolid3}) is used extensively in the context of the high-frequency limit considered in section \ref{highfreqlimit}. We can also use it to simplify the four-operator correlators in Eq. (\ref{4correlators}). In the uniform coupling model, the indices of these correlators are summed over, so we have
\bea
&&\sum_{k\ell pq}\tr\left\{I_p^+I_\ell^-\rho_I(0)I_k^-I_q^+\right\}=\sum_{k\ell pq}\tr\left\{I_p^+I_q^-\rho_I(0)I_k^-I_\ell^+\right\}\nn\\ &&= \sum_{k\ell pq}\tr\left\{I_\ell^+I_k^-\rho_I(0)I_q^-I_p^+\right\}=2N^2c^+c^-,\nn\\ &&\sum_{k\ell pq}\tr\left\{I_\ell^+I_k^-\rho_I(0)I_p^+I_q^-\right\}=N^2c^-,\nn\\ &&\sum_{k\ell pq}\tr\left\{I_q^-I_p^+\rho_I(0)I_k^-I_\ell^+\right\}=N^2c^+.\label{4correlators2}
\eea
These relations were used to obtain Eqs. (\ref{2ndorderbox}) and (\ref{4thorderbox}).

\bibliography{refs_quant}

\begin{thebibliography}{65}
\expandafter\ifx\csname natexlab\endcsname\relax\def\natexlab#1{#1}\fi
\expandafter\ifx\csname bibnamefont\endcsname\relax
  \def\bibnamefont#1{#1}\fi
\expandafter\ifx\csname bibfnamefont\endcsname\relax
  \def\bibfnamefont#1{#1}\fi
\expandafter\ifx\csname citenamefont\endcsname\relax
  \def\citenamefont#1{#1}\fi
\expandafter\ifx\csname url\endcsname\relax
  \def\url#1{\texttt{#1}}\fi
\expandafter\ifx\csname urlprefix\endcsname\relax\def\urlprefix{URL }\fi
\providecommand{\bibinfo}[2]{#2}
\providecommand{\eprint}[2][]{\url{#2}}

\bibitem[{\citenamefont{Hanson et~al.}(2007)\citenamefont{Hanson, Kouwenhoven,
  Petta, Tarucha, and Vandersypen}}]{Hanson_RMP07}
\bibinfo{author}{\bibfnamefont{R.}~\bibnamefont{Hanson}},
  \bibinfo{author}{\bibfnamefont{L.~P.} \bibnamefont{Kouwenhoven}},
  \bibinfo{author}{\bibfnamefont{J.~R.} \bibnamefont{Petta}},
  \bibinfo{author}{\bibfnamefont{S.}~\bibnamefont{Tarucha}}, \bibnamefont{and}
  \bibinfo{author}{\bibfnamefont{L.~M.~K.} \bibnamefont{Vandersypen}},
  \bibinfo{journal}{Rev.\ Mod.\ Phys.} \textbf{\bibinfo{volume}{79}},
  \bibinfo{pages}{1217} (\bibinfo{year}{2007}).

\bibitem[{\citenamefont{Liu et~al.}(2010)\citenamefont{Liu, Yao, and
  Sham}}]{Liu_AP10}
\bibinfo{author}{\bibfnamefont{R.-B.} \bibnamefont{Liu}},
  \bibinfo{author}{\bibfnamefont{W.}~\bibnamefont{Yao}}, \bibnamefont{and}
  \bibinfo{author}{\bibfnamefont{L.~J.} \bibnamefont{Sham}},
  \bibinfo{journal}{Adv. Phys.} \textbf{\bibinfo{volume}{59}},
  \bibinfo{pages}{703} (\bibinfo{year}{2010}).

\bibitem[{\citenamefont{Coish and Baugh}(2009)}]{Coish_pssb09}
\bibinfo{author}{\bibfnamefont{W.~A.} \bibnamefont{Coish}} \bibnamefont{and}
  \bibinfo{author}{\bibfnamefont{J.}~\bibnamefont{Baugh}},
  \bibinfo{journal}{Phys. Status Solidi B} \textbf{\bibinfo{volume}{246}},
  \bibinfo{pages}{2203} (\bibinfo{year}{2009}).

\bibitem[{\citenamefont{Cywi{\'n}ski}(2011)}]{Cywinski_APPA11}
\bibinfo{author}{\bibfnamefont{{\L}.}~\bibnamefont{Cywi{\'n}ski}},
  \bibinfo{journal}{Acta Phys.~Pol.~A} \textbf{\bibinfo{volume}{119}},
  \bibinfo{pages}{576} (\bibinfo{year}{2011}).

\bibitem[{\citenamefont{Khaetskii et~al.}(2002)\citenamefont{Khaetskii, Loss,
  and Glazman}}]{Khaetskii_PRL02}
\bibinfo{author}{\bibfnamefont{A.~V.} \bibnamefont{Khaetskii}},
  \bibinfo{author}{\bibfnamefont{D.}~\bibnamefont{Loss}}, \bibnamefont{and}
  \bibinfo{author}{\bibfnamefont{L.}~\bibnamefont{Glazman}},
  \bibinfo{journal}{Phys. Rev. Lett.} \textbf{\bibinfo{volume}{88}},
  \bibinfo{pages}{186802} (\bibinfo{year}{2002}).

\bibitem[{\citenamefont{Schliemann et~al.}(2002)\citenamefont{Schliemann,
  Khaetskii, and Loss}}]{Schliemann_PRB02}
\bibinfo{author}{\bibfnamefont{J.}~\bibnamefont{Schliemann}},
  \bibinfo{author}{\bibfnamefont{A.~V.} \bibnamefont{Khaetskii}},
  \bibnamefont{and} \bibinfo{author}{\bibfnamefont{D.}~\bibnamefont{Loss}},
  \bibinfo{journal}{Phys. Rev. B} \textbf{\bibinfo{volume}{66}},
  \bibinfo{pages}{245303} (\bibinfo{year}{2002}).

\bibitem[{\citenamefont{Khaetskii et~al.}(2003)\citenamefont{Khaetskii, Loss,
  and Glazman}}]{Khaetskii_PRB03}
\bibinfo{author}{\bibfnamefont{A.}~\bibnamefont{Khaetskii}},
  \bibinfo{author}{\bibfnamefont{D.}~\bibnamefont{Loss}}, \bibnamefont{and}
  \bibinfo{author}{\bibfnamefont{L.}~\bibnamefont{Glazman}},
  \bibinfo{journal}{Phys. Rev. B} \textbf{\bibinfo{volume}{67}},
  \bibinfo{pages}{195329} (\bibinfo{year}{2003}).

\bibitem[{\citenamefont{Dobrovitski and {De Raedt}}(2003)}]{Dobrovitski_PRE03}
\bibinfo{author}{\bibfnamefont{V.~V.} \bibnamefont{Dobrovitski}}
  \bibnamefont{and} \bibinfo{author}{\bibfnamefont{H.~A.} \bibnamefont{{De
  Raedt}}}, \bibinfo{journal}{Phys. Rev. E} \textbf{\bibinfo{volume}{67}},
  \bibinfo{pages}{056702} (\bibinfo{year}{2003}).

\bibitem[{\citenamefont{Coish and Loss}(2004)}]{Coish_PRB04}
\bibinfo{author}{\bibfnamefont{W.~A.} \bibnamefont{Coish}} \bibnamefont{and}
  \bibinfo{author}{\bibfnamefont{D.}~\bibnamefont{Loss}},
  \bibinfo{journal}{Phys.\ Rev.\ B} \textbf{\bibinfo{volume}{70}},
  \bibinfo{eid}{195340} (\bibinfo{year}{2004}).

\bibitem[{\citenamefont{Erlingsson and Nazarov}(2004)}]{Erlingsson_PRB04}
\bibinfo{author}{\bibfnamefont{S.~I.} \bibnamefont{Erlingsson}}
  \bibnamefont{and} \bibinfo{author}{\bibfnamefont{Y.~V.}
  \bibnamefont{Nazarov}}, \bibinfo{journal}{Phys.\ Rev.\ B}
  \textbf{\bibinfo{volume}{70}}, \bibinfo{pages}{205327}
  (\bibinfo{year}{2004}).

\bibitem[{\citenamefont{Shenvi et~al.}(2005)\citenamefont{Shenvi, de~Sousa, and
  Whaley}}]{Shenvi_scaling_PRB05}
\bibinfo{author}{\bibfnamefont{N.}~\bibnamefont{Shenvi}},
  \bibinfo{author}{\bibfnamefont{R.}~\bibnamefont{de~Sousa}}, \bibnamefont{and}
  \bibinfo{author}{\bibfnamefont{K.~B.} \bibnamefont{Whaley}},
  \bibinfo{journal}{Phys.\ Rev.\ B} \textbf{\bibinfo{volume}{71}},
  \bibinfo{eid}{224411} (\bibinfo{year}{2005}).

\bibitem[{\citenamefont{Yao et~al.}(2006)\citenamefont{Yao, Liu, and
  Sham}}]{Yao_PRB06}
\bibinfo{author}{\bibfnamefont{W.}~\bibnamefont{Yao}},
  \bibinfo{author}{\bibfnamefont{R.-B.} \bibnamefont{Liu}}, \bibnamefont{and}
  \bibinfo{author}{\bibfnamefont{L.~J.} \bibnamefont{Sham}},
  \bibinfo{journal}{Phys.\ Rev.\ B} \textbf{\bibinfo{volume}{74}},
  \bibinfo{pages}{195301} (\bibinfo{year}{2006}).

\bibitem[{\citenamefont{Zhang et~al.}(2006)\citenamefont{Zhang, Dobrovitski,
  Al-Hassanieh, Dagotto, and Harmon}}]{Zhang_PRB06}
\bibinfo{author}{\bibfnamefont{W.}~\bibnamefont{Zhang}},
  \bibinfo{author}{\bibfnamefont{V.~V.} \bibnamefont{Dobrovitski}},
  \bibinfo{author}{\bibfnamefont{K.~A.} \bibnamefont{Al-Hassanieh}},
  \bibinfo{author}{\bibfnamefont{E.}~\bibnamefont{Dagotto}}, \bibnamefont{and}
  \bibinfo{author}{\bibfnamefont{B.~N.} \bibnamefont{Harmon}},
  \bibinfo{journal}{Phys.\ Rev.\ B} \textbf{\bibinfo{volume}{74}},
  \bibinfo{pages}{205313} (\bibinfo{year}{2006}).

\bibitem[{\citenamefont{Al-Hassanieh et~al.}(2006)\citenamefont{Al-Hassanieh,
  Dobrovitski, Dagotto, and Harmon}}]{Al_Hassanieh_PRL06}
\bibinfo{author}{\bibfnamefont{K.~A.} \bibnamefont{Al-Hassanieh}},
  \bibinfo{author}{\bibfnamefont{V.}~\bibnamefont{Dobrovitski}},
  \bibinfo{author}{\bibfnamefont{E.}~\bibnamefont{Dagotto}}, \bibnamefont{and}
  \bibinfo{author}{\bibfnamefont{B.~N.} \bibnamefont{Harmon}},
  \bibinfo{journal}{Phys.\ Rev.\ Lett.} \textbf{\bibinfo{volume}{97}},
  \bibinfo{pages}{037204} (\bibinfo{year}{2006}).

\bibitem[{\citenamefont{Deng and Hu}(2006)}]{Deng_PRB06}
\bibinfo{author}{\bibfnamefont{C.}~\bibnamefont{Deng}} \bibnamefont{and}
  \bibinfo{author}{\bibfnamefont{X.}~\bibnamefont{Hu}},
  \bibinfo{journal}{Phys.\ Rev.\ B} \textbf{\bibinfo{volume}{73}},
  \bibinfo{pages}{241303(R)} (\bibinfo{year}{2006}).

\bibitem[{\citenamefont{Liu et~al.}(2007)\citenamefont{Liu, Yao, and
  Sham}}]{Liu_NJP07}
\bibinfo{author}{\bibfnamefont{R.-B.} \bibnamefont{Liu}},
  \bibinfo{author}{\bibfnamefont{W.}~\bibnamefont{Yao}}, \bibnamefont{and}
  \bibinfo{author}{\bibfnamefont{L.~J.} \bibnamefont{Sham}},
  \bibinfo{journal}{New J.~Phys.} \textbf{\bibinfo{volume}{9}},
  \bibinfo{pages}{226} (\bibinfo{year}{2007}).

\bibitem[{\citenamefont{Saikin et~al.}(2007)\citenamefont{Saikin, Yao, and
  Sham}}]{Saikin_PRB07}
\bibinfo{author}{\bibfnamefont{S.~K.} \bibnamefont{Saikin}},
  \bibinfo{author}{\bibfnamefont{W.}~\bibnamefont{Yao}}, \bibnamefont{and}
  \bibinfo{author}{\bibfnamefont{L.~J.} \bibnamefont{Sham}},
  \bibinfo{journal}{Phys.\ Rev.\ B} \textbf{\bibinfo{volume}{75}},
  \bibinfo{pages}{125314} (\bibinfo{year}{2007}).

\bibitem[{\citenamefont{Chen et~al.}(2007)\citenamefont{Chen, Bergman, and
  Balents}}]{Chen_PRB07}
\bibinfo{author}{\bibfnamefont{G.}~\bibnamefont{Chen}},
  \bibinfo{author}{\bibfnamefont{D.~L.} \bibnamefont{Bergman}},
  \bibnamefont{and} \bibinfo{author}{\bibfnamefont{L.}~\bibnamefont{Balents}},
  \bibinfo{journal}{Phys.\ Rev.\ B} \textbf{\bibinfo{volume}{76}},
  \bibinfo{pages}{045312} (\bibinfo{year}{2007}).

\bibitem[{\citenamefont{Bortz and Stolze}(2007{\natexlab{a}})}]{Bortz_PRB07}
\bibinfo{author}{\bibfnamefont{M.}~\bibnamefont{Bortz}} \bibnamefont{and}
  \bibinfo{author}{\bibfnamefont{J.}~\bibnamefont{Stolze}},
  \bibinfo{journal}{Phys.\ Rev.\ B} \textbf{\bibinfo{volume}{76}},
  \bibinfo{pages}{014304} (\bibinfo{year}{2007}{\natexlab{a}}).

\bibitem[{\citenamefont{Bortz and Stolze}(2007{\natexlab{b}})}]{Bortz_JSM07}
\bibinfo{author}{\bibfnamefont{M.}~\bibnamefont{Bortz}} \bibnamefont{and}
  \bibinfo{author}{\bibfnamefont{J.}~\bibnamefont{Stolze}},
  \bibinfo{journal}{J. Stat. Mech.} \textbf{\bibinfo{volume}{2007}},
  \bibinfo{pages}{P06018} (\bibinfo{year}{2007}{\natexlab{b}}).

\bibitem[{\citenamefont{Deng and Hu}(2008)}]{Deng_PRB08}
\bibinfo{author}{\bibfnamefont{C.}~\bibnamefont{Deng}} \bibnamefont{and}
  \bibinfo{author}{\bibfnamefont{X.}~\bibnamefont{Hu}},
  \bibinfo{journal}{Phys.\ Rev.\ B} \textbf{\bibinfo{volume}{78}},
  \bibinfo{pages}{245301} (\bibinfo{year}{2008}).

\bibitem[{\citenamefont{Fischer and Breuer}(2007)}]{Fischer_PRA07}
\bibinfo{author}{\bibfnamefont{J.}~\bibnamefont{Fischer}} \bibnamefont{and}
  \bibinfo{author}{\bibfnamefont{H.-P.} \bibnamefont{Breuer}},
  \bibinfo{journal}{Phys.\ Rev.\ A} \textbf{\bibinfo{volume}{76}},
  \bibinfo{pages}{052119} (\bibinfo{year}{2007}).

\bibitem[{\citenamefont{Coish et~al.}(2008)\citenamefont{Coish, Fischer, and
  Loss}}]{Coish_PRB08}
\bibinfo{author}{\bibfnamefont{W.~A.} \bibnamefont{Coish}},
  \bibinfo{author}{\bibfnamefont{J.}~\bibnamefont{Fischer}}, \bibnamefont{and}
  \bibinfo{author}{\bibfnamefont{D.}~\bibnamefont{Loss}},
  \bibinfo{journal}{Phys.\ Rev.\ B} \textbf{\bibinfo{volume}{77}},
  \bibinfo{pages}{125329} (\bibinfo{year}{2008}).

\bibitem[{\citenamefont{Ferraro et~al.}(2008)\citenamefont{Ferraro, Breuer,
  Napoli, Jivulescu, and Messina}}]{Ferraro_PRB08}
\bibinfo{author}{\bibfnamefont{E.}~\bibnamefont{Ferraro}},
  \bibinfo{author}{\bibfnamefont{H.-P.} \bibnamefont{Breuer}},
  \bibinfo{author}{\bibfnamefont{A.}~\bibnamefont{Napoli}},
  \bibinfo{author}{\bibfnamefont{M.~A.} \bibnamefont{Jivulescu}},
  \bibnamefont{and} \bibinfo{author}{\bibfnamefont{A.}~\bibnamefont{Messina}},
  \bibinfo{journal}{Phys.\ Rev.\ B} \textbf{\bibinfo{volume}{78}},
  \bibinfo{pages}{064309} (\bibinfo{year}{2008}).

\bibitem[{\citenamefont{Cywi{\'n}ski
  et~al.}(2009{\natexlab{a}})\citenamefont{Cywi{\'n}ski, Witzel, and {Das
  Sarma}}}]{Cywinski_PRL09}
\bibinfo{author}{\bibfnamefont{{\L}.}~\bibnamefont{Cywi{\'n}ski}},
  \bibinfo{author}{\bibfnamefont{W.~M.} \bibnamefont{Witzel}},
  \bibnamefont{and} \bibinfo{author}{\bibfnamefont{S.}~\bibnamefont{{Das
  Sarma}}}, \bibinfo{journal}{Phys.\ Rev.\ Lett.}
  \textbf{\bibinfo{volume}{102}}, \bibinfo{pages}{057601}
  (\bibinfo{year}{2009}{\natexlab{a}}).

\bibitem[{\citenamefont{Cywi{\'n}ski
  et~al.}(2009{\natexlab{b}})\citenamefont{Cywi{\'n}ski, Witzel, and {Das
  Sarma}}}]{Cywinski_PRB09}
\bibinfo{author}{\bibfnamefont{{\L}.}~\bibnamefont{Cywi{\'n}ski}},
  \bibinfo{author}{\bibfnamefont{W.~M.} \bibnamefont{Witzel}},
  \bibnamefont{and} \bibinfo{author}{\bibfnamefont{S.}~\bibnamefont{{Das
  Sarma}}}, \bibinfo{journal}{Phys.\ Rev.\ B} \textbf{\bibinfo{volume}{79}},
  \bibinfo{pages}{245314} (\bibinfo{year}{2009}{\natexlab{b}}).

\bibitem[{\citenamefont{Cywi{\'n}ski et~al.}(2010)\citenamefont{Cywi{\'n}ski,
  Dobrovitski, and {Das Sarma}}}]{Cywinski_PRB10}
\bibinfo{author}{\bibfnamefont{{\L}.}~\bibnamefont{Cywi{\'n}ski}},
  \bibinfo{author}{\bibfnamefont{V.~V.} \bibnamefont{Dobrovitski}},
  \bibnamefont{and} \bibinfo{author}{\bibfnamefont{S.}~\bibnamefont{{Das
  Sarma}}}, \bibinfo{journal}{Phys.\ Rev.\ B} \textbf{\bibinfo{volume}{82}},
  \bibinfo{pages}{035315} (\bibinfo{year}{2010}).

\bibitem[{\citenamefont{Coish et~al.}(2010)\citenamefont{Coish, Fischer, and
  Loss}}]{Coish_PRB10}
\bibinfo{author}{\bibfnamefont{W.~A.} \bibnamefont{Coish}},
  \bibinfo{author}{\bibfnamefont{J.}~\bibnamefont{Fischer}}, \bibnamefont{and}
  \bibinfo{author}{\bibfnamefont{D.}~\bibnamefont{Loss}},
  \bibinfo{journal}{Phys.\ Rev.\ B} \textbf{\bibinfo{volume}{81}},
  \bibinfo{pages}{165315} (\bibinfo{year}{2010}).

\bibitem[{\citenamefont{Fischer and Loss}(2010)}]{Fischer_PRL10}
\bibinfo{author}{\bibfnamefont{J.}~\bibnamefont{Fischer}} \bibnamefont{and}
  \bibinfo{author}{\bibfnamefont{D.}~\bibnamefont{Loss}},
  \bibinfo{journal}{Phys.\ Rev.\ Lett.} \textbf{\bibinfo{volume}{105}},
  \bibinfo{pages}{266603} (\bibinfo{year}{2010}).

\bibitem[{\citenamefont{Bortz et~al.}(2010)\citenamefont{Bortz, Eggert,
  Schneider, St\"ubner, and Stolze}}]{Bortz_PRB10}
\bibinfo{author}{\bibfnamefont{M.}~\bibnamefont{Bortz}},
  \bibinfo{author}{\bibfnamefont{S.}~\bibnamefont{Eggert}},
  \bibinfo{author}{\bibfnamefont{C.}~\bibnamefont{Schneider}},
  \bibinfo{author}{\bibfnamefont{R.}~\bibnamefont{St\"ubner}},
  \bibnamefont{and} \bibinfo{author}{\bibfnamefont{J.}~\bibnamefont{Stolze}},
  \bibinfo{journal}{Phys.\ Rev.\ B} \textbf{\bibinfo{volume}{82}},
  \bibinfo{pages}{161308} (\bibinfo{year}{2010}).

\bibitem[{\citenamefont{Neder et~al.}(2011)\citenamefont{Neder, Rudner, Bluhm,
  Foletti, Halperin, and Yacoby}}]{Neder_PRB11}
\bibinfo{author}{\bibfnamefont{I.}~\bibnamefont{Neder}},
  \bibinfo{author}{\bibfnamefont{M.~S.} \bibnamefont{Rudner}},
  \bibinfo{author}{\bibfnamefont{H.}~\bibnamefont{Bluhm}},
  \bibinfo{author}{\bibfnamefont{S.}~\bibnamefont{Foletti}},
  \bibinfo{author}{\bibfnamefont{B.~I.} \bibnamefont{Halperin}},
  \bibnamefont{and} \bibinfo{author}{\bibfnamefont{A.}~\bibnamefont{Yacoby}},
  \bibinfo{journal}{Phys.\ Rev.\ B} \textbf{\bibinfo{volume}{72}},
  \bibinfo{pages}{052113} (\bibinfo{year}{2011}).

\bibitem[{\citenamefont{Gaudin}(1976)}]{Gaudin}
\bibinfo{author}{\bibfnamefont{M.}~\bibnamefont{Gaudin}}, \bibinfo{journal}{J.
  Phys. (France)} \textbf{\bibinfo{volume}{37}}, \bibinfo{pages}{1087}
  (\bibinfo{year}{1976}).

\bibitem[{\citenamefont{Faribault et~al.}(2011)\citenamefont{Faribault,
  El~Araby, Str\"ater, and Gritsev}}]{Faribault_PRB11}
\bibinfo{author}{\bibfnamefont{A.}~\bibnamefont{Faribault}},
  \bibinfo{author}{\bibfnamefont{O.}~\bibnamefont{El~Araby}},
  \bibinfo{author}{\bibfnamefont{C.}~\bibnamefont{Str\"ater}},
  \bibnamefont{and} \bibinfo{author}{\bibfnamefont{V.}~\bibnamefont{Gritsev}},
  \bibinfo{journal}{Phys.\ Rev.\ B} \textbf{\bibinfo{volume}{83}},
  \bibinfo{pages}{235124} (\bibinfo{year}{2011}).

\bibitem[{\citenamefont{Yuzbashyan et~al.}(2005)\citenamefont{Yuzbashyan,
  Altshuler, Kuznetsov, and Enolskii}}]{Yuzbashyan_JPA05}
\bibinfo{author}{\bibfnamefont{E.~A.} \bibnamefont{Yuzbashyan}},
  \bibinfo{author}{\bibfnamefont{B.~L.} \bibnamefont{Altshuler}},
  \bibinfo{author}{\bibfnamefont{V.~B.} \bibnamefont{Kuznetsov}},
  \bibnamefont{and} \bibinfo{author}{\bibfnamefont{V.~Z.}
  \bibnamefont{Enolskii}}, \bibinfo{journal}{J. Phys. A: Math. Gen.}
  \textbf{\bibinfo{volume}{38}}, \bibinfo{pages}{7831} (\bibinfo{year}{2005}).

\bibitem[{\citenamefont{Zhang et~al.}(2007)\citenamefont{Zhang, Konstantinidis,
  Al-Hassanieh, and Dobrovitski}}]{Zhang_JPC07}
\bibinfo{author}{\bibfnamefont{W.}~\bibnamefont{Zhang}},
  \bibinfo{author}{\bibfnamefont{N.}~\bibnamefont{Konstantinidis}},
  \bibinfo{author}{\bibfnamefont{K.~A.} \bibnamefont{Al-Hassanieh}},
  \bibnamefont{and} \bibinfo{author}{\bibfnamefont{V.~V.}
  \bibnamefont{Dobrovitski}}, \bibinfo{journal}{J. Phys.:Condens. Matter}
  \textbf{\bibinfo{volume}{19}}, \bibinfo{pages}{083202}
  (\bibinfo{year}{2007}).

\bibitem[{\citenamefont{Ferraro et~al.}(2010)\citenamefont{Ferraro, Breuer,
  Napoli, and Messina}}]{Ferraro_PS10}
\bibinfo{author}{\bibfnamefont{E.}~\bibnamefont{Ferraro}},
  \bibinfo{author}{\bibfnamefont{H.-P.} \bibnamefont{Breuer}},
  \bibinfo{author}{\bibfnamefont{A.}~\bibnamefont{Napoli}}, \bibnamefont{and}
  \bibinfo{author}{\bibfnamefont{A.}~\bibnamefont{Messina}},
  \bibinfo{journal}{Phys.~Scr.} \textbf{\bibinfo{volume}{T140}},
  \bibinfo{pages}{014021} (\bibinfo{year}{2010}).

\bibitem[{\citenamefont{Greilich et~al.}(2006)\citenamefont{Greilich, Yakovlev,
  Shabaev, Efros, Yugova, Oulton, Stavarache, Reuter, Wieck, and
  Bayer}}]{Greilich_Science06}
\bibinfo{author}{\bibfnamefont{A.}~\bibnamefont{Greilich}},
  \bibinfo{author}{\bibfnamefont{D.~R.} \bibnamefont{Yakovlev}},
  \bibinfo{author}{\bibfnamefont{A.}~\bibnamefont{Shabaev}},
  \bibinfo{author}{\bibfnamefont{A.~L.} \bibnamefont{Efros}},
  \bibinfo{author}{\bibfnamefont{I.~A.} \bibnamefont{Yugova}},
  \bibinfo{author}{\bibfnamefont{R.}~\bibnamefont{Oulton}},
  \bibinfo{author}{\bibfnamefont{V.}~\bibnamefont{Stavarache}},
  \bibinfo{author}{\bibfnamefont{D.}~\bibnamefont{Reuter}},
  \bibinfo{author}{\bibfnamefont{A.}~\bibnamefont{Wieck}}, \bibnamefont{and}
  \bibinfo{author}{\bibfnamefont{M.}~\bibnamefont{Bayer}},
  \bibinfo{journal}{Science} \textbf{\bibinfo{volume}{313}},
  \bibinfo{pages}{341} (\bibinfo{year}{2006}).

\bibitem[{\citenamefont{Latta et~al.}(2009)\citenamefont{Latta, H{\"o}gele,
  Zhao, Vamivakas, Maletinsky, Kroner, Dreiser, Carusotto, Badolato, Schuh
  et~al.}}]{Latta_NP09}
\bibinfo{author}{\bibfnamefont{C.}~\bibnamefont{Latta}},
  \bibinfo{author}{\bibfnamefont{A.}~\bibnamefont{H{\"o}gele}},
  \bibinfo{author}{\bibfnamefont{Y.}~\bibnamefont{Zhao}},
  \bibinfo{author}{\bibfnamefont{A.~N.} \bibnamefont{Vamivakas}},
  \bibinfo{author}{\bibfnamefont{P.}~\bibnamefont{Maletinsky}},
  \bibinfo{author}{\bibfnamefont{M.}~\bibnamefont{Kroner}},
  \bibinfo{author}{\bibfnamefont{J.}~\bibnamefont{Dreiser}},
  \bibinfo{author}{\bibfnamefont{I.}~\bibnamefont{Carusotto}},
  \bibinfo{author}{\bibfnamefont{A.}~\bibnamefont{Badolato}},
  \bibinfo{author}{\bibfnamefont{D.}~\bibnamefont{Schuh}},
  \bibnamefont{et~al.}, \bibinfo{journal}{Nat. Phys.}
  \textbf{\bibinfo{volume}{5}}, \bibinfo{pages}{758} (\bibinfo{year}{2009}).

\bibitem[{\citenamefont{Barthel et~al.}(2009)\citenamefont{Barthel, Reilly,
  Marcus, Hanson, and Gossard}}]{Barthel_PRL09}
\bibinfo{author}{\bibfnamefont{C.}~\bibnamefont{Barthel}},
  \bibinfo{author}{\bibfnamefont{D.~J.} \bibnamefont{Reilly}},
  \bibinfo{author}{\bibfnamefont{C.~M.} \bibnamefont{Marcus}},
  \bibinfo{author}{\bibfnamefont{M.~P.} \bibnamefont{Hanson}},
  \bibnamefont{and} \bibinfo{author}{\bibfnamefont{A.~C.}
  \bibnamefont{Gossard}}, \bibinfo{journal}{Phys.\ Rev.\ Lett.}
  \textbf{\bibinfo{volume}{103}}, \bibinfo{pages}{160503}
  (\bibinfo{year}{2009}).

\bibitem[{\citenamefont{Vink et~al.}(2009)\citenamefont{Vink, Nowack, Koppens,
  Danon, Nazarov, and Vandersypen}}]{Vink_NP09}
\bibinfo{author}{\bibfnamefont{I.~T.} \bibnamefont{Vink}},
  \bibinfo{author}{\bibfnamefont{K.~C.} \bibnamefont{Nowack}},
  \bibinfo{author}{\bibfnamefont{F.}~\bibnamefont{Koppens}},
  \bibinfo{author}{\bibfnamefont{J.}~\bibnamefont{Danon}},
  \bibinfo{author}{\bibfnamefont{Y.~V.} \bibnamefont{Nazarov}},
  \bibnamefont{and} \bibinfo{author}{\bibfnamefont{L.~M.~K.}
  \bibnamefont{Vandersypen}}, \bibinfo{journal}{Nat. Phys.}
  \textbf{\bibinfo{volume}{5}}, \bibinfo{pages}{764} (\bibinfo{year}{2009}).

\bibitem[{\citenamefont{Foletti et~al.}(2009)\citenamefont{Foletti, Bluhm,
  Mahalu, Umansky, and Yacoby}}]{Foletti_NP09}
\bibinfo{author}{\bibfnamefont{S.}~\bibnamefont{Foletti}},
  \bibinfo{author}{\bibfnamefont{H.}~\bibnamefont{Bluhm}},
  \bibinfo{author}{\bibfnamefont{D.}~\bibnamefont{Mahalu}},
  \bibinfo{author}{\bibfnamefont{V.}~\bibnamefont{Umansky}}, \bibnamefont{and}
  \bibinfo{author}{\bibfnamefont{A.}~\bibnamefont{Yacoby}},
  \bibinfo{journal}{Nat. Phys.} \textbf{\bibinfo{volume}{5}},
  \bibinfo{pages}{903} (\bibinfo{year}{2009}).

\bibitem[{\citenamefont{Fick and Sauermann}(1990)}]{Fick}
\bibinfo{author}{\bibfnamefont{E.}~\bibnamefont{Fick}} \bibnamefont{and}
  \bibinfo{author}{\bibfnamefont{G.}~\bibnamefont{Sauermann}},
  \emph{\bibinfo{title}{The Quantum Statistics of Dynamics Processes}}
  (\bibinfo{publisher}{Springer Verlag}, \bibinfo{address}{Berlin Heidelberg},
  \bibinfo{year}{1990}).

\bibitem[{\citenamefont{Breuer and Petruccione}(2007)}]{Breuer}
\bibinfo{author}{\bibfnamefont{H.}~\bibnamefont{Breuer}} \bibnamefont{and}
  \bibinfo{author}{\bibfnamefont{F.}~\bibnamefont{Petruccione}},
  \emph{\bibinfo{title}{The Theory of Open Quantum Systems}}
  (\bibinfo{publisher}{Oxford University Press}, \bibinfo{address}{Oxford},
  \bibinfo{year}{2007}).

\bibitem[{\citenamefont{Makhlin and Shnirman}(2004)}]{Makhlin_PRL04}
\bibinfo{author}{\bibfnamefont{Y.}~\bibnamefont{Makhlin}} \bibnamefont{and}
  \bibinfo{author}{\bibfnamefont{A.}~\bibnamefont{Shnirman}},
  \bibinfo{journal}{Phys. Rev. Lett.} \textbf{\bibinfo{volume}{92}},
  \bibinfo{pages}{178301} (\bibinfo{year}{2004}).

\bibitem[{\citenamefont{Grishin et~al.}(2005)\citenamefont{Grishin, Yurkevich,
  and Lerner}}]{Grishin_PRB05}
\bibinfo{author}{\bibfnamefont{A.}~\bibnamefont{Grishin}},
  \bibinfo{author}{\bibfnamefont{I.~V.} \bibnamefont{Yurkevich}},
  \bibnamefont{and} \bibinfo{author}{\bibfnamefont{I.~V.}
  \bibnamefont{Lerner}}, \bibinfo{journal}{Phys. Rev. B}
  \textbf{\bibinfo{volume}{72}}, \bibinfo{pages}{060509(R)}
  (\bibinfo{year}{2005}).

\bibitem[{\citenamefont{Lutchyn et~al.}(2008)\citenamefont{Lutchyn,
  Cywi{\'n}ski, Nave, and {Das Sarma}}}]{Lutchyn_PRB08}
\bibinfo{author}{\bibfnamefont{R.~M.} \bibnamefont{Lutchyn}},
  \bibinfo{author}{\bibfnamefont{{\L}.}~\bibnamefont{Cywi{\'n}ski}},
  \bibinfo{author}{\bibfnamefont{C.~P.} \bibnamefont{Nave}}, \bibnamefont{and}
  \bibinfo{author}{\bibfnamefont{S.}~\bibnamefont{{Das Sarma}}},
  \bibinfo{journal}{Phys.\ Rev.\ B} \textbf{\bibinfo{volume}{78}},
  \bibinfo{pages}{024508} (\bibinfo{year}{2008}).

\bibitem[{\citenamefont{Yang and Liu}(2008)}]{Yang_PRB08}
\bibinfo{author}{\bibfnamefont{W.}~\bibnamefont{Yang}} \bibnamefont{and}
  \bibinfo{author}{\bibfnamefont{R.~B.} \bibnamefont{Liu}},
  \bibinfo{journal}{Phys.\ Rev.\ B} \textbf{\bibinfo{volume}{77}},
  \bibinfo{pages}{085302} (\bibinfo{year}{2008}).

\bibitem[{\citenamefont{Bluhm et~al.}(2011)\citenamefont{Bluhm, Foletti, Neder,
  Rudner, Mahalu, Umansky, and Yacoby}}]{Bluhm_NP11}
\bibinfo{author}{\bibfnamefont{H.}~\bibnamefont{Bluhm}},
  \bibinfo{author}{\bibfnamefont{S.}~\bibnamefont{Foletti}},
  \bibinfo{author}{\bibfnamefont{I.}~\bibnamefont{Neder}},
  \bibinfo{author}{\bibfnamefont{M.}~\bibnamefont{Rudner}},
  \bibinfo{author}{\bibfnamefont{D.}~\bibnamefont{Mahalu}},
  \bibinfo{author}{\bibfnamefont{V.}~\bibnamefont{Umansky}}, \bibnamefont{and}
  \bibinfo{author}{\bibfnamefont{A.}~\bibnamefont{Yacoby}},
  \bibinfo{journal}{Nat. Phys.} \textbf{\bibinfo{volume}{7}},
  \bibinfo{pages}{109} (\bibinfo{year}{2011}).

\bibitem[{\citenamefont{de~Sousa and {Das Sarma}}(2003)}]{deSousa_PRB03}
\bibinfo{author}{\bibfnamefont{R.}~\bibnamefont{de~Sousa}} \bibnamefont{and}
  \bibinfo{author}{\bibfnamefont{S.}~\bibnamefont{{Das Sarma}}},
  \bibinfo{journal}{Phys. Rev. B} \textbf{\bibinfo{volume}{68}},
  \bibinfo{pages}{115322} (\bibinfo{year}{2003}).

\bibitem[{\citenamefont{Witzel et~al.}(2005)\citenamefont{Witzel, de~Sousa, and
  {Das Sarma}}}]{Witzel_PRB05}
\bibinfo{author}{\bibfnamefont{W.~M.} \bibnamefont{Witzel}},
  \bibinfo{author}{\bibfnamefont{R.}~\bibnamefont{de~Sousa}}, \bibnamefont{and}
  \bibinfo{author}{\bibfnamefont{S.}~\bibnamefont{{Das Sarma}}},
  \bibinfo{journal}{Phys.\ Rev.\ B} \textbf{\bibinfo{volume}{72}},
  \bibinfo{pages}{161306(R)} (\bibinfo{year}{2005}).

\bibitem[{\citenamefont{Witzel and {Das Sarma}}(2006)}]{Witzel_PRB06}
\bibinfo{author}{\bibfnamefont{W.~M.} \bibnamefont{Witzel}} \bibnamefont{and}
  \bibinfo{author}{\bibfnamefont{S.}~\bibnamefont{{Das Sarma}}},
  \bibinfo{journal}{Phys.\ Rev.\ B} \textbf{\bibinfo{volume}{74}},
  \bibinfo{pages}{035322} (\bibinfo{year}{2006}).

\bibitem[{\citenamefont{Witzel and {Das Sarma}}(2008)}]{Witzel_PRB08}
\bibinfo{author}{\bibfnamefont{W.~M.} \bibnamefont{Witzel}} \bibnamefont{and}
  \bibinfo{author}{\bibfnamefont{S.}~\bibnamefont{{Das Sarma}}},
  \bibinfo{journal}{Phys.\ Rev.\ B} \textbf{\bibinfo{volume}{77}},
  \bibinfo{pages}{165319} (\bibinfo{year}{2008}).

\bibitem[{\citenamefont{Witzel et~al.}(2010)\citenamefont{Witzel, Carroll,
  Morello, Cywi{\'n}ski, and {Das Sarma}}}]{Witzel_PRL10}
\bibinfo{author}{\bibfnamefont{W.~M.} \bibnamefont{Witzel}},
  \bibinfo{author}{\bibfnamefont{M.~S.} \bibnamefont{Carroll}},
  \bibinfo{author}{\bibfnamefont{A.}~\bibnamefont{Morello}},
  \bibinfo{author}{\bibfnamefont{{\L}.}~\bibnamefont{Cywi{\'n}ski}},
  \bibnamefont{and} \bibinfo{author}{\bibfnamefont{S.}~\bibnamefont{{Das
  Sarma}}}, \bibinfo{journal}{Phys.\ Rev.\ Lett.}
  \textbf{\bibinfo{volume}{105}}, \bibinfo{pages}{187602}
  (\bibinfo{year}{2010}).

\bibitem[{\citenamefont{Abe et~al.}(2010)\citenamefont{Abe, Tyryshkin, Tojo,
  Morton, Witzel, Fujimoto, Ager, Haller, Isoya, Lyon et~al.}}]{Abe_PRB10}
\bibinfo{author}{\bibfnamefont{E.}~\bibnamefont{Abe}},
  \bibinfo{author}{\bibfnamefont{A.~M.} \bibnamefont{Tyryshkin}},
  \bibinfo{author}{\bibfnamefont{S.}~\bibnamefont{Tojo}},
  \bibinfo{author}{\bibfnamefont{J.~J.~L.} \bibnamefont{Morton}},
  \bibinfo{author}{\bibfnamefont{W.~M.} \bibnamefont{Witzel}},
  \bibinfo{author}{\bibfnamefont{A.}~\bibnamefont{Fujimoto}},
  \bibinfo{author}{\bibfnamefont{J.~W.} \bibnamefont{Ager}},
  \bibinfo{author}{\bibfnamefont{E.~E.} \bibnamefont{Haller}},
  \bibinfo{author}{\bibfnamefont{J.}~\bibnamefont{Isoya}},
  \bibinfo{author}{\bibfnamefont{S.~A.} \bibnamefont{Lyon}},
  \bibnamefont{et~al.}, \bibinfo{journal}{Phys.\ Rev.\ B}
  \textbf{\bibinfo{volume}{82}}, \bibinfo{pages}{121201}
  (\bibinfo{year}{2010}).

\bibitem[{\citenamefont{George et~al.}(2010)\citenamefont{George, Witzel,
  Riemann, Abrosimov, N\"otzel, Thewalt, and Morton}}]{George_PRL10}
\bibinfo{author}{\bibfnamefont{R.~E.} \bibnamefont{George}},
  \bibinfo{author}{\bibfnamefont{W.}~\bibnamefont{Witzel}},
  \bibinfo{author}{\bibfnamefont{H.}~\bibnamefont{Riemann}},
  \bibinfo{author}{\bibfnamefont{N.~V.} \bibnamefont{Abrosimov}},
  \bibinfo{author}{\bibfnamefont{N.}~\bibnamefont{N\"otzel}},
  \bibinfo{author}{\bibfnamefont{M.~L.~W.} \bibnamefont{Thewalt}},
  \bibnamefont{and} \bibinfo{author}{\bibfnamefont{J.~J.~L.}
  \bibnamefont{Morton}}, \bibinfo{journal}{Phys.\ Rev.\ Lett.}
  \textbf{\bibinfo{volume}{105}}, \bibinfo{pages}{067601}
  (\bibinfo{year}{2010}).

\bibitem[{\citenamefont{Latta et~al.}(2011)\citenamefont{Latta, Srivastava, and
  Imamoglu}}]{Latta_arxiv11}
\bibinfo{author}{\bibfnamefont{C.}~\bibnamefont{Latta}},
  \bibinfo{author}{\bibfnamefont{A.}~\bibnamefont{Srivastava}},
  \bibnamefont{and} \bibinfo{author}{\bibfnamefont{A.}~\bibnamefont{Imamoglu}},
  \bibinfo{journal}{arXiv:1104.1111}  (\bibinfo{year}{2011}).

\bibitem[{\citenamefont{Coish et~al.}(2007)\citenamefont{Coish, Loss,
  Yuzbashyan, and Altshuler}}]{Coish_JAP07}
\bibinfo{author}{\bibfnamefont{W.~A.} \bibnamefont{Coish}},
  \bibinfo{author}{\bibfnamefont{D.}~\bibnamefont{Loss}},
  \bibinfo{author}{\bibfnamefont{E.~A.} \bibnamefont{Yuzbashyan}},
  \bibnamefont{and} \bibinfo{author}{\bibfnamefont{B.~L.}
  \bibnamefont{Altshuler}}, \bibinfo{journal}{J.\ Appl.\ Phys.}
  \textbf{\bibinfo{volume}{101}}, \bibinfo{pages}{081715}
  (\bibinfo{year}{2007}).

\bibitem[{\citenamefont{Melikidze et~al.}(2004)\citenamefont{Melikidze,
  Dobrovitski, {De Raedt}, Katsnelson, and Harmon}}]{Melikidze_PRB04}
\bibinfo{author}{\bibfnamefont{A.}~\bibnamefont{Melikidze}},
  \bibinfo{author}{\bibfnamefont{V.~V.} \bibnamefont{Dobrovitski}},
  \bibinfo{author}{\bibfnamefont{H.~A.} \bibnamefont{{De Raedt}}},
  \bibinfo{author}{\bibfnamefont{M.~I.} \bibnamefont{Katsnelson}},
  \bibnamefont{and} \bibinfo{author}{\bibfnamefont{B.~N.}
  \bibnamefont{Harmon}}, \bibinfo{journal}{Phys.\ Rev.\ B}
  \textbf{\bibinfo{volume}{70}}, \bibinfo{pages}{014435}
  (\bibinfo{year}{2004}).

\bibitem[{\citenamefont{Krovi et~al.}(2007)\citenamefont{Krovi, Oreshkov,
  Ryazanov, and Lidar}}]{Krovi_PRA07}
\bibinfo{author}{\bibfnamefont{H.}~\bibnamefont{Krovi}},
  \bibinfo{author}{\bibfnamefont{O.}~\bibnamefont{Oreshkov}},
  \bibinfo{author}{\bibfnamefont{M.}~\bibnamefont{Ryazanov}}, \bibnamefont{and}
  \bibinfo{author}{\bibfnamefont{D.~A.} \bibnamefont{Lidar}},
  \bibinfo{journal}{Phys.\ Rev.\ A} \textbf{\bibinfo{volume}{76}},
  \bibinfo{pages}{052117} (\bibinfo{year}{2007}).

\bibitem[{\citenamefont{Merkulov et~al.}(2002)\citenamefont{Merkulov, Efros,
  and Rosen}}]{Merkulov_PRB02}
\bibinfo{author}{\bibfnamefont{I.~A.} \bibnamefont{Merkulov}},
  \bibinfo{author}{\bibfnamefont{A.~L.} \bibnamefont{Efros}}, \bibnamefont{and}
  \bibinfo{author}{\bibfnamefont{M.}~\bibnamefont{Rosen}},
  \bibinfo{journal}{Phys. Rev. B} \textbf{\bibinfo{volume}{65}},
  \bibinfo{pages}{205309} (\bibinfo{year}{2002}).

\bibitem[{\citenamefont{Doll et~al.}(2008)\citenamefont{Doll, Zueco, Wubs,
  Kohler, and Hanggi}}]{Doll_CP08}
\bibinfo{author}{\bibfnamefont{R.}~\bibnamefont{Doll}},
  \bibinfo{author}{\bibfnamefont{D.}~\bibnamefont{Zueco}},
  \bibinfo{author}{\bibfnamefont{M.}~\bibnamefont{Wubs}},
  \bibinfo{author}{\bibfnamefont{S.}~\bibnamefont{Kohler}}, \bibnamefont{and}
  \bibinfo{author}{\bibfnamefont{P.}~\bibnamefont{Hanggi}},
  \bibinfo{journal}{Chem.~Phys.} \textbf{\bibinfo{volume}{347}},
  \bibinfo{pages}{243} (\bibinfo{year}{2008}).

\bibitem[{\citenamefont{Dobrovitski et~al.}(2001)\citenamefont{Dobrovitski, {De
  Raedt}, Katsnelson, and Harmon}}]{Dobrovitski_01}
\bibinfo{author}{\bibfnamefont{V.~V.} \bibnamefont{Dobrovitski}},
  \bibinfo{author}{\bibfnamefont{H.~A.} \bibnamefont{{De Raedt}}},
  \bibinfo{author}{\bibfnamefont{M.~I.} \bibnamefont{Katsnelson}},
  \bibnamefont{and} \bibinfo{author}{\bibfnamefont{B.~N.}
  \bibnamefont{Harmon}}, \bibinfo{journal}{arXiv:quant-ph/0112053}
  (\bibinfo{year}{2001}).

\bibitem[{\citenamefont{Taylor and Lukin}(2006)}]{Taylor_QIP06}
\bibinfo{author}{\bibfnamefont{J.~M.} \bibnamefont{Taylor}} \bibnamefont{and}
  \bibinfo{author}{\bibfnamefont{M.~D.} \bibnamefont{Lukin}},
  \bibinfo{journal}{Quant.~Info.~Process.} \textbf{\bibinfo{volume}{5}},
  \bibinfo{pages}{503} (\bibinfo{year}{2006}).

\bibitem[{\citenamefont{Koppens et~al.}(2007)\citenamefont{Koppens, Klauser,
  Coish, Nowack, Kouwenhoven, Loss, and Vandersypen}}]{Koppens_PRL07}
\bibinfo{author}{\bibfnamefont{F.~H.~L.} \bibnamefont{Koppens}},
  \bibinfo{author}{\bibfnamefont{D.}~\bibnamefont{Klauser}},
  \bibinfo{author}{\bibfnamefont{W.~A.} \bibnamefont{Coish}},
  \bibinfo{author}{\bibfnamefont{K.~C.} \bibnamefont{Nowack}},
  \bibinfo{author}{\bibfnamefont{L.~P.} \bibnamefont{Kouwenhoven}},
  \bibinfo{author}{\bibfnamefont{D.}~\bibnamefont{Loss}}, \bibnamefont{and}
  \bibinfo{author}{\bibfnamefont{L.~M.~K.} \bibnamefont{Vandersypen}},
  \bibinfo{journal}{Phys.\ Rev.\ Lett.} \textbf{\bibinfo{volume}{99}},
  \bibinfo{pages}{106803} (\bibinfo{year}{2007}).

\bibitem[{\citenamefont{Hanson et~al.}(2008)\citenamefont{Hanson, Dobrovitski,
  Feiguin, Gywat, and Awschalom}}]{Hanson_Science08}
\bibinfo{author}{\bibfnamefont{R.}~\bibnamefont{Hanson}},
  \bibinfo{author}{\bibfnamefont{V.~V.} \bibnamefont{Dobrovitski}},
  \bibinfo{author}{\bibfnamefont{A.~E.} \bibnamefont{Feiguin}},
  \bibinfo{author}{\bibfnamefont{O.}~\bibnamefont{Gywat}}, \bibnamefont{and}
  \bibinfo{author}{\bibfnamefont{D.~D.} \bibnamefont{Awschalom}},
  \bibinfo{journal}{Science} \textbf{\bibinfo{volume}{320}},
  \bibinfo{pages}{352} (\bibinfo{year}{2008}).

\end{thebibliography}

\end{document}